\documentclass[twocolumn]{autart}
\usepackage{graphicx}
\usepackage{amsmath,amssymb,amscd}
\usepackage{bm}
\usepackage{float}
\usepackage{boondox-cal}
\usepackage{threeparttable}
\usepackage{color}
\usepackage{hyperref}
\usepackage{stfloats}
\usepackage{subfigure}
\newtheorem{theorem}{Theorem}
\newtheorem{proposition}{Proposition}
\newtheorem{lemma}{Lemma}
\newtheorem{corollary}{Corollary}
\newtheorem{problem}{\indent Problem}
\begin{document}

\begin{frontmatter}

\title{  Optimal and two-step adaptive quantum detector tomography \thanksref{footnoteinfo}} 

\thanks[footnoteinfo]{ This research was supported by the National Natural Science Foundation
of China (62173229),  the Australian Government via AUSMURI Grant No. AUSMURI000002 and the Australian Research Council Discovery Projects Funding Scheme under
Project DP190101566. The material in this paper was partially
presented at the 60th IEEE conference on Decision and Control, Austin, Texas, USA, December 13-17, 2021.}

\author[1,3]{Shuixin Xiao}\ead{xiaoshuixin@sjtu.edu.cn},    
\author[2]{Yuanlong Wang}\ead{yuanlong.wang.qc@gmail.com}, 
\author[3,a]{Daoyi Dong}\ead{daoyidong@gmail.com},             
\author[1,a]{Jun Zhang}\ead{zhangjun12@sjtu.edu.cn}

\thanks[a]{Corresponding author.}

\address[1]{UMich-SJTU Joint Institute, Shanghai Jiao Tong
	University, Shanghai 200240, China}   
\address[2]{Centre for Quantum Computation and Communication Technology (Australian Research Council), Centre for Quantum Dynamics, Griffith University, Brisbane, Queensland 4111, Australia}
\address[3]{School of Engineering and Information Technology, University of New South Wales, Canberra ACT 2600, Australia}


\begin{abstract}  
Quantum detector tomography is a fundamental technique for calibrating quantum devices and
performing quantum engineering tasks. In this paper,  we design optimal probe states for  detector estimation based on the minimum upper bound of the mean squared error (UMSE) and the maximum robustness. We establish the minimum UMSE and the minimum condition number for quantum detectors and provide concrete examples that can achieve optimal detector tomography. In order to enhance the estimation precision,  we also propose a two-step adaptive detector tomography algorithm to optimize the probe states adaptively based on a modified fidelity index. We present a sufficient condition on when the estimation error of our two-step strategy scales inversely proportional to the number of state copies. Moreover, the superposition of coherent states is used as probe states for quantum detector tomography and the estimation error is analyzed. Numerical results demonstrate the effectiveness of both the proposed optimal and adaptive quantum detector tomography methods. 
\end{abstract}

\end{frontmatter}

\endNoHyper

\section{Introduction}
In the past decades, we have witnessed significant progress in a variety of fields of quantum science and technology, including quantum computation \cite{DiVincenzo255}, quantum communication \cite{qci} and quantum sensing \cite{RevModPhys.89.035002}. In these applications, a fundamental task is to develop efficient estimation and identification methods to acquire information of quantum states, system parameters and quantum detectors. There are three typical classes of quantum estimation and identification tasks: (i) quantum state tomography (QST) which aims to estimate unknown states \cite{Qi2013,Hou2016,MU2020108837}; (ii) quantum process tomography which targets in identifying parameters of evolution operators \cite{WANG2019269,PhysRevA.63.020101,9283060,YU2021109612,xiao} (e.g., the system Hamiltonian \cite{8022944,9026783,PhysRevLett.113.080401,PhysRevA.95.022335,zhang2,PhysRevA.96.062334}); and (iii) quantum detector tomography (QDT) which aims to identify and calibrate quantum measurement devices. In this paper, we focus on QDT and aim to present optimal and adaptive strategies for enhancing the efficiency and precision of QDT.

For general QDT protocols, the first solution was proposed in \cite{PhysRevA.64.024102} using maximum likelihood estimation (MLE). Subsequent works divide quantum detectors into  phase-insensitive detectors and phase-sensitive detectors. Phase-insensitive detectors only have  diagonal elements in the photon number basis, and therefore are relatively straightforward to be characterized using linear regression \cite{Grandi_2017}, function fitting  \cite{Renema:12}, or convex optimization \cite{Feito_2009,Lundeen2009,natarajan2013quantum}. In experiment, a regularized least-square method was  used in \cite{add1,add2} for phase-insensitive detectors. For phase-sensitive detectors, non-diagonal elements can be nonzero and they are usually more challenging to reconstruct. The work in \cite{zhang2012mapping,Zhang_2012} formulated QDT as a convex quadratic optimization problem for this type of detectors.  Ref. \cite{wang2019twostage} proposed a two-stage solution with an analytical computational complexity and error upper bound. Ref. \cite{9029759} further studied a binary detector tomography method with lower computational complexity by projection. 
Self-characterization of one-qubit QDT was proposed in \cite{PhysRevLett.124.040402} which does not rely on  precisely calibrated probe states.

	\begin{figure*}[ht]
	\centering
	\includegraphics[width=0.8\linewidth]{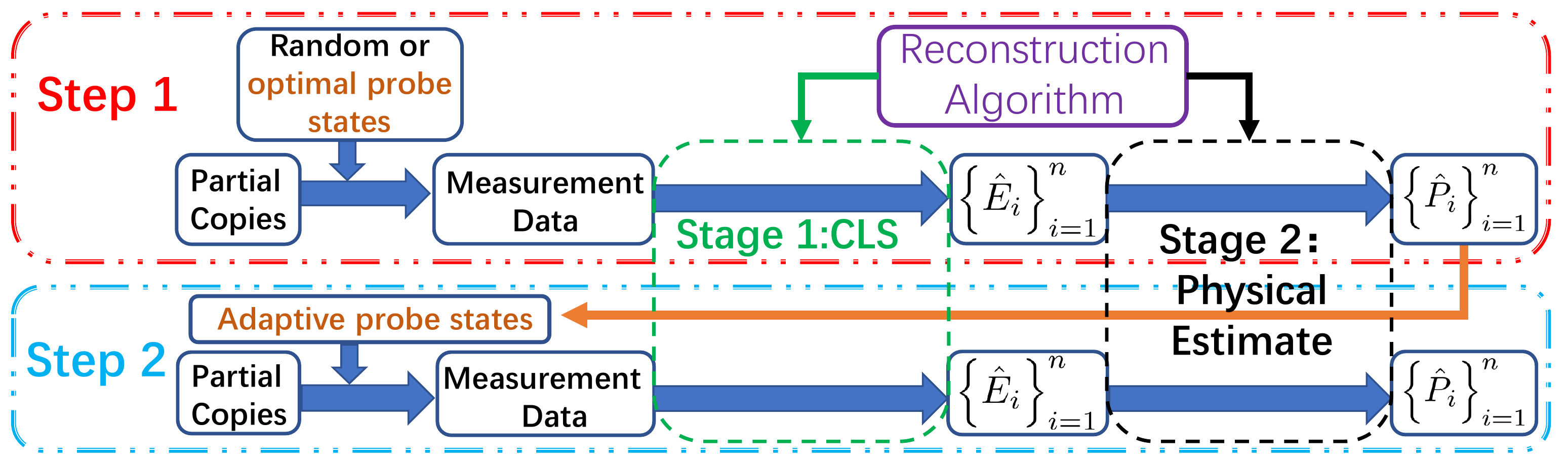}
	\caption{General procedures for optimal detector tomography and adaptive detector tomography. For optimal detector tomography, we only consider Step 1 and it is non-adaptive.  We focus on finding the optimal probe states. For adaptive detector tomography, we choose adaptive probe states in Step 2 based on Step 1 estimation to obtain a more accurate estimate.}
	\label{algorithm}
\end{figure*}

In this paper,  we consider optimal QDT and adaptive QDT which are applicable to both phase insensitive and phase sensitive detectors. The general tomography procedures are shown in Fig. \ref{algorithm}. For optimal QDT, we only consider Step 1 where we obtain measurement data and then use the two-stage reconstruction algorithm in \cite{wang2019twostage} to identify the unknown detectors. A natural question
would be which  input probe states are optimal, according to some
requirements or criteria. Here we use the upper bound of the mean squared error (UMSE) \cite{wang2019twostage} and the robustness described by the condition number  against measurement errors as two criteria \cite{cdc}. We prove that the minimum UMSE  is $\frac{(n-1)(d^{4}+ d^3-d^{2})}{4 N}  $ where $d$ is the dimension of detector matrices, $ n $ is the number of detector matrices and $ N $ is the resource number (i.e., number of copies of probe states). We also prove the minimum condition number is $ \sqrt{d+1} $. We then provide two examples of optimal probe states--SIC (symmetric informationally complete) states with the smallest $ M=d^2 $ and MUB (mutually unbiased) states for $ M=d(d+1) $  where $ M $ is the type of probe states. When restricted to  product states, we prove the minimum of UMSE is $ \frac{20^m(n-1)}{4N}$ and that of the condition number is $ 3^\frac{m}{2} $ where $ m $ is the number of qubits.

Another focus we consider is to develop adaptive QDT to enhance the identification accuracy of quantum detectors. Adaptive strategies have been employed in QST and existing results show that adaptive QST has great potential to enhance the estimation precision of quantum states \cite{b1,b2,b3,PhysRevLett.111.183601,Qi2017,PhysRevA.98.012339}. Inspired by adaptive QST, we propose a two-step adaptive QDT as shown in Fig. \ref{algorithm}. Step 1 is the same as optimal QDT, and adaptive probe states are chosen in Step 2 based on  a rough detector estimation in Step 1. With these adaptive probe states, we can improve the infidelity  from $O(1 / \sqrt{N})$ to optimal value $O(1 / N)$. We also use the superposition of coherent states to realize QDT and use numerical examples to demonstrate the effectiveness of our optimal and adaptive QDT.

This paper is organized as follows. In Section 2, we present background knowledge and  reconstruction algorithm.  In Section 3, we propose optimal QDT and provide concrete examples. In Section 4, we present a two-step adaptive QDT. In Section 5, we give numerical examples of optimal and adaptive QDT. Conclusions are presented in Section 6.

Notation: For a matrix $ A $, $A \geq 0$ means $A$ is positive semidefinite.  The conjugation and transpose $(T)$
of $A$ is $A^{\dagger}$. The trace
of $A$ is $\operatorname{Tr}(A)$. The identity matrix is $I$. The real and complex domains are $\mathbb{R}$ and $\mathbb{C}$, respectively. The
tensor product is $\otimes$. The set of all $d$-dimensional
complex/real vectors is $ \mathbb{C}^{d}/\mathbb{R}^{d} $. Row and column vectors also denoted as $  \langle\psi| $ and $|\psi\rangle$, respectively. The Frobenius norm for matrix and 2-norm for vector are $\|\cdot\|$.  The Kronecker delta function is $\delta$. $\mathrm{i}=\sqrt{-1}$. The column vectorization function is $\operatorname{vec}$. The diagonal elements of 
a diagonal matrix $\operatorname{diag}(X)$  are the elements in $X$ and $X$ is a vector.   The estimation of
variable $X$ denotes $\hat{X}$. For any positive semidefinite $X_{d \times d}$ with spectral decomposition $X=U P U^{\dagger},$ define $\sqrt{X}$ or $X^{\frac{1}{2}}$ as $U \operatorname{diag}\left(\sqrt{P_{11}}, \sqrt{P_{22}}, \ldots, \sqrt{P_{d d}}\right) U^{\dagger}$.  Pauli matrices are $ \sigma_{x}=\left(\begin{array}{ll}
0 & 1 \\
1 & 0
\end{array}\right) $, $ \sigma_y=\left(\begin{array}{cc}
0 & -\mathrm i \\
\mathrm i & 0
\end{array}\right)$, $ \sigma_z=\left(\begin{array}{cc}
1 & 0 \\
0 & -1
\end{array}\right) $. The fidelity between quantum states $ \rho $ and $\hat{\rho}  $ is $ F_s(\hat{\rho},\rho)=[\operatorname{Tr} \sqrt{\sqrt{\hat{\rho}} \rho \sqrt{\hat{\rho}}}]^{2} $.
\section{Preliminaries}
Here we present the background knowledge and  briefly introduce the two-stage QDT reconstruction algorithm in \cite{wang2019twostage}, which will be employed as a critical part for developing optimal QDT.

\subsection{Quantum state and measurement}
For a $d$-dimensional quantum system, its state can be described by a $d \times d$ Hermitian
matrix $\rho$, satisfying $\rho\geq0,\operatorname{Tr}(\rho)=1$. When $\rho=|\psi\rangle\langle\psi|$ for some $|\psi\rangle\in\mathbb{C}^{d}$, we call $ \rho $ a pure state, and its purity $ \operatorname{Tr}(\rho^2) $ reaches the maximum value $ 1 $.  Otherwise, $\rho$ is called a
mixed state, and can be represented using pure states $\left\{\left|\psi_{i}\right\rangle\right\}: \rho=\sum_{i} c_{i}\left|\psi_{i}\right\rangle\left\langle\psi_{i}\right|$ where $c_{i} \in \mathbb{R}$ and
$\sum_{i} c_{i}=1 $.

A quantum detector connects the classical and quantum world through a set of operators known as positive-operator-valued measure (POVM). A set of POVM elements is a set of Hermitian and positive semidefinite operators $\left\{P_{i}\right\}$, which is the mathematical representation of quantum detectors, satisfying the completeness
constraint $\sum_{i} P_{i}=I$. We directly call $\left\{P_{i}\right\}$ a \emph{POVM element} in this paper. The operators $ P_i $ may be finite or infinite dimensional in theory. When infinite dimensional, we usually truncate them at a finite dimension $d$ in practice. When $ \rho $ is measured using $\left\{P_{i}\right\}$, the probability of obtaining the $i$-th result is given by the Born's rule
\begin{equation}
	p_{i}=\operatorname{Tr}\left(P_{i} \rho\right).
\end{equation}
Because of the completeness constraint, we have $\sum_{i} p_{i}=1$. In practical experiments, suppose $ N $ (called  \emph{resource number})   identical copies of $\rho$  are prepared and the $i$-th result occurs $N_{i}$ times. Then $\hat p_{i}=N_{i} / N$ is the experimental estimation of the true value $p_{i}$. 

\subsection{Problem formulation of QDT}
By applying known quantum states to an unknown quantum detector and obtaining the measurement results, one can estimate the detector, which is called quantum detector tomography (QDT). Denote the true value of the detector
as $\left\{P_{i}\right\}_{i=1}^{n}$. We design $ M $ different types of quantum states (called \emph{probe states}), where each state $ \rho_j $ use the same resource number $ N/M $ thus their total number of copies is $ N $.
 One can formulate the problem of QDT as \cite{wang2019twostage}:
\begin{problem}\label{problem1}
	Given experimental data $\left\{\hat{p}_{i j}\right\}$, solve $$ \min _{\left\{\hat{P}_{i}\right\}_{i=1}^n} \sum_{i=1}^{n} \sum_{j=1}^{M}\left[\hat{p}_{i j}-\operatorname{Tr}\left(\hat{P}_{i} \rho_{j}\right)\right]^{2} $$ such
	that $\hat P_i=\hat P_i^{\dagger},\hat{P}_{i} \geq 0 $  for  $1 \leq i \leq n$ and $\sum_{i=1}^{n} \hat{P}_{i}=I$.
\end{problem}

There are various methods to formulate and solve the problem of QDT. For maximum likelihood estimation (MLE) \cite{PhysRevA.64.024102}, if large data are given, the MLE can asymptotically reach the Cram$\acute{e}$r-Rao bound for parameter estimation while the computational complexity is usually high. There also exist several convex optimization approaches \cite{Feito_2009,Lundeen2009,natarajan2013quantum} which might be more  efficient. However, an analytical error upper bound is missing for MLE and convex optimization approaches, making it difficult to optimize the input probe states in the general case without available prior information about the detector. Here we choose the linear regression method due to its simplicity and computational efficiency. Moreover, Ref. \cite{wang2019twostage} presents the two-stage QDT solution to Problem 1, giving an analytical error upper bound in favor of optimizing general input states. We thus review this solving procedure as follows.

 Let $\left\{\Omega_{i}\right\}_{i=1}^{d^{2}}$ be a complete
basis set of orthonormal operators with $d$-dimension. Without loss of
generality, let $\operatorname{Tr}\left(\Omega_{i}^{\dagger} \Omega_{j}\right)=\delta_{i j}$, $\Omega_{i}=\Omega_{i}^{\dagger}$ where $\operatorname{Tr}\left(\Omega_{i}\right)=0$ except $ \Omega_{1}=I / \sqrt{d} $. Then we parameterize the detector
and probe states as
\begin{equation}
	\label{para}
	\begin{aligned}
		P_{i} &=\sum_{a=1}^{d^{2}} \theta_{i}^{a} \Omega_{a}, \\
		\rho_{j} &=\sum_{b=1}^{d^{2}} \phi_{j}^{b} \Omega_{b},
	\end{aligned}
\end{equation}
where $\theta_{i}^{a} \triangleq \operatorname{Tr}\left(P_{i} \Omega_{a}\right)$ and $\phi_{j}^{b} \triangleq \operatorname{Tr}\left(\rho_{j} \Omega_{b}\right)$ are real. Using Born's rule, we can obtain
\begin{equation}
	p_{i j}=\sum_{a=1}^{d^{2}} \phi_{j}^{a} \theta_{i}^{a} \triangleq \phi_{j}^{T} \theta_{i},
\end{equation}
where $\phi_{j} $, $ \theta_{i} $ is the parameterization of  $ \rho_{j} $ and   $ P_i $, respectively.
Suppose the outcome for $P_{i}$ appears $n_{i j}$ times, then $\hat{p}_{i j}=n_{i j} /(N / M) $. Denote
the estimation error as $e_{i j}=\hat{p}_{i j}-p_{i j} $. According to the central limit theorem, $ e_{ij} $ converges in distribution
to a normal distribution with mean zero and variance $ \left(p_{i j}-p_{i j}^{2}\right) /(N / M) $. We thus have the linear regression equation 
\begin{equation}
	\hat{p}_{i j}=\phi_{j}^{T} \theta_{i}+e_{i j}.
\end{equation}
Let $\Theta=\left(\theta_{1}^{T}, \theta_{2}^{T}, \ldots, \theta_{n}^{T}\right)^{T}$ be the vector of all the unknown parameters to be estimated. Collect the parameterization of the probe states as $X=\left(\phi_{1}, \phi_{2}, \ldots, \phi_{M}\right)^{T}$. Let $\hat{\mathcal y}=$
$(\hat{p}_{11}, \hat{p}_{12}, \ldots, \hat{p}_{1 M}, \hat{p}_{21}, \hat{p}_{22}$, $\ldots, \hat{p}_{2 M}, \ldots, \hat{p}_{n M})^{T}$, $\mathcal X=I_{n} \otimes X$, $\mathcal{e}=(e_{11}, e_{12}, \ldots, e_{1 M}$, $e_{21}, e_{22}, \ldots, e_{2 M}, \ldots, e_{n M})^{T}$,
$\mathcal H=(1,1, \ldots, 1)_{1 \times n} \otimes I_{d^{2}}, \mathcal d_{d^{2} \times 1}=(\sqrt{d}, 0, \ldots, 0)^{T} $. Then the regression equations can be rewritten
in a compact form \cite{wang2019twostage}:
\begin{equation}\label{cls2}
	\hat{\mathcal y}=\mathcal X \Theta+\mathcal e,
\end{equation}
with a linear constraint
\begin{equation}
	\mathcal H \Theta=\mathcal d.
\end{equation}
Now Problem 1 can be transformed into the following equivalent form:
\begin{problem}
	Given experimental data $ \hat{\mathcal y} $, solve \\$ \min _{\left\{\hat{P}_{i}\right\}_{i=1}^n}\|\hat{\mathcal y}-\mathcal X \hat{\Theta}\|^{2}  $ such that $\mathcal H \hat{\Theta}=\mathcal d$ and $\hat{P}_{i} \geq 0 $ for $ 1 \leq i \leq n $, where $ \hat{\Theta} $ is the parametrization of $ \left\{\hat{P}_{i}\right\} $.
\end{problem}
In \cite{wang2019twostage}, Problem 2 is split  into two approximate subproblems:
\addtocounter{problem}{-1}
\renewcommand{\theproblem}{\arabic{problem}$.\bm{1}$}
\begin{problem}
	\label{pro1}
	Given experimental data $ \hat{\mathcal y} $, solve $ \min _{\left\{\hat{E}_{i}\right\}_{i=1}^n}\|\hat{\mathcal y}-\mathcal X \hat{\Theta}\|^{2}  $ such that $\mathcal H \hat{\Theta}=\mathcal d$ where $ \hat{\Theta} $ is the parametrization of $ \left\{\hat{E}_{i}\right\} $.
\end{problem}
\renewcommand{\theproblem}{\arabic{problem}}
\addtocounter{problem}{-1}
\renewcommand{\theproblem}{\arabic{problem}$.\bm{2}$}
\begin{problem}
	\label{pro2}
	Given $ \sum_{i=1}^{n} \hat{E}_{i}=I $, solve \\
	$ \min _{\left\{\hat{P}_{i}\right\}_{i=1}^{n}} \sum_{i=1}^{n}\left\|\hat{E}_{i}-\hat{P}_{i}\right\|^{2} $ such that $ \sum_{i=1}^{n} \hat{P}_{i}=I $ and $ \hat{P}_{i} \geq 0 $ for $ 1 \leq i \leq n $.
\end{problem}

The reconstruction algorithm has two stages, as shown in Fig.~\ref{algorithm}. In Stage 1,  Problem 2.1 is solved by Constrained Least Squares (CLS) method and the form of the solution is further simplified.
In Stage 2,  Problem 2.2 is solved by matrix decomposition. We briefly review the two-stage QDT reconstruction algorithm in \cite{wang2019twostage} in the next subsection.
\subsection{Two-stage QDT reconstruction algorithm}
In Stage 1, we directly give the simplified form of the  CLS solution to Problem 2.1 as
\begin{equation}
	\label{big}
	\hat{\Theta}_\text{CLS}
	=\left(\begin{array}{c}
		\left(X^{T} X\right)^{-1} X^{T}\left(\hat{y}_{1}-\frac{1}{n} y_{0}\right)+\frac{1}{n} \mathcal d \\
		\vdots \\
		\left(X^{T} X\right)^{-1} X^{T}\left(\hat{y}_{n}-\frac{1}{n} y_{0}\right)+\frac{1}{n} \mathcal d
	\end{array}\right),
\end{equation}
where $\hat{y}_{i}=\left(\hat{p}_{i 1}, \hat{p}_{i 2}, \ldots, \hat{p}_{i M}\right)^{T}  $ for $1 \leq i \leq n  $ and $ y_{0}=\left((1, \ldots, 1)_{1 \times M}\right)^{T}=\sum_{i} \hat{y}_{i} $.
Let the $ i $-th block be 
 $ \hat{\Theta}_{i,\text{CLS}}=\left(\hat{\theta}_{i}^{1}, \ldots, \hat{\theta}_{i}^{d^2}\right)^{T} $ and the Stage 1 estimate is $\hat E_i=\sum_{a=1}^{d^2}\hat \theta_i^{a}\Omega_a$ which  may have negative eigenvalues. Thus, Stage 2 is designed to obtain a physical estimate $\hat{P_{i}}$ by eigenvalue correction with three substages.
Firstly, we decompose $ \hat{E}_{i}=\hat{F}_{i}-\hat{G}_{i} $ with $ \hat{F}_{i} \geq 0, \hat{G}_{i} \geq 0 $. We then perform a spectral decomposition to obtain $ \hat{E}_{i}=\hat{R}_{i} \hat{K}_{i} \hat{R}_{i}^{\dagger} $. Assume there are $\hat{n}_{i}$ nonpositive eigenvalues for $\hat{E}_{i}$, and they are in decreasing order. Thus, the best decomposition in the sense of minimizing $ \|\hat{G}_{i}\| $ is
$\hat F_i=\hat R_i\text{diag}((\hat K_i)_{11},(\hat K_i)_{22},...,(\hat K_i)_{(d-\hat n_i)(d-\hat n_i)},0,...,0)\hat R_i^\dagger$, and
		$\hat G_i=-\hat R_i\text{diag}(0,...,0,(\hat K_i)_{(d-\hat n_i+1)(d-\hat n_i+1)}$, \\$(\hat K_i)_{(d-\hat n_i+2)(d-\hat n_i+2)},...,(\hat K_i)_{dd})\hat R_i^\dagger$.
Secondly, we apply decomposition
$I+\sum_{i} \hat{G}_{i}=\sum_{i} \hat{F}_{i}=\hat{C}\hat{C}^{\dagger}$,
and hence $ \sum_{i} \hat{C}^{-1} \hat{F}_{i} \hat{C}^{-\dagger}=I $.
Let $ \hat{B}_{i}=\hat{C}^{-1} \hat{F}_{i} \hat{C}^{-\dagger} $ which is positive semidefinite and  $ \sum_{i} \hat{B}_{i}=I $. For any unitary $\hat{U}$,  $\hat{C} \hat{C}^{\dagger}=\hat{C} \hat{U} \hat{U}^{\dagger} \hat{C}^{\dagger}$ holds. Therefore, $\hat{U}^{\dagger} \hat{B}_{i} \hat{U}$ can also be an estimate
of the detector. To neutralize the effect of $\hat{U}^{\dagger} \hat{C}^{-1}(\cdot)(\hat{C}^{\dagger})^{-1} \hat{U}$ on $\hat{F}_{i}$, we define an optimal unitary matrix $\hat{U}_{\text{op}}$ to minimize $\|\hat{C} \hat{U}_{\text{op}}-I\|$ \cite{wang2019twostage} and its analytical expression is $ \hat{U}_{\text{op}}=\sqrt{\hat{C}^{\dagger} \hat{C}} \hat{C}^{-1} $.
Therefore, the final estimate is $ \hat{P}_{i}=\hat{U}_{\text{op}}^{\dagger} \hat{B}_{i} \hat{U}_{\text{op}} $. This general  algorithm can be used for arbitrary $ n\geq2 $.

\section{Optimal quantum detector tomography}

In detector tomography as shown in Step 1 in Fig.~\ref{algorithm}, a natural way to reduce the tomography error is to carefully choose the probe states according to a certain optimality criterion, which should be independent of the detector. In this section, we propose a criterion based on two non-conflicting indices (UMSE and condition number), subsequently present the  conditions on achieving optimal detector tomography and provide illustrative examples.
\subsection{Optimality criterion}

One index to evaluate probe states is to score their worst performance, i.e., their upper bounds of estimation errors. Let $ \mathbb{E} $ denote the expectation w.r.t. all possible measurement
results.  The Stage 1 error $\left\|\hat{E}_{i}-P_{i}\right\|$ between the estimate $\hat{E}_{i}$ and its true value $P_{i}$ is bounded by  UMSE (upper bound of the mean squared error )  \cite{wang2019twostage}
\begin{equation}
	\label{upmse}
	\begin{aligned}
		\mathbb{E}\left(\sum_{i}\left\|\hat{E}_{i}-P_{i}\right\|^{2}\right) 
		&=\mathbb{E}\left(\left\|\hat{\Theta}_{\text{CLS}}-\Theta\right\|^{2}\right) \\
		& \leq \frac{(n-1) M}{4 N} \operatorname{Tr}\big[(X^{T} X)^{-1}\big],
	\end{aligned}
\end{equation}
and  the final  estimation error for all detectors \\ $ \mathbb{E}\left(\Sigma_{i}\left\|\hat{P}_{i}-P_{i}\right\|^{2}\right) $   is bounded by  \cite{wang2019twostage}
\begin{equation}
	\label{final}
	\begin{aligned}
		& \mathbb{E}\left(\sum_{i}\left\|\hat{P}_{i}-P_{i}\right\|^{2}\right) \\
		\leq & \frac{(d n+2 \sqrt{d} n+1)(n-1) M}{4 N} \operatorname{Tr}\big[(X^{T} X)^{-1}\big]+o\left(\frac{1}{N}\right).
	\end{aligned}
\end{equation}
Since in \eqref{upmse} and \eqref{final} the parts  dependent on probe states are both $ M\operatorname{Tr}\big[(X^{T} X)^{-1}\big] $, we take it as our first criterion. For general QDT, we have no special prior knowledge on the detectors, and the following conditions are equivalent: (i) the detector can not be uniquely identified; (ii)  the probe states do not span the space of all $d$-dimensional states; (iii) $X^TX$ is singular; (iv) the UMSE is infinite. From (i) and (iv), it is thus reasonable to take $M\operatorname{Tr}(X^TX)^{-1}$ as an evaluation index. We require the optimal probe states to minimize  $M\operatorname{Tr}(X^TX)^{-1}$.   This criterion is  conservative and its limitation is that the UMSE might not be tight.   The relation \eqref{upmse} is only tight for the case $P_1=P_2=I/2$ and \eqref{final} is always loose. Therefore, minimizing UMSE is not equivalent to minimizing MSE. In addition, UMSE depends on the assumption that there only exists statistics error in the measurement data from finite state copies (discussed in Remark \ref{rem4}). Despite these limitations, numerical results in \cite{wang2019twostage} (e.g., Fig. 4 therein) indicate a very similar behavior between UMSE and MSE when changing the probe states while maintaining the other parameters $n$, $M$, $N$ fixed. Hence, UMSE is also a useful index evaluating the probe state set.

The second index we consider is, for a set of probe states, how robust the generated estimation result is w.r.t. measurement noise. Note that our CLS estimation \eqref{big} is in fact equivalent to the least squares estimation of the linear regression problem 
\begin{equation}
	\label{bigreg}
	\left(\begin{array}{c}
		\hat{y}_{1}-\frac{1}{n} y_{0} \\
		\vdots \\
		\hat{y}_{n}-\frac{1}{n} y_{0}
	\end{array}\right)=(\begin{array}{l}
		I_{n} \otimes X
	\end{array})\left(\begin{array}{c}
		\Theta_{1}-\frac{1}{n} \mathcal d \\
		\vdots \\
		\Theta_{n} -\frac{1}{n}\mathcal d
	\end{array}\right).
\end{equation}
Typically, the sensitivity of a linear system solution to perturbations in the data  is evaluated by the condition number of the coefficient matrix, defined (among several possible choices) in this paper as $ \operatorname{cond}(A)=\frac{\sigma_{\max }(A)}{\sigma_{\min }(A)}$, where $\sigma_{\max/\min}(A)$ is the maximum/minimum singular value of $A$. Hence, to maximize the estimation robustness w.r.t. measurement noise amounts to minimizing the condition number in \eqref{bigreg} 
$\operatorname{cond}(I_n \otimes X)=\operatorname{cond}(I_n)\operatorname{cond}(X)=\operatorname{cond}(X)$. Therefore, the second evaluation index is chosen as $\operatorname{cond}(X)$. To sum up, the optimal probe states should minimize  $M\operatorname{Tr}(X^TX)^{-1}$ and minimize $\operatorname{cond}(X)$ \emph{simultaneously}. In the following we  give specific characterization and show that the two optimal indices can be achieved simultaneously for several examples.

\subsection{Optimal probe states}
We now give the condition on optimal probe states (\emph{OPS}).

 \begin{theorem}
	\label{theorem1}
	For a $ d $-dimensional detector with $ n $ matrices, assume each type of probe states has the same number of copies, altogether summed to $ N $ copies. Then
	the minimum of  UMSE is $\frac{(n-1)(d^{4}+ d^3-d^{2})}{4 N}  $ and the minimum of $\operatorname{cond}(X)$ is $ \sqrt{d+1} $. These minima are achieved simultaneously if and only if there exist $ M $ different types of probe states such that $ X^{T} X$ is  diagonal and its  eigenvalues  are  $ \lambda_{1}=\frac{M}{d} $ and $\lambda_{2}=\dots=\lambda_{d^2}=\frac{M}{d(d+1)}  $.
\end{theorem}

\begin{pf} 
 We assume that $ M\geq d^2 $ different types of probe states are used, since clearly $ M<d^2 $ is not optimal. Denote  the eigenvalues of $ X^TX $ as $  \lambda_{1} \geq \lambda_{2} \geq \cdots \geq \lambda_{d^2}  $. 
For minimizing UMSE, it can be formulated as
\begin{equation}
	\label{mse}
	\begin{aligned}
		\text { min } & \sum_{i=1}^{d^2} \frac{M}{\lambda_{i}} \\
		\text { s.t. } & \sum_{i=1}^{d^2} \lambda_{i}\leq M,\lambda_{1}\geq \frac{M}{d}.
	\end{aligned}
\end{equation}
For maximizing robustness, it can be formulated as
\begin{equation}
\label{robust}
\begin{aligned}
\text { min } & \sqrt{ \frac{\lambda_{1}}{\lambda_{d^2}}} \\
\text { s.t. } & \sum_{i=1}^{d^2} \lambda_{i}\leq M,\lambda_{1}\geq \frac{M}{d}.
\end{aligned}
\end{equation}
The constraint $ \sum_{i=1}^{d^2}\lambda_{i} \leq M$ comes from the purity requirement. For the parameterization of each probe state, we have
\begin{equation}
\begin{aligned}
	\left\|\phi_{j}\right\|^{2}&=\sum_{b=1}^{d^{2}}\left|\phi_{j}^{b}\right|^{2} \\
	&=\operatorname{Tr}\left[\left(\sum_{b=1}^{d^{2}} \phi_{j}^{b} \Omega_{b}\right)^{\dagger}\left(\sum_{b=1}^{d^{2}} \phi_{j}^{b} \Omega_{b}\right)\right] \\
&=\operatorname{Tr}\left(\rho_{j}^{\dagger} \rho_{j}\right)=\operatorname{Tr}\left(\rho_{j}^{2}\right)\leq 1.
\end{aligned}
\end{equation}
Thus for the parameterization matrix of probe states, we have
\begin{equation}
	\label{con1}
	\begin{aligned}
		\sum_{i=1}^{d^2}\lambda_{i} &= \operatorname{Tr}\left(X^{T} X\right)=\|X\|^2=\sum_{j=1}^{M}\left\|\phi_{j}\right\|^2\\
		&=\sum_{j=1}^{M} \operatorname{Tr}\left(\rho_j^{2}\right)\leq M.
	\end{aligned}
\end{equation}

 The second constraint $ \lambda_{1}\geq \frac{M}{d} $  is from the matrix element satisfying 
 \begin{equation}
 	\left(X^{T} X\right)_{11}=\sum_{j=1}^{M}\left(\phi_{j}^{1}\right)^{2}=M\left(\frac{1}{\sqrt{d}}\right)^2=\frac{M}{d}
 \end{equation}
 in our chosen basis $ \left\{\Omega_{i}\right\}_{i=1}^{d^{2}} $.
 
  For \eqref{mse} and \eqref{robust}, using Lagrange multiplier method as in \cite{Qi2013}, the minimum of $\sum_{i=1}^{d^2} \frac{M}{\lambda_{i}}$ is $  d^{4}+ d^3-d^{2}  $, achieved when $ \lambda_{1}=\frac{M}{d} $ and $\lambda_{2}=\dots=\lambda_{d^2}=\frac{M}{d(d+1)}  $. Hence, the minimum of UMSE is 
  \begin{equation}
  \begin{aligned}
  &\quad\frac{(n-1)M}{4 N} \operatorname{Tr}\left[\left(X^{T} X\right)^{-1}\right]\\
  &\geq\frac{(n-1)M}{4 N}\left[\frac{d}{M}+\frac{d(d+1)}{M}(d^2-1)\right]\\
  &=\frac{(n-1)(d^{4}+ d^3-d^{2})}{4 N} \sim O\left(\frac{ nd^{4}}{N}\right),
  \end{aligned}
  \end{equation}
  which does not rely on the type number of probe states $ M $.
   The minimum of $ \sqrt{\frac{\lambda_{1}}{\lambda_{d^2}}}  $ is $ \sqrt{d+1} $ because $ \lambda_{1}\geq\frac{M}{d} $ and $ \lambda_{d^2}\leq\frac{M}{d(d+1)} $. These minima can be achieved at the same time if and only if $ \lambda_{1}=\frac{M}{d} $ and $\lambda_{2}=\dots=\lambda_{d^2}=\frac{M}{d(d+1)}  $.

   Now, we show that $ X^TX $ is a diagonal matrix. Assume $ X^TX =V\operatorname{diag}\left(\frac{M}{d},\frac{M}{d(d+1)}I_{d-1}\right) V^{T} $. For $\left(X^{T} X\right)_{11}$, we have
   \begin{equation}
   	\begin{aligned}
   	\frac{M}{d}&=\left(X^{T} X\right)_{11}\\
   	&= v^{T}V\operatorname{diag}\left(\frac{M}{d}, \frac{M}{d(d+1)}I_{d-1}\right) V^{T}v\\
   &=\frac{M}{d}\left|V_{11}\right|^{2}+\frac{M}{d(d+1)} \sum_{i=2}^{d}\left|V_{1 i}\right|^{2}\\
   &=\frac{M}{d(d+1)}+\frac{M}{d+1}\left|V_{11}\right|^{2}
   	\end{aligned}
   \end{equation}
where $ v=\left[1, 0, \cdots, 0\right]^T $.
Thus, $ V_{11}=1 $ and $ V=\operatorname{diag}(1,\tilde V_{d-1}) $ where $ \tilde V_{d-1} $ is a $ d-1 $ dimensional  orthogonal matrix. Therefore, we have 
\begin{equation}
	\begin{aligned}	
	&\quad X^{T} X\\&=\operatorname{diag}\left(1, \tilde V_{d-1}\right) \operatorname{diag}\left(\frac{M}{d}, \frac{M}{d(d+1)} I_{d-1}\right) \operatorname{diag}\left(1, \tilde V_{d-1}^{T}\right)\\
&=\operatorname{diag}\left(\frac{M}{d}, \frac{M}{d(d+1)} I_{d-1}\right),
	\end{aligned}
\end{equation}
which is a diagonal matrix.
  \hfill $\Box$
\end{pf}

Our results also show that
the description of OPS in Assumption $ 1 $ in \cite{wang2019twostage} needs to be more precise.
For certain $ M $, if there exist OPS, all of them must be  pure states because $\sum_{j=1}^{M}\operatorname{Tr}(\rho_j^2)=\sum_{j=1}^{d^2}\lambda_{i}=M $ from \eqref{con1}. This indicates that pure states are better than mixed states. 	However, it is not clear whether these OPS exist for arbitrary $ M\geq d^2 $. Thus, whether we can always find a pure state set which is better than  arbitrary mixed state set for given $ M $ is still an open problem.

\begin{rem}
	In system identification, the similar problem called input design problem has been widely discussed. 	There are many existing results, e.g., D,A,E-optimal
		input design \cite{Boyd2004Convex}. The common idea behind the problem of our optimal probe states and  optimal input design is that both problems consider minimizing the trace of the covariance matrix, which is A-optimal input design \cite{Boyd2004Convex}.  The difference is that we also consider robustness and other physical eigenvalue constraints  (e.g., purity) for probe states. Thus, we cannot directly adapt classical input design problem for the quantum case.
	\end{rem}

\begin{rem}
	If we only want to reach the minimum condition number $ \sqrt{d+1} $ without considering UMSE, we  need to ensure $ \frac{\lambda_{1}}{\lambda_{d^2}} = d+1$ which can be satisfied even for mixed states. For example, for a concentric sphere inside the Bloch sphere, we can also find the corresponding platonic solid on it which has the smallest condition number $ \sqrt{d+1} $ (this case will be discussed later). Therefore, if we only consider minimum condition number, we cannot obtain the  minimum UMSE. However, if we only consider minimum UMSE, the eigenvalues also satisfy the requirement of minimum condition number.  Hence, when we solve the optimization problem and analyze  optimal probe states, it suffices to only consider  minimum UMSE.
\end{rem}

\begin{rem}
In quantum state tomography, the condition number for the optimal measurement is $ 1 $, achieved by special measurement such as the optimal generalized Pauli operators \cite{PhysRevA.90.062123}. However, in QDT, OPS need to satisfy the unit-trace constraint. Therefore, the largest eigenvalue $ \lambda_{1} $ must be equal to or larger than $ \frac{M}{d} $ and  the minimum condition number is $ \sqrt{d+1} $. 
\end{rem}

\begin{rem}\label{rem4}
We consider two criteria for optimization--upper bound of the mean squared error (UMSE) and the robustness described by the condition number. For UMSE, we assume there only exists statistics error from finite state copies and the analytical upper bound depends on this assumption. If this assumption is not satisfied, the  UMSE should be adjusted by adding the unmodeled  noise (e.g., apparatus noise) to the error $ \mathcal{e} $ in \eqref{cls2}. For robustness, condition number characterizes the sensitivity of the estimation result to errors in measurement data. Hence, this criterion is unrelated to the specific sources of the errors. 

\end{rem}

 We then give two examples of OPS for $ M=d^2 $ and $ M=d(d+1) $ which are motivated by projection measurements.

The first example is motivated from SIC-POVM. 
To reconstruct an unknown quantum state  $ \rho $, a generalized measurement must have at least $  d^2$  linear independent elements, which is called informationally complete. Furthermore, if  the measurement results are maximally independent, the POVM is called symmetric informationally complete POVM (SIC-POVM). The simplest mathematical  definition of an SIC-POVM is a set of $d^{2}$ normalized vectors $\left|\phi_{k}\right\rangle$ in $\mathbb{C}^{d}$ satisfying \cite{sic}
\begin{equation}
\label{sic}
\left|\left\langle\phi_{j} | \phi_{k}\right\rangle\right|^{2}=\frac{1}{d+1}, \quad j \neq k.
\end{equation}
It has been conjectured that SIC-POVMs exist for all dimensions \cite{sic} and their existence has been given for $ d\leq121$, and some other sporadic values \cite{scott2017sics}.  When SIC-POVMs exist in $ d $ dimension, we define the $ d^2 $ pure states $ |\phi_{k}\rangle\in\mathbb{C}^{d}(1\leq k \leq d^2) $ to be \emph{SIC states} if they satisfy \eqref{sic}.
 \begin{proposition}
 	\label{lem1}
 $ d $-dimensional 	SIC  states (when they exist) are a set of OPS with the smallest $ M $ as $ M=d^2 $.	
\end{proposition}
\begin{pf}
 Note that choosing different orthogonal bases does not change the inner product between vectors. Therefore, for parameterization of the SIC states, we have
\begin{equation}
\label{sic2}
\left(XX^{T} \right)_{i j}=\left\{\begin{array}{c}
1, i=j \\
\frac{1}{d+1}, i \neq j
\end{array}\right.
\end{equation}
and $ M=d^2 $.  We assume the standard singular value decomposition of $ X $ is $ X=U\Sigma V^{T} $ where $\Sigma  $ is a diagonal matrix, and $ U $ and $ V $ are unitary matrices. According to \eqref{sic2}, we have
\begin{equation}
X X^{T}=\left(\begin{array}{ccc}
1 & \cdots & \frac{1}{d+1} \\
\vdots & \ddots & \vdots \\
\frac{1}{d+1} & \cdots & 1
\end{array}\right)=U\Sigma^2 U^{T}.
\end{equation}
 Thus, $\Sigma=\operatorname{diag}(d,{\frac{d}{d+1}I_{d^{2}-1}})$ and
\begin{small}
	\begin{equation}
	\begin{aligned}
	X^{T} X&=V\Sigma U^{T} U\Sigma V^T\\
	&=V\operatorname{diag}(d,{\frac{d}{d+1}I_{d^{2}-1}}) V^{T}.\\
	\end{aligned}
	\end{equation}
\end{small}
The largest eigenvalue of $ X^{T}X $ is $ d $ and the other eigenvalues are $ \frac{d}{d+1} $, which satisfies Theorem \ref{theorem1}. Therefore, SIC states are OPS. Also, among all $ d $-dimensional OPS, SIC states have the smallest $ M $ as $ M=d^2 $. If $ M<d^2 $, it is not informationally complete, and the detector cannot be uniquely determined without other prior knowledge. 
\hfill $\Box$
\end{pf}
The second example is motivated from  MUB measurement. Two sets of orthogonal
bases $ \mathcal{B}^{k}=\{\left|\psi_{i}^{k}\right\rangle: i=1, \ldots$$, d\} $ and $ \mathcal{B}^{\ell}=\left\{\left|\psi_{j}^{\ell}\right\rangle: j=1, \ldots, d\right\} $ are called
mutually unbiased if and only if \cite{PhysRevLett.105.030406}
\begin{equation}
\label{mub}
\left|\left\langle\psi_{i}^{k} | \psi_{j}^{\ell}\right\rangle\right|^{2}=\left\{\begin{array}{ll}
1 / d & \text { for } k \neq \ell, \\
\delta_{i, j} & \text { for } k=\ell.
\end{array}\right.
\end{equation}
 In particular, one can find maximally  $d+1$ sets of mutually unbiased bases in Hilbert spaces of prime-power dimension $d=p^{k},$ with $p$ a prime and ${k}$ a positive integer \cite{mubreview}.  When $ d+1 $ sets of MUB measurement exist in $ \mathbb{C}^{d} $, we view each projective MUB measurement operator as a pure state and we call them \emph{MUB  states}. Thus, MUB states always exist for $ m $-qubit systems $ (d=2^m) $.
 \begin{proposition}
 	\label{lem2}
	$ d $-dimensional MUB states (when they exist) are a set of OPS for $ M=d(d+1) $.	
\end{proposition}
\begin{pf}
For MUB states in \eqref{mub}, we have $ M=d(d+1) $. For their parameterization, we have 
\begin{equation}
\left(X X^{T}\right)_{i j}=\left\{\begin{array}{c}\delta_{i j}, k d+1 \leq i, j \leq(k+1) d, 0\leq k \leq d,
	 \\ \frac{1}{d}, \text { otherwise. }\end{array}\right.
\end{equation}
We assume the standard singular value decomposition of $ X $ is $ X=U\Sigma V^{T} $. Thus, we have
\begin{equation}
X X^{T}=\left(\begin{array}{cccc}
I_{d} & \left(\frac{1}{d}\right)_{d} & \cdots & \left(\frac{1}{d}\right)_{d} \\
\left(\frac{1}{d}\right)_{d} & I_{d} & \cdots & \left(\frac{1}{d}\right)_{d} \\
\vdots & \vdots & \vdots & \vdots \\
\left(\frac{1}{d}\right)_{d} & \left(\frac{1}{d}\right)_{d} & \cdots & I_{d}
\end{array}\right)=U\Sigma\Sigma^T U^{T},
\end{equation}
where $ \left(\frac{1}{d}\right)_{d} $ denotes a $ d\times d $ matrix and all the elements are $ \frac{1}{d} $. We have
	\begin{equation}
		\begin{aligned}
			X^{T} X&=V\Sigma^T U^{T} U\Sigma V^{T}\\
			&=V\operatorname{diag}(d+1,I_{d^2-1}) V^{T},\\
		\end{aligned}
	\end{equation}
where the largest eigenvalue is $ d+1 $ and the remaining $ d^2-1 $ eigenvalues are $ 1 $. They are OPS because their eigenvalues satisfy Theorem \ref{theorem1}. 
\hfill $\Box$
\end{pf}
For $ M\neq d^2 $ and $ M\neq d(d+1) $, we leave it an open problem when there always exist OPS satisfying the two indices simultaneously.
For two-qubit detectors, from the above results we know the optimal probe states can be constructed using $ 4 $-dimensional MUB states and SIC states as shown in \ref{appenA}. A similar two-qubit problem for QST was discussed in \cite{Qi2013} to determine the optimal measurement based on UMSE and in \cite{PhysRevA.90.062123} based on condition number.

 For one-qubit probe states, they have simple geometric property; i.e.,  they are in the Bloch sphere. The pure states are on the surface and the mixed states are inside the sphere. Hence, an alternative method to search for one-qubit OPS is based on geometric symmetry. For $ M=4 $, when the probe states are on the surface of Bloch sphere and become the four vertices of a tetrahedron concentric with Bloch sphere, they are OPS. In fact, they are also SIC states. For $ M=6 $, one can construct OPS similarly using octahedron, and they are also MUB  states.  We also have cube, icosahedron, dodecahedron for $ M=8,12,20 $, respectively, which are OPS. We conjecture all one-qubit OPS are constructed from the five platonic solids on the Bloch sphere in this way, and we show there do not exist OPS for $ M=5 $ in \ref{appenB}. For multi-qubit states, we still do not fully know the geometric property and it is thus an open problem to find other optimal probe states.

\subsection{Product probe state}\label{product}
In experiment, product states are among the ones most straightforwardly to be implemented. In this subsection we consider the case $d=2^m$ for some integer $m$ and each probe state is an $m$-qubit tensor product state as  $ \rho_{j}=\rho^{(1)}_{j_{1}} \otimes \cdots \otimes \rho^{(m)}_{j_{m}} $ where ${1 \leq j_{1} \leq M_{1}, \cdots, 1 \leq j_{m} \leq M_{m}} $ and there are $ M_i $ different types of one-qubits for the $ i $-th qubit of the product states. The total number of these $ m $-qubit tensor product states is thus $\prod_{i=1}^{m}{M_i}$.
 We then give their optimal UMSE and condition number.

\begin{theorem}\label{theorem2}
	The minimum UMSE of $ m $-qubit product probe state is  $\frac{20^m(n-1)}{4N}$ and the minimum condition number is $ \sqrt{3^m}$.   These minima are achieved simultaneously if and only if each qubit is among optimal one-qubit states.
\end{theorem}
\begin{pf}
Assume the parameterization of all the $ i $-th single-qubit states $ \left\{\rho_{j_{i}}^{(i)}\right\}_{j_{i}=1}^{M_i} $ is $ X_i $. We thus have
\begin{equation}
\begin{aligned}
X^{T} X&=\left(X_{1} \otimes X_{2} \cdots \otimes X_{m}\right)^{T}\left(X_{1} \otimes X_{2} \cdots \otimes X_{m}\right)\\
&=X_{1}^{T} X_{1} \otimes \cdots \otimes X_{m}^{T} X_{m},
\end{aligned}
\end{equation}
and therefore,
\begin{equation}
\begin{aligned}
\operatorname{Tr}\left[\left(X^{T} X\right)^{-1}\right]&=\prod_{i=1}^{m}\operatorname{Tr}\left[\left(X_{i}^{T} X_{i}\right)^{-1}\right]\geq\prod_{i=1}^{m}\frac{20}{M_i},
\end{aligned}
\end{equation}
where $ 20 $ comes from $ d^4+d^3-d^2 $ for $ d=2 $.
Thus, the UMSE is $ \frac{(n-1)M}{4N}\operatorname{Tr}\left[\left(X^{T} X\right)^{-1}\right] \geq\frac{20^m(n-1)}{4N}$. For condition number, 
\begin{equation}
\operatorname{cond}(X)=\prod_{i=1}^{m} \operatorname{cond}\left(X_{i}\right) \geq \prod_{i=1}^{m} \sqrt{3}=\sqrt{3^{m}},
\end{equation}
where $3 $ comes from $ d+1 $ for $ d=2 $.
We can reach these minima if and only if for each $ i $, $ \left\{\rho_{j_{i}}^{(i)}\right\}_{j_{i}=1}^{M_i} $ is an OPS set.
\hfill $\Box$
\end{pf}

Here we compare the two criteria--UMSE and condition number between  OPS and optimal product states for an $ m $-qubit detector with the dimension $ d=2^m $. The UMSEs are $\frac{(n-1)(16^{m}+ 8^m-4^{m})}{4 N}  $ for OPS and $\frac{20^m(n-1)}{4N}$ for optimal product states.  UMSE of optimal probe states is always smaller than that of optimal product states for $ m\geq2 $. For $ m=1 $, they are both $ \frac{20(n-1)}{4N} $. For condition number, it is $ \sqrt{2^m+1} $ for OPS and $ \sqrt{3^m} $ for optimal product states. Thus,  for both  UMSE and condition number, OPS play better than optimal product states in multi-qubit systems. Because most of the OPS are entangled states, this is an example showcasing the advantage of entanglement in QDT.

We give an example of two-qubit product probe states. Let the parameterization of the $ j $-th two-qubit product probe state $ \rho_{j} $ be
	\begin{equation}
		\begin{aligned}
			\phi_{j}&=\phi_{j_1}^{(1)} \otimes \phi_{j_2}^{(2)}=\left[\frac{1}{2}, \frac{1}{\sqrt{2}}\psi_{j_2}^{(2)}, \frac{1}{\sqrt{2}}\psi_{j_1}^{(1)}, \psi_{j_1}^{(1)} \otimes \psi_{j_2}^{(2)}\right]^{T},	
		\end{aligned}
	\end{equation}
	where $ \phi_{j_i}^{(i)}=\left[\frac{1}{\sqrt{2}}, \psi_{j_i}^{(i)}\right]^{T} $ is the parameterization of the $ i $-th  qubit of $ \rho_{j} $. Since optimal one-qubit probe states are pure states, $ \left\|\psi_{j_1}^{(1)}\right\|=\left\|\psi_{j_2}^{(2)}\right\|=\frac{1}{\sqrt{2}} $. Denote  the eigenvalues of $ X^TX $ as $  \lambda_{1} \geq \lambda_{2} \geq \cdots \geq \lambda_{d^2} $. We can obtain three constraints as
	\begin{equation}
		\left(X^{T} X\right)_{11}=\sum_{j=1}^{M}\left(\phi_{j}^{1}\right)^{2}=M\left(\frac{1}{2}\right)^2=\frac{M}{4},
	\end{equation}
	\begin{equation}
		\sum_{i=1}^{4} \lambda_{i} \geq \sum_{i=1}^{4}\left(X^{T} X\right)_{i i}=M\left(\frac{1}{2}\right)^{2}+\sum_{j=1}^{M}\left\|\frac{1}{\sqrt{2}} \psi_{j_1}^{(2)}\right\|^{2}=\frac{M}{2},
	\end{equation}
	\begin{equation}
		\begin{aligned}
		&\sum_{i=1}^{7} \lambda_{i} \geq \sum_{i=1}^{7}\left(X^{T} X\right)_{i i}\\
		&=M\left(\frac{1}{2}\right)^{2}+\sum_{j=1}^{M}\left\|\frac{1}{\sqrt{2}} \psi_{j_1}^{(1)}\right\|^{2}+\sum_{j=1}^{M}\left\|\frac{1}{\sqrt{2}} \psi_{j_2}^{(2)}\right\|^{2}=\frac{3 M}{4}.
		\end{aligned}
	\end{equation}Thus, the optimization problem is
\begin{equation}
\begin{aligned}
\text { min } & \sum_{i=1}^{16} \frac{M}{\lambda_{i}} \\
\text { s.t. } & \sum_{i=1}^{16} \lambda_{i}= M,\lambda_{1}\geq \frac{M}{4},\\
&\sum_{i=1}^{4} \lambda_{i}\geq \frac{M}{2},\sum_{i=1}^{7} \lambda_{i}\geq \frac{3M}{4}.\\
\end{aligned}
\end{equation}
It can be proven that $ \sum_{i=1}^{16} \frac{M}{\lambda_{i}} $ reaches its minimum $ 400 $ and the minimum UMSE is $ \frac{100(n-1)}{N} $ when $ \lambda_{1}=\frac{M}{4}$$,\lambda_{2}=\cdots=\lambda_{7}=\frac{M}{12}$, $\lambda_{8}=\cdots=\lambda_{16}=\frac{M}{36}$. It can be verified that this minimum UMSE  can
be reached by using the tensor of platonic solid states such as $ M=36 $ (Cube). The minimum condition number is $ \sqrt{\frac{\lambda_{1}}{\lambda_{16}}}=\sqrt{\frac{M / 4}{M / 36}}=3 $.

\subsection{Superposition of coherent probe state}\label{coherent}
In quantum optics experiments, the preparation of number states (or Fock states) $|k\rangle(k \in \mathbb{N})$ is usually a difficult task,
especially when $k$ is large. Therefore, it is also difficult to prepare SIC and MUB states. Coherent
states, more straightforward to be prepared, are more commonly used as probe states for QDT in practice.
Thus, a good approach is to use the superposition of several coherent states to approximate SIC states and MUB states.

A coherent state is denoted as $|\alpha\rangle$ where $\alpha \in \mathbb{C}$ and it can be expanded using
number states as
\begin{equation}
|\alpha\rangle=e^{-\frac{|\alpha|^{2}}{2}} \sum_{i=0}^{\infty} \frac{\alpha^{i}}{\sqrt{i !}}|i\rangle.
\end{equation}
The inner product relationship between two states $ |\alpha\rangle $ and $ |\beta\rangle $ is
\begin{equation}
\langle\beta | \alpha\rangle=\mathrm{e}^{-\frac{1}{2}\left(|\beta|^{2}+|\alpha|^{2}-2 \beta^{*} \alpha\right)}.
\end{equation}
Let $ |\alpha_{d}\rangle=\mathrm{e}^{-\frac{|\alpha|^{2}}{2}} \sum_{i=0}^{d-1} \frac{\alpha^{i}}{\sqrt{i} !}|i\rangle $. Coherent states are in essence infinite dimensional. To estimate a $d$-dimensional detector,  we employ $|\alpha_{d}\rangle$ as the approximate  description of $|\alpha\rangle$, and we assume that the discarded part $|\alpha\rangle-|\alpha_d\rangle$ has a small enough influence  such that it can be neglected.  A matrix or vector with subscript $d$ means it is truncated in $d$ dimension.

\begin{rem}
	It can lead to significant error to approximate a general pure state using only one coherent state instead of the superposition of many.  This problem is often  referred to  as the nonclassicality of states. As an example, we now consider a Fock state $|n\rangle$. The infidelity is 
	\begin{equation*}
	1-F_s(|n\rangle\langle n|,|
	\alpha\rangle\langle \alpha|) 
	=\left[2\left(1-\exp (-|\alpha|^2) \frac{|\alpha|^{2n}}{n !}\right)\right]^{1 / 2}.
	\end{equation*}
	The minimum infidelity is obtained by $ |\alpha|^2=n $ and for $ n=0,1,2 $, the minimum infidelity is $ 0,0.6321,0.7293 $, respectively. Thus, the distance between one Fock state and one coherent state may be quite large. This also shows that the Fock state $|n\rangle$ becomes more and more non-classical as the value $n$ increases  \cite{nonclassicality}. 
\end{rem}
In \emph{quantum state engineering}, a central problem is how to construct a pure state by  superposition of coherent states. There are two main approaches to do this. One is to write the pure state as a superposition of coherent states along the real axis in phase space
\begin{equation}
|\psi\rangle=\int_{-\infty}^{\infty} \mathcal F(\alpha)|\alpha\rangle d \alpha,
\end{equation}
where $\mathcal F(\alpha)  $ is the distribution function. The other choice 
is the superposition on a circle
\begin{equation}
|\psi\rangle=\int_{0}^{2 \pi} \mathcal F_{R}(\phi)\left|R e^{i \phi}\right\rangle d \phi,
\end{equation}
where $R$ is the radius of the circle and $  \mathcal F_{R}(\phi) $ a circle distribution function. The analytical solutions of $\mathcal F(\alpha)  $ and $  \mathcal F_{R}(\phi) $ were given in \cite{PhysRevA.53.2698}. They also gave the discrete superposition of coherent states to construct pure states by discretizing the above equations.   Ref. \cite{PhysRevA.51.4191} evaluated the performance to construct squeezed displaced number states and \cite{PhysRevA.53.2698} gave a representation of a Fock state $ |n\rangle $ by $ n+1 $ coherent states. If we can construct all Fock states with high accuracy, we can use these Fock states to construct all the ideal pure states  we need. However, in practice, the technique to superpose many coherent states arbitrarily is still developing.

In this paper, we  use the finite superposition of $s$ coherent states  $ |\tilde\psi\rangle = \sum\limits_{k = 1}^s {{c_k}} \left| {{\alpha _d}} \right\rangle_k$ to approximate an ideal pure state $ |\psi\rangle$ where  $  |\tilde\psi\rangle$ indicates that it has not been normalized. Hence, $ |\tilde\psi\rangle  $ is not yet a quantum state in the most strict sense. 	
To find the  superposition state $  |\tilde\psi\rangle$,  at first sight it can be formulated as an optimization problem to maximize fidelity as
\begin{equation}
\max_{{|\tilde\psi\rangle}} |\langle\psi|\tilde\psi\rangle|^2,
\end{equation}
Since the fidelity is $ F_s(\hat{\rho}, \rho)=[\operatorname{Tr} \sqrt{\sqrt{\hat{\rho}} \rho \sqrt{\hat{\rho}}}]^{2}=|\langle\psi | \tilde{\psi}\rangle|^{2}$.
However, for the purpose of QDT, this cost function is not the most suitable. The most important part of the superposed state $|\tilde\psi\rangle$ is its direction. We hope to align it in the same direction as $|\psi\rangle$, even if the norm$ \||\tilde\psi\rangle\|$ can be different from $\||\psi\rangle\|=1$. Hence, we need the normalized state $\frac{|\tilde\psi\rangle}{\sqrt{\langle\tilde\psi|\tilde\psi\rangle}}$ to be a good approximation to $|\psi\rangle$, which thus leads to the numerator of  \eqref{newcost}. Also, if $ \||\tilde\psi\rangle\| $ is too small, we cannot neglect elements in the dimension larger than $ d $ and the corresponding measurement data $ \operatorname{Tr}\left(\left|\tilde\psi\right\rangle\left\langle\tilde\psi\right| P_{i}\right) $ will be small, leading to low measurement accuracy for given the same resource number $ N $. Therefore, we add a penalty $ \langle\tilde\psi|\tilde\psi\rangle $ to prevent the norm $|\tilde\psi\rangle  $ from being too small. 
Thus, we design a new cost function as
\begin{equation}
\label{newcost}
\begin{aligned}
\min_{{|\tilde\psi\rangle}} \frac{\| \frac{|\tilde\psi\rangle}{\sqrt{\langle\tilde\psi|\tilde\psi\rangle}}-|\psi\rangle\|^{2}}{\langle\tilde\psi|\tilde\psi\rangle}.
\end{aligned}
\end{equation}
 This optimization problem is usually non-convex, and thus we select different initial points and numerically search for the best solution.
\subsection{Error analysis  for state preparation}
When we prepare  probe states
such SIC states and MUB states in experiment, there usually exist state preparation errors. For example, if we use superposition of coherent states, there exists approximation error between the ideal target probe state and the superposition of coherent states when the number of coherent states for superposition is not large enough. We  give the error analysis on UMSE and condition number when there exists state preparation error.
\begin{theorem}\label{theorem3}
	Given two probe state sets $ \left\{\rho_{j}\right\}_{1}^{M} $ and $ \left\{\hat\rho_{j}\right\}_{1}^{M} $,	 if  $ \max_{j}\left\|\rho_{j}-\hat{\rho}_{j}\right\| \leq \varepsilon $ and $ \varepsilon\leq \frac{\lambda_{d^2}}{2M} $, the corresponding error in $ \frac{(n-1)M}{4 N}\big|\operatorname{Tr}\left(X^{T} X\right)^{-1}\!-\!\operatorname{Tr}\left(\hat{X}^{T} \hat{X}\right)^{-1}\big| $ is upper bounded by $ \dfrac{2(n-1)d^2M^2\varepsilon}{4N{\left(\lambda_{d^{2}}-2 M \varepsilon\right)^{2}}} $ where $ \lambda_{d^{2}}>0 $ is the smallest eigenvalue of  the parameterization matrix $ X^TX $ and the corresponding error on condition number is upper bounded by $\dfrac{ M \left(\lambda_{1}+\lambda_{d^{2}}\right)\varepsilon}{\left(\lambda_{d^{2}}-2 M \varepsilon\right)^2}$ where $ \lambda_{1} $ is the largest eigenvalue of  $ X^TX $.
\end{theorem}
\begin{pf}
	See \ref{appenC}.
	\end{pf}

\section{Adaptive detector tomography}\label{adaptive}
As shown in Fig.~\ref{algorithm}, after employing partial resources to obtain a rough estimate of the detector through Step 1, one can design new probe states dependent on the specific estimation value of the detector to further improve the accuracy in Step 2. Our adaptive QDT scheme is applicable for arbitrary reconstruction algorithm.
\subsection{Evaluation index}
In quantum information, fidelity (or infidelity) has profound physical meaning to characterize the distance and similarity between two quantum states (or operations). It has been a widely used metric \cite{qci,inf1,HUBNER1992239}. In developing adaptive QDT, we also employ infidelity as the evaluation index.

The fidelity between two arbitrary states $ \hat{\rho} $ and $ \rho $ is defined by
\begin{equation}
	F_s(\hat{\rho},\rho)\triangleq[\operatorname{Tr} \sqrt{\sqrt{\hat{\rho}} \rho \sqrt{\hat{\rho}}}]^{2},
\end{equation}
which has three basic properties:
\begin{equation}
	(i)\ \  F_s(\hat{\rho},\rho)=F_s(\rho,\hat{\rho});
\end{equation}
\begin{equation}
	(ii)\ \  0\leq F_s(\hat{\rho},\rho) \leq 1;
\end{equation}
\begin{equation}\label{property3}
	(iii)\ \  F_s(\hat{\rho},\rho) = 1 \ \ \Leftrightarrow \ \ \hat{\rho}=\rho.
\end{equation}
To extend the fidelity definition from states to detectors, a natural idea is to normalize the POVM element such that it has the same mathematical property as a quantum state. This leads to the definition
\begin{equation}
	\label{infide}
	F_0\left(\hat{P}_{i}, P_{i}\right)\triangleq\left[\operatorname{Tr}\sqrt{\sqrt{\hat P_{i}} {P}_{i} \sqrt{\hat P_{i}}}\right]^{2} /\left[\operatorname{Tr}\left(P_{i}\right) \operatorname{Tr}\left(\hat{P}_{i}\right)\right].
\end{equation}
This definition has been widely used in QDT \cite{Feito_2009,Lundeen2009,Zhang_2012} to evaluate the estimation performance, and corresponding infidelity is defined as $ 1-F_0\left(\hat{P}_{i}, P_{i}\right) $. However,  we find that this definition is not always appropriate, because in certain circumstances the third property (\ref{property3}) dose not hold for (\ref{infide}) (we call this phenomenon \emph{distortion}). More specifically, for certain detector $\left\{{P}_{i}\right\}_{i=1}^{n}$, \emph{distortion} means there exists $\left\{\hat{P}_{i}\right\}_{i=1}^{n}$ such that $F_0\left(\hat{P}_{i}, P_{i}\right)=1$ for all $1\leq i\leq n$ while $\hat{P}_{i}= P_{i}$ fails for at least one $i$. For example, suppose the detector is $ P_{1}=P_{2}=P_{3}=\frac{I}{3} $ and the estimations are $ \hat P_1=a_1P_1$, $\hat P_2=a_2P_2 $, $\hat P_3=a_3P_3 $ where $ a_1 $, $ a_2 $ and $ a_3 $ are three  arbitrary positive numbers satisfying $ a_1+a_2+a_3=3 $. The fidelities for the three detectors are all maximum $ 1 $, but the estimation is in fact usually not accurate.
We characterize when the evaluation index \eqref{infide} will distort in the following proposition, and present its proof in \ref{appenD}.

\begin{proposition}
	\label{distort}
	The evaluation index \eqref{infide} will distort if and only if $ \left\{ P_{i}\right\}_{i=1}^{n} $ are linearly dependent, i.e., there exists nonzero  $ c\in\mathbb{R}^n$ such that $ \sum_{i=1}^{n} c_{i}  P_{i}=0 $. Specially, if $ d^2<n $, there must exist distortion.
\end{proposition}

When distortion happens, $\lim_{F_0(\hat{P}_{i}, P_{i})\rightarrow 1} \hat{P}_{i}=P_{i}$ fails. That is, the estimation can  be on a wrong track even if the fidelity approaches to 1. To fix this problem, we add  $\big[\operatorname{Tr}(P_i-\hat{P}_i)\big]^2/d^2$ and propose a  new detector fidelity as
\begin{equation}
\label{infide2}
\begin{aligned}
F\left(\hat{P}_{i}, P_{i}\right)=&\left[\operatorname{Tr}(\sqrt{\sqrt{\hat P_{i}} {P}_{i} \sqrt{\hat P_{i}}})\right]^{2} /\left[\operatorname{Tr}\left(P_{i}\right) \operatorname{Tr}\left(\hat{P}_{i}\right)\right]\\
&-\left[\operatorname{Tr}\left(P_{i}-\hat{P}_{i}\right)\right]^{2} / d^{2}.
\end{aligned}
\end{equation}
The fidelity \eqref{infide2} takes values in $(\frac{1}{d}-1,1]$ and we give the proof of the lower bound in \ref{appenE}. When $F\left(\hat{P}_{i}, P_{i}\right)=1$, we must have $\hat{P}_i=P_i$ (and vice versa), solving the distortion problem of \eqref{infide}. For the rest of this paper, we refer to ``fidelity" as \eqref{infide2}, unless otherwise declared. The infidelity is defined as $ 1-F\left(\hat{P}_{i}, P_{i}\right) $.

 \subsection{Two-step adaptive quantum detector tomography}
Let $\left\{\left|\lambda_{t}\right\rangle\right\}$ be the eigenvectors of $P_i$ for given $ i $,  where the zero eigenvalues are  $\lambda_{r+1}= \cdots= \lambda_{d}=0$.  We can view $ \frac{P_{i}}{\operatorname{Tr}\left(P_{i}\right)} $ and $ \frac{\hat P_{i}}{\operatorname{Tr}\left(\hat P_{i}\right)}  $ as two quantum states. Thus, we may use the analysis in \cite{PhysRevA.98.012339} to obtain the Taylor series expansion of the  new infidelity $1-F(\hat P_i,  P_i)$ based on \eqref{infide2} up to the second order as
\begin{equation}
\label{infidelity}
\begin{aligned}
&\mathbb{E}\left(1-F(\hat P_i,  P_i)\right)\\
=&\mathbb{E}\left(1-F_{0}\left(\hat{P}_{i}, P_{i}\right)+\left[\operatorname{Tr}\left(P_{i}-\hat{P}_{i}\right)\right]^{2}/d^2\right)\\
=&\mathbb{E}\left(1-F_{0}\left(\hat{P}_{i}, P_{i}\right)+\left(\sum_{t=1}^{d}\left\langle\lambda_{t}|\Delta_2| \lambda_{t}\right\rangle\right)^{2}/d^2\right)\\
=&\mathbb{E}\left(\sum_{t=r+1}^{d}\left\langle\lambda_{t}|\Delta_1| \lambda_{t}\right\rangle\right)+\mathbb{E}\left(\frac{1}{2} \sum_{t, k=1}^{r} \frac{\left|\left\langle\lambda_{t}|\Delta_1| \lambda_{k}\right\rangle\right|^{2}}{\lambda_{t}+\lambda_{k}}\right) \\
&-\mathbb{E}\left(\frac{1}{4}\left[\sum_{t=r+1}^{d}\left\langle\lambda_{t}|\Delta_1| \lambda_{t}\right\rangle\right]^{2}\right)\\
&+\mathbb{E}\left(\sum_{t=1}^{d} \sum_{k=1}^{d}\left\langle\lambda_{t}|\Delta_2| \lambda_{t}\right\rangle\left\langle\lambda_{k}|\Delta_2| \lambda_{k}\right\rangle/d^2\right)+O\left(\|\Delta_1\|^{3}\right),
\end{aligned}
\end{equation}
where $ \Delta_1=\frac{P_{i}}{\operatorname{Tr}\left(P_{i}\right)}-\frac{\hat P_{i}}{\operatorname{Tr}\left(\hat P_{i}\right)}, \Delta_2=P_i-\hat P_i $. Crucially, the new term $ -\left[\operatorname{Tr}\left(P_{i}-\hat{P}_{i}\right)\right]^{2} / d^{2}$ is in the second order instead of in the first order, and we can thus imitate the analysis for QST in \cite{PhysRevA.98.012339}. Note that we use a perturbation method thus \eqref{infidelity} is valid only around $P_i$.  The condition can be guaranteed because we are analyzing the asymptotic behavior as $N$ tends to infinity (large enough).

According to the Taylor series,  the scaling performance of the infidelity depends on the rank of the detector. For instance,  for a full-rank POVM element, the first-order term vanishes and the infidelity scales as $O(1 / N)$ using just  non-adaptive  QDT. For a rank deficient $ P_i $, the first-order term dominates and thus the infidelity scales as $O(1 / \sqrt{N}) $ by QDT which only has Step 1 in Fig. \ref{algorithm}. The optimal scaling of infidelity $ 1-	F_s(\hat{\rho},\rho) $ is $ O(1/N) $ for unbiased estimate in QST \cite{zhu} and  the optimal scaling of $ -\left[\operatorname{Tr}\left(P_{i}-\hat{P}_{i}\right)\right]^{2} / d^{2}$ is always not worse than $ O(1/N)  $. Thus, the optimal scaling of infidelity $ 1-F_0(\hat P_i,  P_i) $ or   $ 1-F(\hat P_i,  P_i) $ is also $ O(1/N) $. From \cite{PhysRevA.98.012339}, we know the second-order terms always scale as $ O(1/N) $. Therefore,  for rank-deficient detectors,
the behavior of the infidelity can be corrected to $O(1 / N)  $ if one can eliminate the influence of the first-order term \cite{PhysRevA.98.012339}. This depends on the diagonal coefficients of $\Delta_1$ in the kernel of $\frac{P_{i}}{\operatorname{Tr}\left(P_{i}\right)}$, which suggests performing QDT in a basis aligning with the eigenvectors of $P_i$ \cite{PhysRevA.98.012339}. 

The detailed adaptive procedure is as follows. To simplify the expression, we use $ P $ to represent each POVM element $ P_i $. For $ P $, given a  probe state set $ \{\rho_{j}\}_{j=1}^M $, we choose one among them (e.g., $ \rho_a $) and its spectral decomposition is $ \rho_{a}=V_a\Gamma_aV_a^{\dagger} $. The spectral decomposition of $ P_i $ is $ P=U\Lambda U^{\dagger} $. To perform QDT in a basis that agrees with the eigenvectors of $P$ \cite{PhysRevA.98.012339}, we change each probe state $ \rho_{j} $ to $ \tilde\rho_{j} $ by a common conjugation as
\begin{equation}
\tilde\rho_{j}=UV_a^{\dagger}\rho_{j}V_aU^{\dagger}.	
\end{equation}
  However, in practice, we do not know $U_i$. Therefore, we have two steps as shown in Fig. \ref{algorithm}. In Step 1, we obtain an estimator $ \left\{\tilde{P}\right\} $ by applying non-adaptive QDT on an ensemble of size $ N_0 $. Then in Step 2, we use new probe states $ \tilde \rho_{j} $ as
  \begin{equation}
  		\label{newprobe}
  	\tilde\rho_{j}=\tilde UV_a^{\dagger}\rho_{j}V_a\tilde U^{\dagger},
  \end{equation}
   for an ensemble of size $N- N_0 $, where $\tilde P=\tilde U\tilde \Lambda\tilde U^{\dagger} $. For each POVM element, we need to repeat the above procedure.

Furthermore, to guarantee that the infidelity scaling is improved to $ O(1/N) $, one needs to carefully choose the probe states. Before showcasing how to do this, we first introduce several basic definitions.  For a $ d $-dimensional Hermitian space, we  define  bases $ \mathcal B\left(\left\{|i\rangle_{i=1}^{d}\right\}\right) $ consisting of the following elements,
 \begin{equation}
 	\label{op1}
 	\sigma_{i}^{z}=|i\rangle\langle i| \quad(1\leq i\leq d),
 \end{equation}
 \begin{equation}
 	\label{op2}
 	\sigma_{i j}^{x}=(|i\rangle\langle j|+| j\rangle\langle i|) / \sqrt{2}\quad(1\leq i<j\leq d),
 \end{equation}
 \begin{equation}
 	\label{op3}
 	\sigma_{i j}^{y}=(-\mathrm i|i\rangle\langle j|+\mathrm i| j\rangle\langle i|) / \sqrt{2} \quad(1\leq i<j\leq d),
 \end{equation}
 where  the set $\left\{|i\rangle_{i=1}^{d}\right\}$  is an arbitrary orthogonal basis. Note that all the operators are orthogonal w.r.t. the inner product $ \langle A, B\rangle=\operatorname{Tr}\left(A^{\dagger} B\right) $. For a given basis $\{\xi_k\}_{k=1}^{K_1}  $, denote  $ \operatorname{span}\left\{\{\xi_k\}_{k=1}^{K_1}\right\} $ as the set of all finite real linear combinations of $ \xi_k$ $(1\leq k \leq K_1) $. If $ \operatorname{span}\left\{\{\xi_k\}_{k=1}^{K_1}\right\} = \operatorname{span}\left\{\{\mu_k\}_{k=1}^{K_2}\right\} $,  we say  $\{\xi_k\}_{k=1}^{K_1}$ and $\{\mu_k\}_{k=1}^{K_2}$ are equivalent. If the span of the probe state set contains all $ d $-dimensional Hermitian matrices and $ M=d^2 $, we say the probe state set is complete. If we further have $ M>d^2 $, we say it is over-complete.

In this section, we assume the true value of a rank $ r $ POVM element is $ P $, the Step 1 estimation is $ \tilde{P} $ and the Step 2 estimation is $ \hat{P} $. The  spectral decomposition of $ P $ is $ {P}=\sum_{i=1}^{d} {\lambda}_{i}\left|{\lambda}_{i}\right\rangle\left\langle{\lambda}_{i}\right| $ where $\lambda _1\geq \lambda _2\geq\cdots\geq \lambda _r>0   $, $ \lambda _{r+1}=\cdots=\lambda _d =0$. In \eqref{op1}-\eqref{op3},  if we change  $\{|i\rangle\}$ to  $\{|\lambda_{i}\rangle\}$ for $i=1, \ldots, d$, we call this new basis $ \mathcal B\left(\left\{|\lambda_i\rangle_{i=1}^{d}\right\}\right) $ \emph{ideal bases}. If we change  $\{|i\rangle\}$ to  $\{|\lambda_{i}\rangle\}$ with $ i $  restricted in $[r+1, d]$, we call the set $ \mathcal B\left(\left\{|\lambda_i\rangle_{i=r+1}^{d}\right\}\right) $ of these new $ (d-r)^2 $ elements as the \emph{null bases} of $ P $. If we change  $\{|i\rangle\}$ to  $\{|\lambda_{i}\rangle\}$ with $ i $  restricted in $[1, r]$, we call the set $ \mathcal B\left(\left\{|\lambda_i\rangle_{i=1}^{r}\right\}\right) $ of these new $ r^2 $ elements as the \emph{range bases} of $ P $.
 After Step 1, we obtain an estimate $ \tilde P $ and the spectral decomposition is $ \tilde{P}=\sum_{i=1}^{d} \tilde{\lambda}_{i}\left|\tilde{\lambda}_{i}\right\rangle\left\langle\tilde{\lambda}_{i}\right| $.  If we change  $\{|i\rangle\}$ to  $\{|\tilde\lambda_{i}\rangle\}$ for $i=1, \ldots, d$, we call this new basis $ \mathcal B\left(\left\{|\tilde \lambda_i\rangle_{i=1}^{d}\right\}\right) $ \emph{estimated bases}. If we change  $\{|i\rangle\}$ to  $\{|\tilde\lambda_{i}\rangle\}$ with $ i $  restricted in $[r+1, d]$, we call the set $ \mathcal B\left(\left\{|\tilde \lambda_i\rangle_{i=r+1}^{d}\right\}\right) $ of these new $ (d- r)^2 $ elements \emph{estimated null bases}.

 Then we give the following theorem as a guideline to design the probe state set.
 \begin{theorem}
 	\label{theorem4}
 	For two-step adaptive QDT using arbitrary reconstruction algorithm with an $O(1/N)$ scaling for the MSE, suppose the resource number is $ N_0 $ in Step 1 and $ N-N_0 $ in Step 2, both evenly distributed for each probe state. The infidelity $ \mathbb{E}\left(1-F(\hat{P},P)\right) $ of any rank-deficient POVM element scales as $ O\left(\frac{1}{ \sqrt{N_{0}(N-N_{0})}}\right)+O\left(\frac{1}{N-N_0}\right) $ if
 	\begin{enumerate}
 		\item[c1)]  the probe states  in Step 1 are complete or over-complete;
 		\item[c2)]  the probe states in Step 2 are complete or over-complete;
 		\item[c3)] the probe state set in Step 2 includes a subset equivalent to the $ (d- r)^2 $ estimated null basis $ \mathcal B\left(\left\{|\tilde \lambda_i\rangle_{i=r+1}^{d}\right\}\right) $ from Step 1.
 	\end{enumerate}
 \end{theorem}
 \begin{pf}
 From Condition c1), we obtain a rank $ \tilde{r} $ estimation $ \tilde P $ and the spectral decomposition is $ \tilde{P}=\sum_{i=1}^{d} \tilde{\lambda}_{i}\left|\tilde{\lambda}_{i}\right\rangle\left\langle\tilde{\lambda}_{i}\right| $ where $\tilde\lambda _1\geq \tilde\lambda _2\geq\cdots\geq \tilde\lambda_{\tilde r}>0   $, $\tilde \lambda _{\tilde r+1}=\cdots=\tilde \lambda _d =0$. 
 Thereby, the inner product between the eigenvectors of $P$ and their estimates obtained from $\tilde P$  \cite{PhysRevA.98.012339} is
 \begin{equation}
 	\label{erroreig}
 	\begin{aligned}
 		\mathbb{E}\left(\left\langle\lambda_{i} | \tilde{\lambda}_{j}\right\rangle\left\langle\tilde{\lambda}_{j} | \lambda_{i}\right\rangle\right)&=O\left(\frac{1}{N_{0}}\right),\\
 		\mathbb{E}\left(\left\langle\lambda_{i} | \tilde{\lambda}_{i}\right\rangle\left\langle\tilde{\lambda}_{i} | \lambda_{i}\right\rangle\right)&=1+O\left(\frac{1}{N_{0}}\right).
 	\end{aligned}
 \end{equation}

 In the estimated basis $ \mathcal B\left(\left\{|\tilde \lambda_i\rangle_{i=1}^{d}\right\}\right) $, the true value matrix $ P $ can be uniquely represented as
 \begin{equation}
 	\begin{aligned}
 		{P}=\sum_{1 \leq i<j \leq d}\left(S_{i j}^{x} \tilde\sigma_{i j}^{x}+S_{i j}^{y} \tilde\sigma_{i j}^{y}\right)+\sum_{k=1}^{d}S_{k}^{z} \tilde\sigma_{k}^{z}.
 	\end{aligned}
 \end{equation}
 Then in Step 2, according to Conditions c2) and c3), we can obtain an estimate 
 \begin{equation}
 	\begin{aligned}
 		{\hat P}=\sum_{1 \leq i<j \leq d}\left(\hat S_{i j}^{x} \tilde\sigma_{i j}^{x}+\hat S_{i j}^{y} \tilde\sigma_{i j}^{y}\right)+\sum_{k=1}^{d}\hat S_{k}^{z} \tilde\sigma_{k}^{z}.
 	\end{aligned}
 \end{equation}
 
 For Step 2 estimation,  the first-order term is 
 \begin{equation}
 	\label{first}
 	\begin{aligned}
 	&\mathbb{E}\left(\sum_{t=r+1}^{d}\left\langle\lambda_{t}|\Delta_1| \lambda_{t}\right\rangle\right)	\\
 		&=\mathbb{E}\left[\sum_{1 \leq i< j \leq d} \sum_{t=r+1}^{d}\left(\frac{\hat{S}_{i j}^{x}}{\operatorname{Tr}(\hat P)}-\frac{S_{i j}^{x}}{\operatorname{Tr}( P)}\right)\left\langle\lambda_{t}\left|\tilde\sigma_{i j}^{x}\right| \lambda_{t}\right\rangle\right.\\
 		&+ \sum_{1 \leq i< j \leq d} \sum_{t=r+1}^{d}\left(\frac{\hat{S}_{i j}^{y}}{\operatorname{Tr}(\hat P)}-\frac{S_{i j}^{y}}{\operatorname{Tr}( P)}\right)\left\langle\lambda_{t}\left|\tilde\sigma_{i j}^{y}\right| \lambda_{t}\right\rangle\\
 		&+\left.\sum_{k=1}^{d} \sum_{t=r+1}^{d}\left(\frac{\hat{S}_{k}^{z}}{\operatorname{Tr}(\hat P)}-\frac{S_{k}^{z}}{\operatorname{Tr}( P)}\right)\left\langle\lambda_{t}\left|\tilde\sigma_{k}^{z}\right| \lambda_{t}\right\rangle\right]\\
 		&\leq\mathbb{E}\left(\sum_{1 \leq i< j \leq d} \sum_{t=r+1}^{d}\left|\frac{\hat{S}_{i j}^{x}}{\operatorname{Tr}(\hat P)}-\frac{S_{i j}^{x}}{\operatorname{Tr}( P)}\right|\left|\left\langle\lambda_{t}\left|\tilde\sigma_{i j}^{x}\right| \lambda_{t}\right\rangle\right|\right)\\
 		&+\mathbb{E}\left(\sum_{1 \leq i< j \leq d} \sum_{t=r+1}^{d}\left|\frac{\hat{S}_{i j}^{x}}{\operatorname{Tr}(\hat P)}-\frac{S_{i j}^{y}}{\operatorname{Tr}( P)}\right|\left|\left\langle\lambda_{t}\left|\tilde\sigma_{i j}^{y}\right| \lambda_{t}\right\rangle\right|\right) \\
 		&+\mathbb{E}\left(\sum_{k=1}^{d} \sum_{t=r+1}^{d}\left|\frac{\hat{S}_{k}^{z}}{\operatorname{Tr}(\hat P)}-\frac{S_{k}^{z}}{\operatorname{Tr}( P)}\right|\left|\left\langle\lambda_{t}\left|\tilde\sigma_{k}^{z}\right| \lambda_{t}\right\rangle\right|\right).		
 	\end{aligned}
 \end{equation}
 Then we show the first-order term scales as
\begin{equation}
 \begin{aligned}
 \mathbb{E}\left(\sum_{t=r+1}^{d}\left\langle\lambda_{t}|\Delta_1| \lambda_{t}\right\rangle\right)=O\left(\frac{1}{ \sqrt{N_{0}(N-N_{0})}}\right),
 \end{aligned}
 \end{equation}
and the detailed calculation can be found in \ref{appenF}.
 
 Note that the $ (d- r)^2 $ estimated null bases are not probe states. Hence, we cannot directly use them. We need to use the linear combinations of these estimated null bases to construct quantum states. According to Condition c3), the probe state set $\left\{\tilde\rho_{l}\right\}_{l=1}^{M}$ includes a subset $ \left\{\tilde\rho_{l}\right\}_{l=1}^{(d-r)^{2}} $ equivalent to the $ (d- r)^2 $ estimated null bases $ \mathcal B\left(\left\{|\tilde \lambda_i\rangle_{i=r+1}^{d}\right\}\right) $. We assume the practical measurement results are $ \hat{p}_{l} $ and the ideal measurement results are $ \tilde{p}_{l} $ for these probe states. For $ 1\leq l\leq (d-r)^2 $, the measurement results $ \tilde{p}_l $ are used to estimate the null bases of $ P $, and for the other $ M-(d-r)^2 $ measurement results, they are used to estimate the range bases of $ P $. Even though we do not know ideal measurement results $ \tilde{p}_{l} $, we can still use them like \eqref{sx1} and \eqref{sx2} because we focus on infidelity.
 Using the linear combinations of $ \left\{\tilde{p}_{l}\right\}_{l=1}^{(d-r)^{2}} $, we can obtain $ \hat{S}_{i j}^{x}=\sum_{l=1}^{(d-r)^{2}} \hat a_l^{ij}\tilde{p}_{l} $ and $ {S}_{i j}^{x}=\sum_{l=1}^{(d-r)^{2}}  a_l^{ij}\tilde{p}_{l} $ for $ r+1\leq i<j \leq d $ for certain real numbers $ \hat{a}_{l}^{ij} $ and $ {a}_{l}^{ij} $. The relation \eqref{sx2} also holds because $ \left|\hat{a}_{l}^{ij}-{a}_{l}^{ij}\right|=O\left(\frac{1}{\sqrt{N-N_{0}}}\right) $. Then we  assume $ \hat{S}_{k}^{z}=\sum_{l=1}^{(d-r)^{2}} c_l^{k}\hat{p}_{l} $ for certain real numbers $ {c}_{l}^{k} $. From condition c3), $\left\{\tilde\rho_{l}\right\}_{l=1}^{(d-r)^{2}}  $  are linear combinations of the estimated null bases $ \mathcal B\left(\left\{|\tilde \lambda_i\rangle_{i=r+1}^{d}\right\}\right) $,  and thus $ \tilde{p}_{l}=O\left(\frac{1}{N_{0}}\right)$. Hence, like \eqref{varsz}, we have
\begin{equation}
	\label{sz2}
	\begin{aligned}
		&\operatorname{var}\left({\hat{S}_{k}^{z}}\right)=\operatorname{var}\left(\sum_{l=1}^{(d-r)^2} c_{l}^{k} \hat{p}_{l}\right)\\
		&=\frac{1}{N_{k}^{z}}\left[\sum_{l=1}^{(d-r)^2} (c_{l}^{k})^2 \tilde{p}_{l}\left(1-\tilde{p}_{l}\right)+\operatorname{cov}_{i \neq j}\left(c_{i}^{k} \tilde{p}_{i}, c_{j}^{k} \tilde{p}_{j}\right)\right]\\
		&=O\left(\frac{1}{N_{0}\left(N-N_{0}\right)}\right).
	\end{aligned}
\end{equation}
Therefore, using probe states satisfying Conditions c1)-c3), the first-order term scales as 
\begin{equation}
\label{firstinf}
\begin{aligned}
\mathbb{E}\left(\sum_{t=r+1}^{d}\left\langle\lambda_{t}|\Delta_1| \lambda_{t}\right\rangle\right)=O\left(\frac{1}{ \sqrt{N_{0}(N-N_{0})}}\right).
\end{aligned}
\end{equation}
Finally, the second-order term scales as 
\begin{equation}
	\label{secondinf}
	\begin{aligned}
	&\mathbb{E}\left(	\frac{1}{2} \sum_{t, k=1}^{r} \frac{\left|\left\langle\lambda_{t}\left|\Delta_{1}\right| \lambda_{k}\right\rangle\right|^{2}}{\lambda_{t}+\lambda_{k}}\right) \sim O\left(\frac{1}{N-N_{0}}\right),\\
	&\mathbb{E}	\left(\frac{1}{4}\left[\sum_{t=r+1}^{d}\left\langle\lambda_{t}\left|\Delta_{1}\right| \lambda_{t}\right\rangle\right]^{2}\right) \sim O\left(\frac{1}{N_{0}\left(N-N_{0}\right)}\right),\\
	&\mathbb{E}\left(	\sum_{t=1}^{d} \sum_{k=1}^{d}\left\langle\lambda_{t}\left|\Delta_{2}\right| \lambda_{t}\right\rangle\left\langle\lambda_{k}\left|\Delta_{2}\right| \lambda_{k}\right\rangle / d^{2}\right) \sim O\left(\frac{1}{N-N_{0}}\right).
	\end{aligned}
\end{equation}
 
 Using \eqref{firstinf} and \eqref{secondinf}, the infidelity $ \mathbb{E}\left(1-F(\hat{P},P)\right) $ of any rank-deficient POVM element scales as \\$ O\left(\frac{1}{ \sqrt{N_{0}(N-N_{0})}}\right)+O\left(\frac{1}{N-N_0}\right) $.
 \hfill $\Box$
\end{pf}

We have the following corollary if we use GPB (generalized Pauli basis) states shown in \eqref{standard1}--\eqref{standard3} in Step 2,

\begin{corollary}\label{corollary1}
 For a POVM element $ P $,	using GPB states in Step 2, the infidelity reaches the optimal scaling $ O(1/N) $ if $ N_0=\alpha N $ for certain $ 0<\alpha<1 $. 
\end{corollary}
The proof is straightforward. If the POVM element $ P $ is full rank, it does not have first-order term and the infidelity scales as $ O(1/N) $. If $ P $ is rank deficient with an unknown rank $ r $, GPB states always satisfy Conditions c1)-c3) and thus the infidelity still scales as $ O(1/N) $ by choosing $ N_0=\alpha N $ for certain $ 0<\alpha<1 $. Because GPB states are effective for all cases, we will show their performance in Numerical Examples (Section \ref{numerical}).

We may use three methods to further make the estimation from Step 1 more accurate if the rank is not precise.  The first one is to use Corollary \ref{corollary1}. We use $ d^2 $ GPB states which always span the null basis. The second one is the possible scenario where we know the rank value from prior information. The third method is, from the rough estimation of Step 1, we may know an exact rank interval of the to-be-estimated POVM element. Since  we know the upper bound of the estimated error as \eqref{final} here,  we thus know the variation range of all the eigenvalues from Lemma \ref{lemma1}. Assume the rank interval is  $ [a,b] $ where $ 1\leq a < b \leq d $, and $r\in [a,b]$ is the unknown rank. The more accurate is the estimation result from Step 1, the smaller is the interval length $b-a$.	
	Then to ensure the infidelity scaling is $ O(1/N) $, a conservative method is to  add $ (d-a)^2 $ quantum states spanning the estimated null basis $ \mathcal B\left(\left\{|\tilde \lambda_i\rangle_{i=a+1}^{d}\right\}\right) $.
\begin{rem}
The key in proving Theorem \ref{theorem4} is to characterize the first-order term in \eqref{first}, which is unaffected by the added term $-\left[\operatorname{Tr}\left(P_{i}-\hat{P}_{i}\right)\right]^{2}/d^2  $	comparing \eqref{infide2} with \eqref{infide}. Hence, when \eqref{infide} holds without distortion (see Proposition \ref{distort}), one can still use \eqref{infide} as the definition of fidelity, and GPB states still reach $ O(1/N) $ scaling for the infidelity.
\end{rem}

The choice of $N_{0}$ plays a key role in the performance of the two-step adaptive QDT. In this paper, we choose $ N_0=\frac{N}{2} $ and the infidelity scales as  $ O(1/N) $. Note that the infidelity behavior for different detector matrices can be different, even for binary detectors. For example, if $P_{1}=U_{1} \operatorname{diag}\left(1, 0 ,0,0\right) U_{1}^{\dagger}$ for certain unitary $ U_1 $ and $P_{2}=I-P_1 $, $ P_2 $'s eigenvalues are $ 1,1,1,0 $. Since $ P_2 $ has one zero eigenvalue, the infidelity reaches $ O(1 / \sqrt{N}) $ by non-adaptive QDT. However, if $ P_1=U_{1} \operatorname{diag}\left(0.1,0 ,0,0\right) U_{1}^{\dagger} $, $ P_2 $'s eigenvalues are $ 1,1,1,0.9 $. Since $ P_2 $ has no zero eigenvalues, the infidelity can reach $ O(1 / {N}) $ by non-adaptive QDT. Therefore, for a complete characterization and analysis, we need to calculate the infidelity for every POVM element.

\section{Numerical Examples}\label{numerical}
Both the non-adaptive and adaptive QDT protocols need to prepare certain (in this paper pure) states $\{|\psi\rangle\}$ (we call them \emph{ideal} states) as the probe states, which can be difficult to achieve in practice. As stated in Sec. \ref{coherent}, a realistic way in quantum optics experiments is to use the superposition of coherent states to approximate the ideal states. In this section we demonstrate the performance of our optimal and adaptive protocols both using ideal probe states and using superposed coherent probe states. We use the two-stage QDT reconstruction algorithm for numerical simulations.
\subsection{Optimal detector tomography}
\begin{table*}[h]
	\centering
	TABLE I.  Comparison of  various four-dimensional QDT protocols.
	\begin{threeparttable}
	\begin{tabular}{llllllllll}
		\cline{1-7}
		Protocol & Probe states & Number($ M $)  &  $ M\operatorname{Tr}\big[(X^{T} X)^{-1}\big] $ & $ \operatorname{cond}(X) $  & Eqs or Ref \\ \cline{1-7}
		1        &  SIC     &        16                           &     $304^*$   & $\sqrt{5}^*$ & \eqref{sicm} \\
		2	&   MUB       &    20    &      $ 304^{*} $                &   $\sqrt{5}^*$    &  \eqref{mubbase}   \\
		3	&  $\text{Cube}^{**} $       &  36                     &       $400$           &  $3$   &   \eqref{cubebase}  \\ 
		4	&  GPB        &    16                   &       
	      $640$   & 	$ \sqrt{\dfrac{9+\sqrt{73}}{9-\sqrt{73}}} $      &  \eqref{standard1},\eqref{standard2},\eqref{standard3}   \\
		5	&   Random Pure       &  32                  &             $629.16$  &   $2.39$    &  \cite{Zyczkowski_1994,MISZCZAK2012118} \\ 
		\hline
		6	&  1-coherent SIC        &    16                      &             ${2.48\times10^{4}} $     &  $ 307.11 $ &  \eqref{newcost}  \\
		
		7	&   1-coherent MUB       &   20                  &       $2.46\times10^3 $         & $82.37$    &  \eqref{newcost}  \\ 
		8	&   1-coherent Random       &   32                  &     $ 9.81\times10^3 $            &  $26.03$   & \cite{wang2019twostage}  \\ 
		\hline
		9	&  2-coherent SIC        &  16                     &     $ 400.61 $            &  $3.44$    &   \eqref{newcost}  \\
		10	&   2-coherent MUB       &   20                   &      $ 470.67 $        &  $3.27$   &  \eqref{newcost}  \\ 
		11	&   2-coherent Random       &    32                  &         $ 607.35$        &  $5.48$   & \cite{wang2019twostage}  \\ 
		\hline
		12	&  3-coherent SIC        & 16         &$ 352.82$                     &  $2.79$    &       \eqref{newcost}        &         \\
		13	&   3-coherent MUB       &   20                     &        $ 364.01 $         &  $2.60 $  & \eqref{newcost}   \\ 
		14	&   3-coherent Random       &    32                    &      $532.85$          &  $4.95$   & \cite{wang2019twostage}  \\ 
		\hline
	\end{tabular}
\begin{tablenotes}
	\footnotesize
	\item[*] This is the optimal value.
	\item[**] Cube states are product probe states.
\end{tablenotes}
\end{threeparttable}
\end{table*}
For non-adaptive  QDT in $d=4$ systems, we test  $ 14 $ different protocols   using different probe states in Table I. In protocols 1-5, we compare $M\operatorname{Tr}\big[(X^{T} X)^{-1}\big]$ and condition numbers of ideal pure states such as MUB, Cube, SIC, GPB probe states which are constructed as in  \ref{appenA}. ``Cube" states here are product states of one-qubit MUB states, assuming that the $d=4$ system here is the composition of two $d=2$ systems.  GPB states are shown in  \eqref{standard1}--\eqref{standard3}, similar to the optimal measurement-generalized Pauli operators in \cite{PhysRevA.90.062123}. ``Random Pure" means that we  generate $ 32 $ random pure states using the algorithm in \cite{Zyczkowski_1994,MISZCZAK2012118} where $M\operatorname{Tr}\big[(X^{T} X)^{-1}\big]$ and condition number are obtained by the average of $ 1000 $ results. We find SIC states and MUB states have the minimum value  of $M\operatorname{Tr}\big[(X^{T} X)^{-1}\big]$ as $ 304 $ and minimum condition number as $ \sqrt{5} $, satisfying Theorem \ref{theorem1}. For two-qubit product states--Cube states which are easier to generate in experiment, the values of UMSE and conidtion number are $400$ and $3$, respectively, a little larger than the optimal values and satisfying Theorem \ref{theorem2}.

In protocols 6-14, we use superposition of $ n_c$ $(n_c=1,2,3) $ coherent states (denoted as $ n_c $-coherent SIC/MUB) to approximate the four-dimensional SIC and MUB states by solving the optimization problem  \eqref{newcost}. The case of $ n_c=1 $ is using one coherent state without superposition.  As a comparison, we also  add  the protocols of using the  superposition of $ n_c $ random coherent states (denoted as $ n_c $-coherent Random). The random algorithm is the same as the coherent states preparation procedure in \cite{wang2019twostage}.  The optimization landscape might have local minima. Therefore, we run the optimization algorithm with $ 100 $ different initial values and choose the best result.
 
 We find for one coherent state, it cannot approximate SIC and MUB states well and both the condition number and UMSE are large. For the superposition of two and three coherent states, they can approximate SIC and MUB states well and the corresponding  condition number and UMSE are close to the optimal values, and smaller than 2,3-coherent Random protocols. Also, with $ n_c $ increasing, the superposition result becomes close to ideal probe states and thus the UMSE and condition number decrease. 

\subsection{Adaptive QDT using ideal probe states }
\subsubsection{Binary detectors}
For binary detectors $ P_1+P_2=I $, $ P_1 $ and $ P_2 $ can be simultaneously diagonalized by a common unitary \cite{9029759}. Hence,  the eigenvalues of $ P_1 $ will affect the eigenvalues of $ P_2 $. This determines  the rank of $ P_2 $, which will further influence the scaling of non-adaptive tomography. Let
	\begin{equation}\label{muall}
		\begin{aligned}
			&P_{1}=U_{1} \operatorname{diag}\left(\mu, 0,0,0\right) U_{1}^{\dagger}.\\
		\end{aligned}
	\end{equation}	
	With non-adaptive tomography, when $\mu<1$, $P_{2}$ is full-ranked and the infidelity of estimating $P_{2}$ scales as $O(1 / N)$, while for $\mu=1$, $P_{2}$ is rank-deficient and the infidelity of estimating $P_{2}$ scales as $O(1 / \sqrt{N})$.

Therefore, we firstly consider a  binary detector for $ \mu=1 $ where
\begin{equation}
\label{binary1}
\begin{aligned}
&P_{1}=U_{1} \operatorname{diag}\left(1, 0,0,0\right) U_{1}^{\dagger}.\\
\end{aligned}
\end{equation}
This detector is fully specified by the projection measurement  $ P_1 $. The matrix $ U_1 $ is randomly generated using the algorithm in \cite{Zyczkowski_1994,qetlab}. For each resource number, we run the algorithm $ 100 $ times and obtain the average infidelity and standard deviation.
\begin{figure}
	\centering
	\subfigure{
		\begin{minipage}[b]{0.5\textwidth}
			\includegraphics[width=1\textwidth]{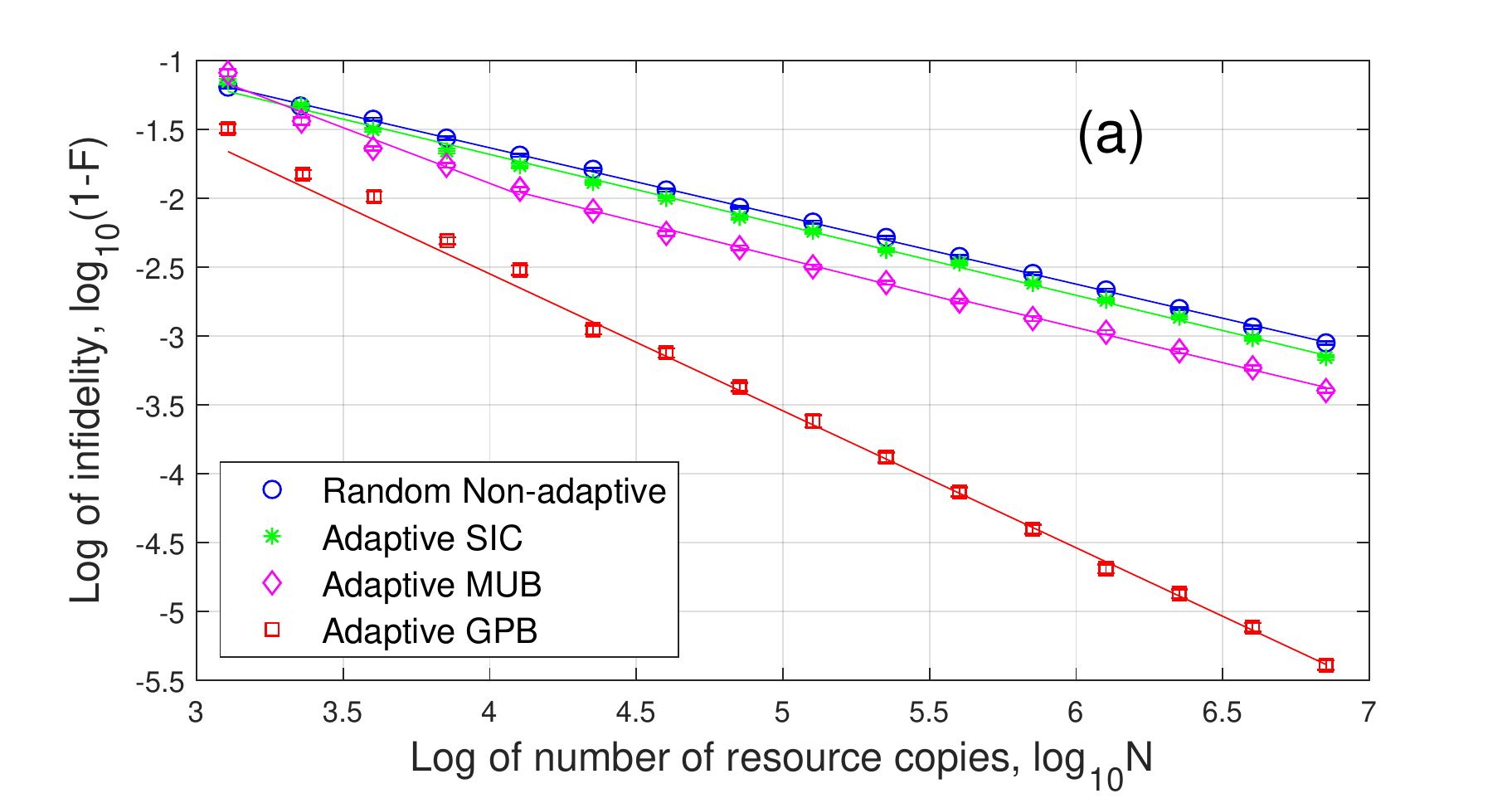}
		\end{minipage}
	}
	\subfigure{
		\begin{minipage}[b]{0.5\textwidth}
			\includegraphics[width=1\textwidth]{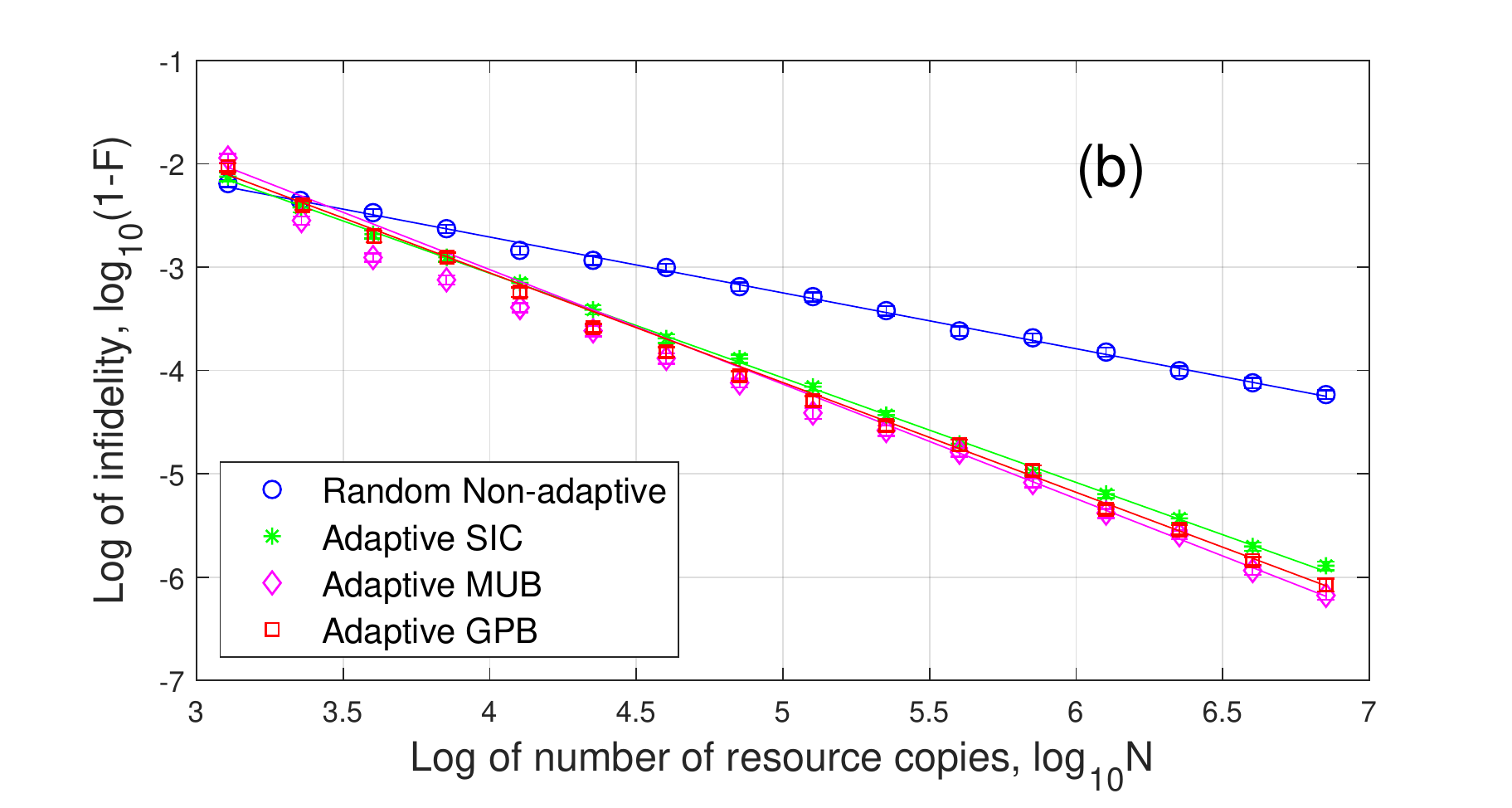}
		\end{minipage}
	}
	\caption{Infidelity for binary detectors with given unitary matrix $ U_1 $ in \eqref{binary1}. (a) $ P_1 $; (b) $ P_2 $.} \label{dis21}
\end{figure}

The four curves in Fig. \ref{dis21} are as follows:
\begin{itemize}
	\item Random Non-adaptive: We only have Step 1 and choose 48 random pure states.
	\item Adaptive SIC: In Step 1, we use $ 16 $ SIC states $ \left\{\rho_{j}^{\text{(SIC)}}\right\} $ and in Step 2, we use  $ 32 $ new states as \eqref{newprobe}.
	\item Adaptive MUB: In Step 1, we use $ 20 $ MUB states $ \left\{\rho_{j}^{\text{(MUB)}}\right\} $ and in Step 2, we use  $ 40 $ new states as \eqref{newprobe}.
	\item Adaptive GPB: In Step 1, we use $ 16 $ GPB states  and in Step 2, we use $ 32 $ new GBP states by  replacing the set $\{|i\rangle\}$ by $\left\{\left|\tilde \lambda_{i}\right\rangle\right\}$ in \eqref{standard1}, \eqref{standard2} and \eqref{standard3}.
\end{itemize}

As shown in Fig. \ref{dis21}, Random Non-adaptive tomography  only reaches  $1-F=O(1 / \sqrt{N})$ for both $ P_1 $ and $ P_2 $ because they both have zero eigenvalues and the first-order term scales as $ O(1 / \sqrt{N}) $. Adaptive GPB tomography can reach  $1-F=O(1 / {N})$ for $ P_1 $ and $ P_2 $ as proved in Corollary \ref{corollary1}. For  adaptive SIC and MUB tomography, they can only reach  $1-F=O(1 / \sqrt{N})$ for $ P_1 $ probably because adaptive SIC and MUB states do not have a subset equivalent to the $ 9 $ estimated null bases and does not satisfy Condition c3) in Theorem \ref{theorem4}. They can reach  $1-F=O(1 / {N})$ for $ P_2 $ because adaptive SIC and MUB states have a subset equivalent to the  estimated null basis and  satisfy Condition c3).

\begin{figure}
	\centering
	\subfigure{
		\begin{minipage}[b]{0.5\textwidth}
			\includegraphics[width=1\textwidth]{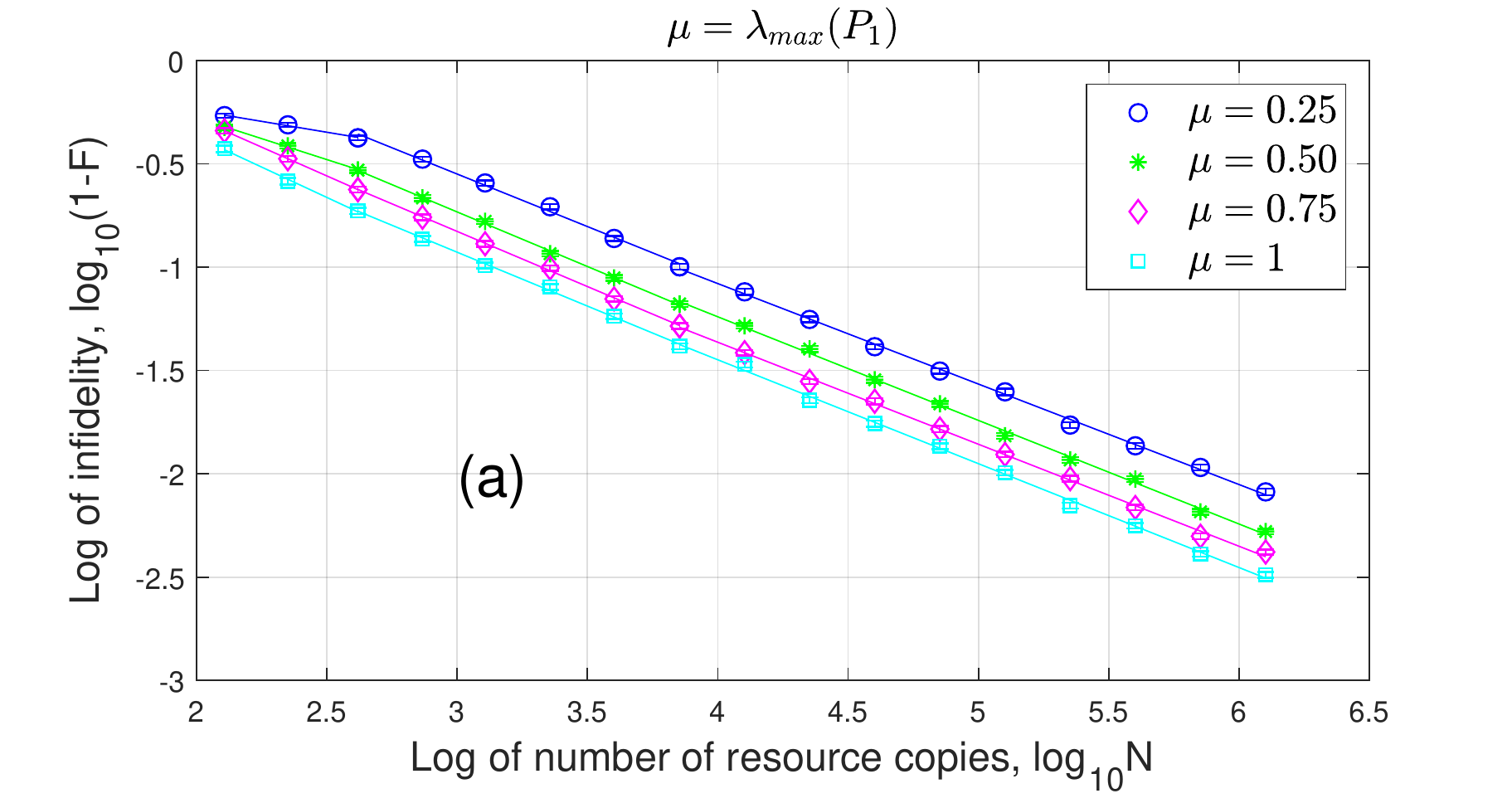}
		\end{minipage}
	}
	\subfigure{
		\begin{minipage}[b]{0.5\textwidth}
			\includegraphics[width=1\textwidth]{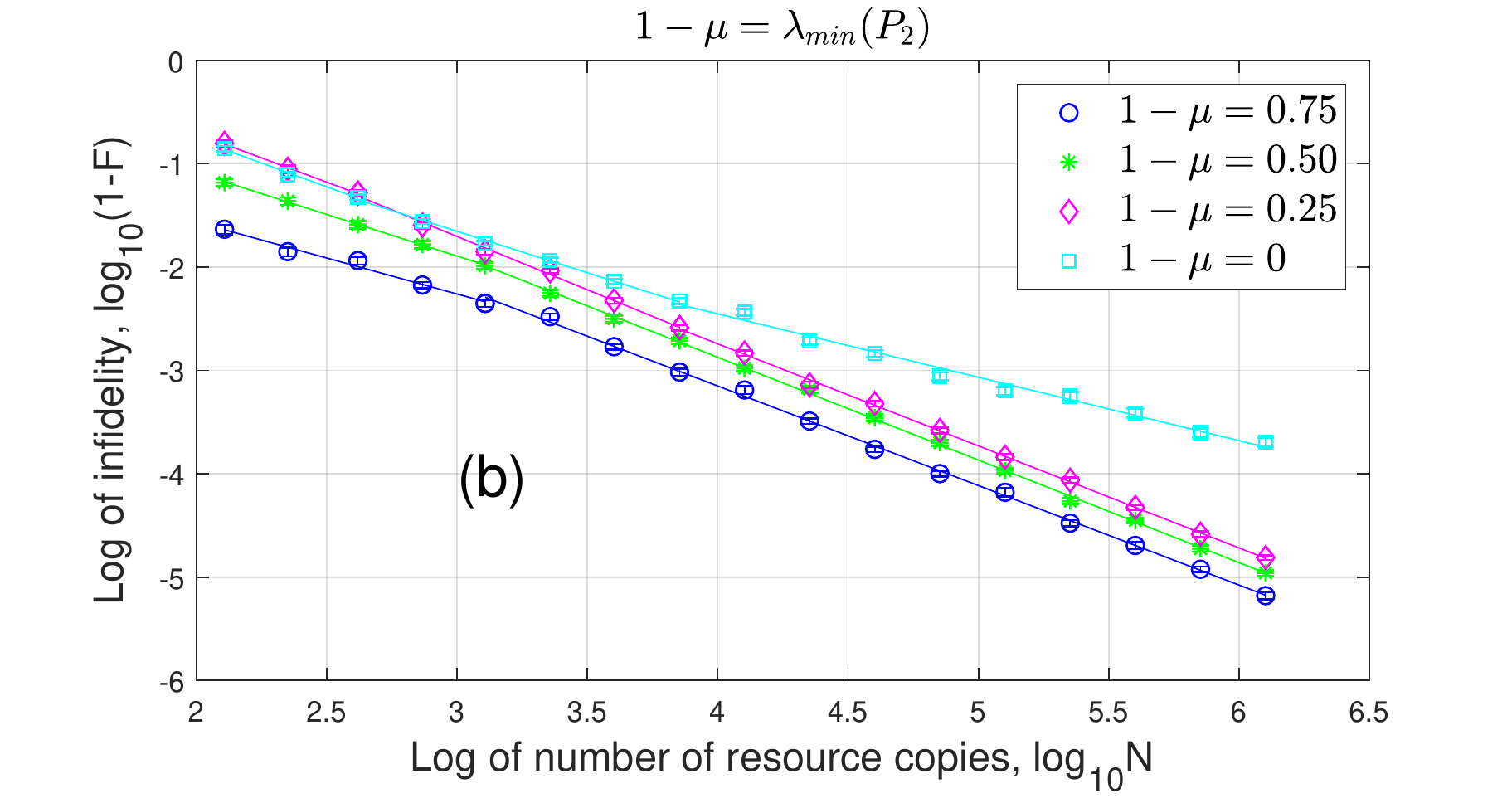}
		\end{minipage}
	}
	\caption{Infidelity for non-adaptive tomography with different eigenvalues $ \mu $ of binary detectors. (a) $ P_1 $; (b) $ P_2 $.}
	\label{mu}
\end{figure}
Then we show infidelity for non-adaptive tomography with different eigenvalues $ \mu=0.25,0.5,0.75,1 $ in \eqref{muall} of binary detectors in Fig. \ref{mu}. For each resource number, we run the algorithm $ 100$ times and obtain the average infidelity and the standard deviation. 
	Because $ P_1 $ is always rank deficient, the infidelity for $ P_1 $ scales as $ O(1 / \sqrt{N} )$. As $ \mu $  increases,  the measurement accuracy increases and thus the infidelity of $ P_1 $ becomes smaller for a given $ N $. Because $ P_2 $ is full rank for $ \mu<1 $, its infidelity scales as $ O(1 / N )$. In addition, the infidelity of $ P_2 $ becomes larger as $ \mu $  increases for a given  $ N $. When $ \mu $  increases to $ 1 $, both the  infidelities of $ P_1 $ and $ P_2 $ scale as $ O(1 / \sqrt{N} )$ because they are both rank deficient.

\begin{figure}
	\centering
	\subfigure{
		\begin{minipage}[b]{0.5\textwidth}
			\includegraphics[width=1\textwidth]{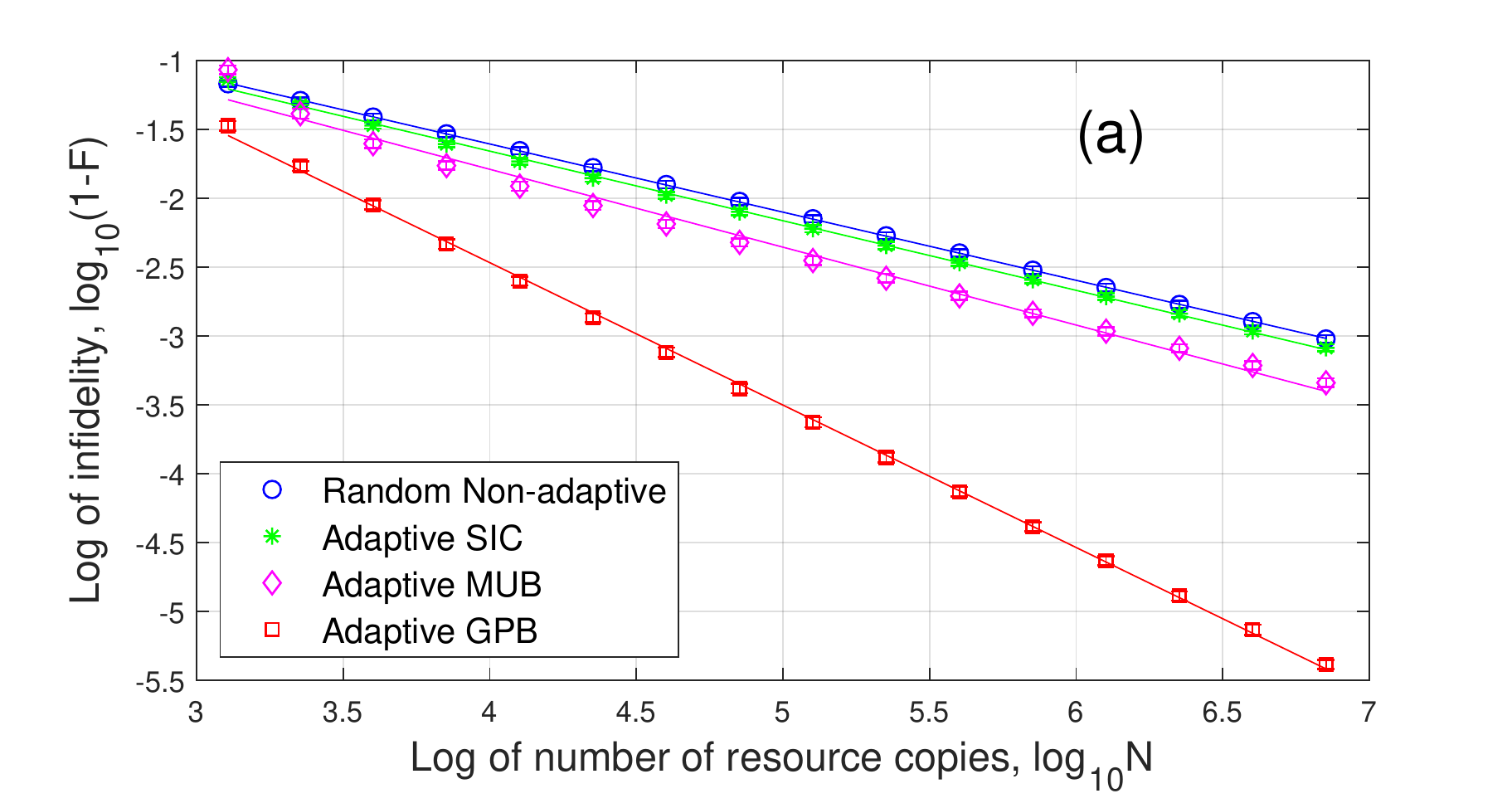}
		\end{minipage}
	}
	\subfigure{
		\begin{minipage}[b]{0.5\textwidth}
			\includegraphics[width=1\textwidth]{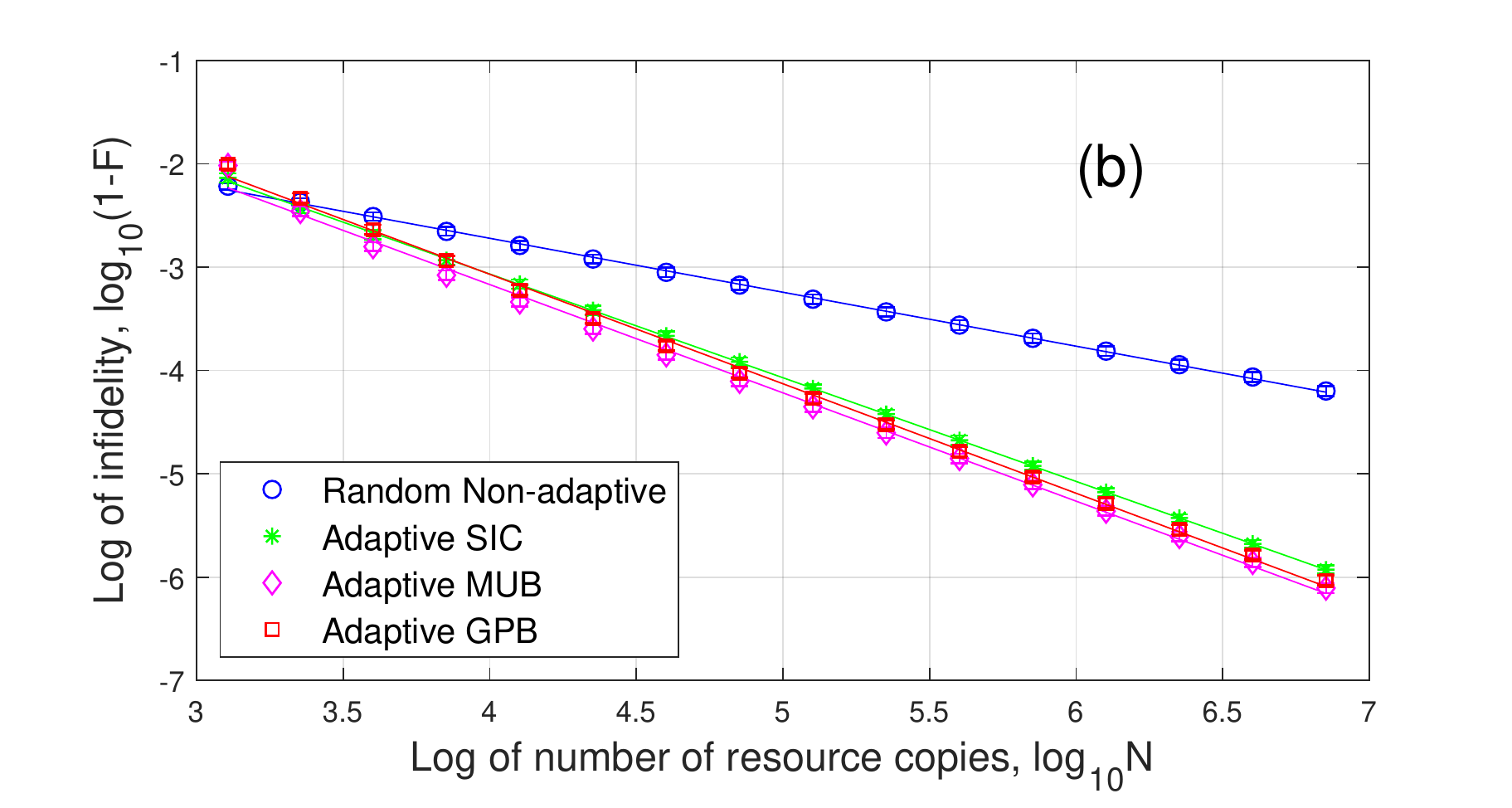}
		\end{minipage}
	}
	\caption{Infidelity for $ 200 $ different binary detectors by changing  $ U_1 $ in \eqref{binary1}. (a) $ P_1 $; (b) $ P_2 $.} \label{change2}
\end{figure}

To test the robustness of our adaptive protocol, we perform adaptive QDT on $ 200 $ random binary detectors in the form of  \eqref{binary1} by changing the unitary matrix $ U_1 $ which is  randomly created using the algorithm in \cite{Zyczkowski_1994,qetlab}. For each $ U_1 $, we run our tomography algorithm $ 100 $ times and obtain the mean infidelities for given resource number $ N $. Then we calculate the mean values and the standard deviations of these $ 200 $ mean infidelities. The result is shown in Fig. \ref{change2}. It is clear that Adaptive GPB tomography can reach  $1-F=O(1 / {N})$ for $ P_1 $ and $ P_2 $. In addition, all the  standard deviations are small, which demonstrates that our Adaptive GPB protocol is robust.

We then consider the case that $ P_1 $ is a perturbed projection measurement,
\begin{equation}
\label{binary2}
\begin{aligned}
&P_{1}+P_{2}=I,\\
&P_{1}=U_{1} \operatorname{diag}\left(0.6,0.001,0.001,0.001\right) U_{1}^{\dagger}.\\
\end{aligned}
\end{equation}
\begin{figure}
	\centering
	\subfigure{
		\begin{minipage}[b]{0.5\textwidth}
			\includegraphics[width=1\textwidth]{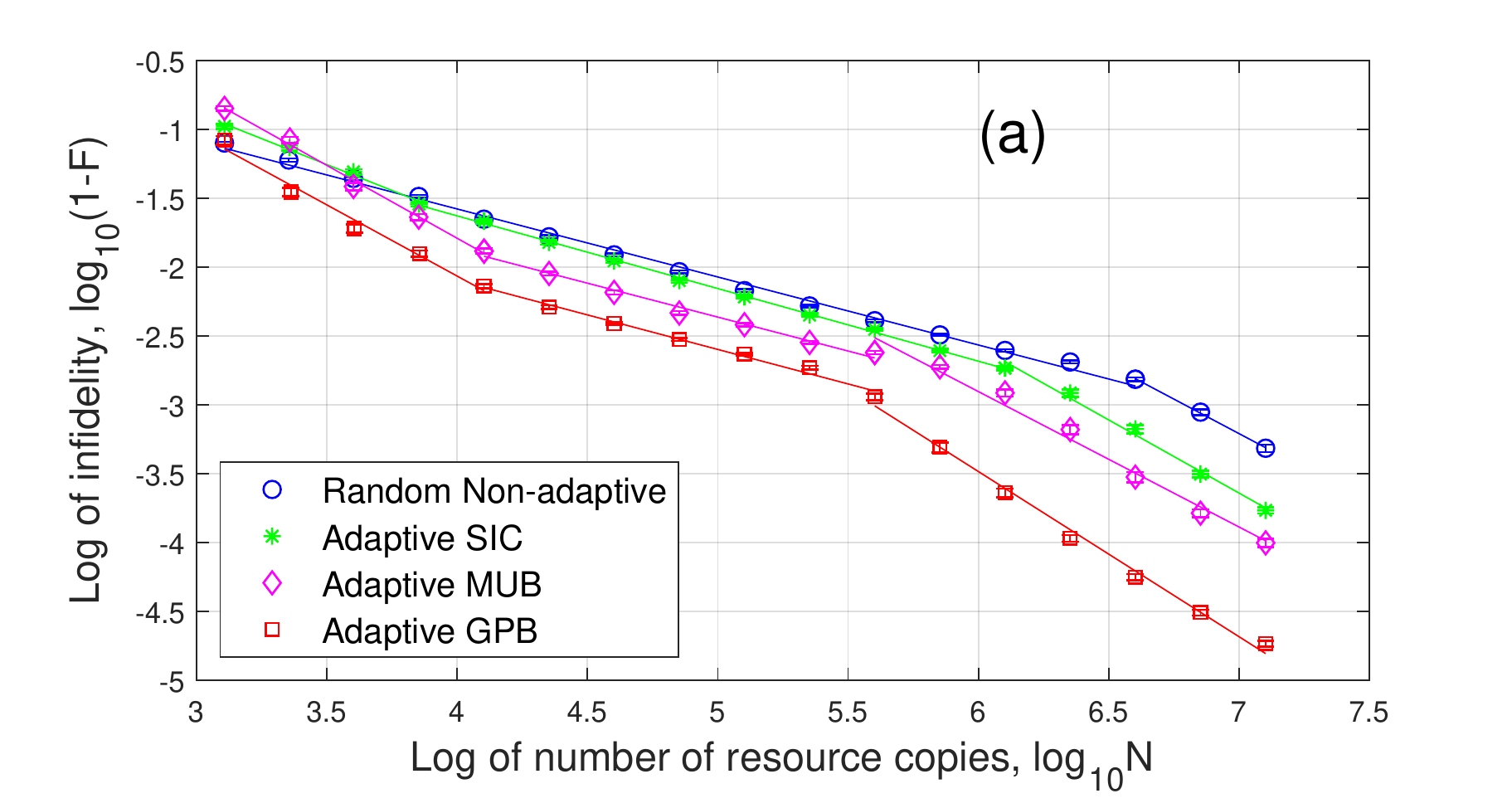}
		\end{minipage}
	}
	\subfigure{
		\begin{minipage}[b]{0.5\textwidth}
			\includegraphics[width=1\textwidth]{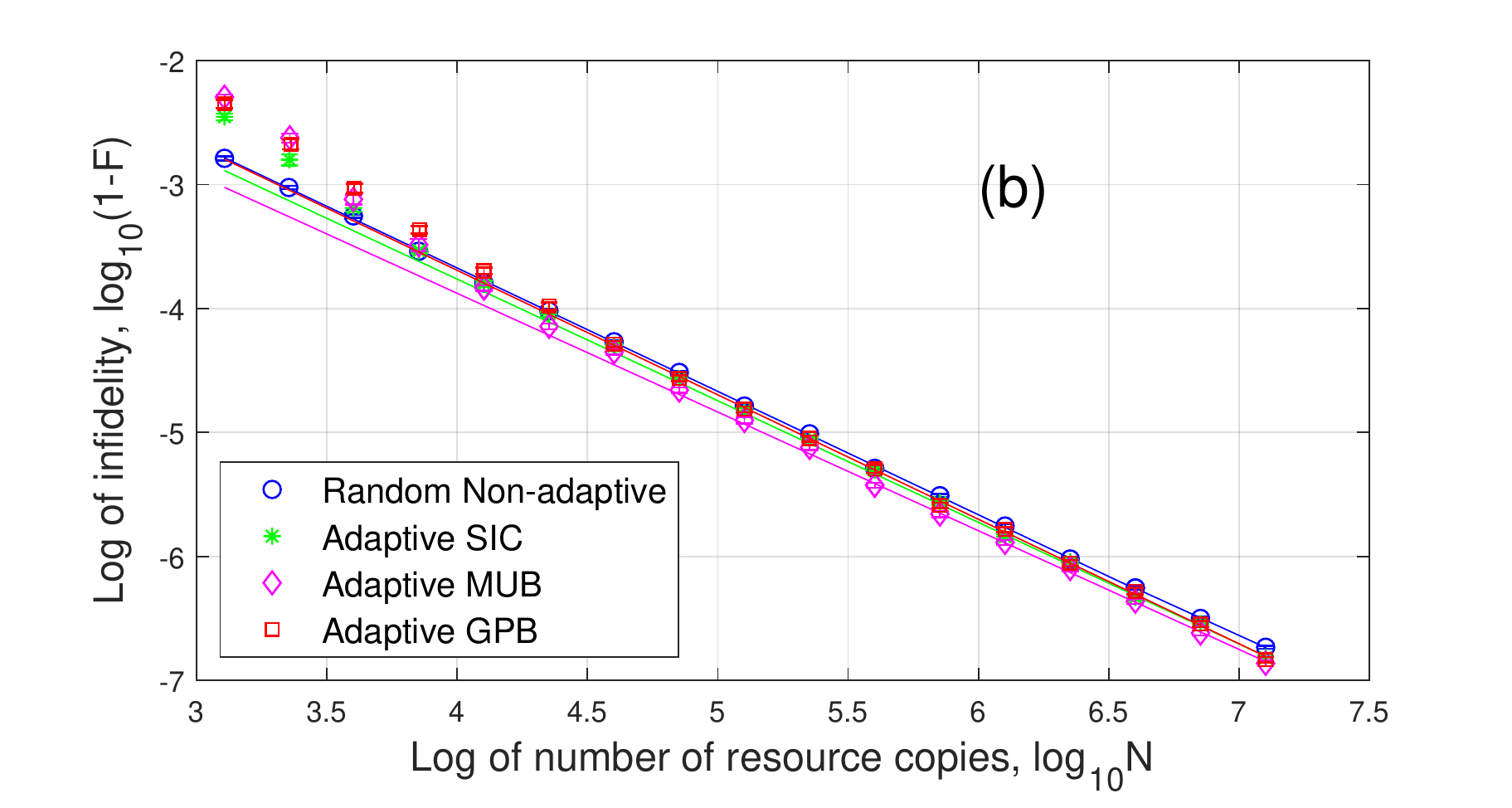}
		\end{minipage}
	}
	\caption{Infidelity for binary detectors in \eqref{binary2}. (a) $ P_1 $; (b) $ P_2 $.}
\label{dis22}
\end{figure}
The tomography errors are shown in Fig. \ref{dis22}. From Fig. \ref{dis22}(a), 
we can see that the curves of adaptive QDT can be roughly divided into three segments from left to right for the detector $ P_1 $, which is similar to the phenomenon in QST \cite{Qi2017}. In the first segment, the resource number is not large enough ($ N\leq10^4 $) and the near-zero eigenvalues are not strong enough to make a difference from zero. The performance is thus the same as projection measurement. Hence,
the infidelity decreases as $ O(1 / {N})$ firstly. When the resource number increases ($ 10^4\leq N\leq10^{5.5} $), the near-zero eigenvalues start to take effect. We cannot
distinguish them from zero accurately and thus the infidelity scales as $ O(1/\sqrt{N})$. Finally, when the resource number is large enough ($ N\geq10^{5.5} $) to clearly distinguish between the near-zero eigenvalues and
zero, we are performing full rank detector tomography actually, which has $ O(1 / {N})$
decay rate for infidelity.
For Random Non-adaptive tomography, it can be divided into two segments. When the resource number is not enough ($ N\leq10^{6.5} $) to estimate the near-zero eigenvalues accurately, the infidelity decreases as $ O(1/\sqrt{N})$. When the resource number is large enough ($ N\geq10^{6.5} $) to clearly distinguish between the near-zero eigenvalues and
zero, the infidelity scales as $ O(1/{N})$. Overall, the Adaptive GPB tomography is the best among these methods.

For detector $ P_2 $, all the eigenvalues are significantly larger than zero, and $ P_2 $ is full-rank. Hence, the infidelity decreases as $ O(1 / {N})$ for both non-adaptive and adaptive tomography.

\subsubsection{Three-valued detectors}
Three-valued detectors is different from binary detectors because three-valued detector matrices generally cannot be diagonalized by the same unitary matrix like binary detectors. We consider a three-valued detector as
\begin{equation}
	\label{three1}
\begin{aligned}
&P_{1}+P_{2}+P_{3}=I,\\
&P_{1}=U_{1} \operatorname{diag}\left(0.4, 0,0,0\right) U_{1}^{\dagger}=0.4 U_{1}(|00\rangle\langle 00|) U_{1}^{\dagger},\\
&P_{2}=U_{2} \operatorname{diag}\left(0, 0.5,0,0\right) U_{2}^{\dagger}=0.5 U_{2}(|01\rangle\langle 01|) U_{2}^{\dagger}.
\end{aligned}
\end{equation}
This detector is constructed from two-qubit MUB measurement (see \ref{appenA}). The first detector $ P_1 $ is from $ |00\rangle $ where we product a coefficient $ 0.4 $ and conjugate a unitary  rotation $ U_1 $. In a similar way,  $ P_2 $ is from $ |01\rangle $ where we product a coefficient $ 0.5 $ and  conjugate a unitary  rotation $ U_2 $. We can prove that  $ P_3=I-P_1-P_2 $ is always positive semidefinite. The unitary  rotations $ U_1 $ and $U_2  $ are generated by the random unitary algorithm in \cite{Zyczkowski_1994,qetlab}. For each resource number, we run the algorithm $ 100 $ times and obtain the average infidelity and standard deviation.
	\begin{figure*}[ht]
	\centering
\includegraphics[width=1\linewidth]{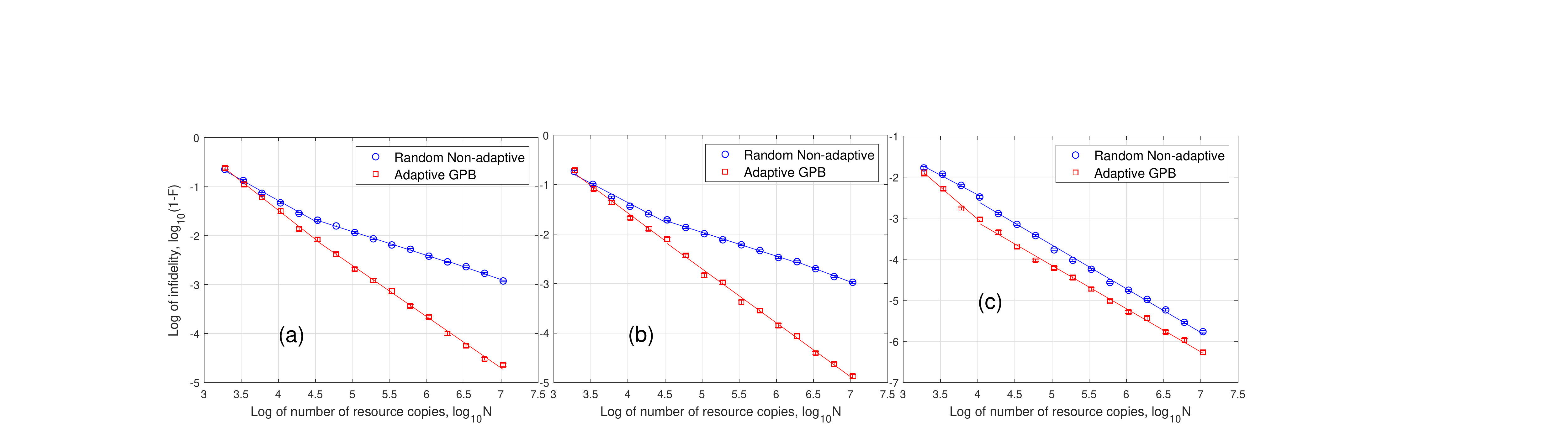}
\caption{Infidelity for three-valued detectors with given unitary matrices $ U_1 $ and $ U_2 $ in \eqref{three1}. (a) $ P_1 $; (b) $ P_2 $; (c) $ P_3 $.}
\label{dis31}
\end{figure*}
We focus on the comparison  between Adaptive GPB tomography and Random Non-adaptive tomography. The simulation result is in Fig. \ref{dis31} where our adaptive tomography can reach $ O(1/N) $ for $ P_1 $ and $ P_2 $ as proved in Corollary \ref{corollary1},  improving the $ O(1/\sqrt{N}) $ scaling of  non-adaptive tomography. For $ P_3 $, all the eigenvalues are far from  zero and both tomography methods can reach $ O(1/N) $.
We also test robustness by performing adaptive QDT on $ 200 $ random three-valued detectors in the form of  \eqref{three1} by changing unitary matrices $ U_1 $ and $ U_2 $, which is similar as binary detectors. The result is shown in Fig. \ref{change3}. For $ P_1 $ and $ P_2 $,  the adaptive tomography  is robust (with small standard deviation) and their infidelities can reach $ O(1/N) $. For full rank $ P_3 $, both tomography methods can reach $ O(1/N) $.
	\begin{figure*}[ht]
	\centering
	\includegraphics[width=1\linewidth]{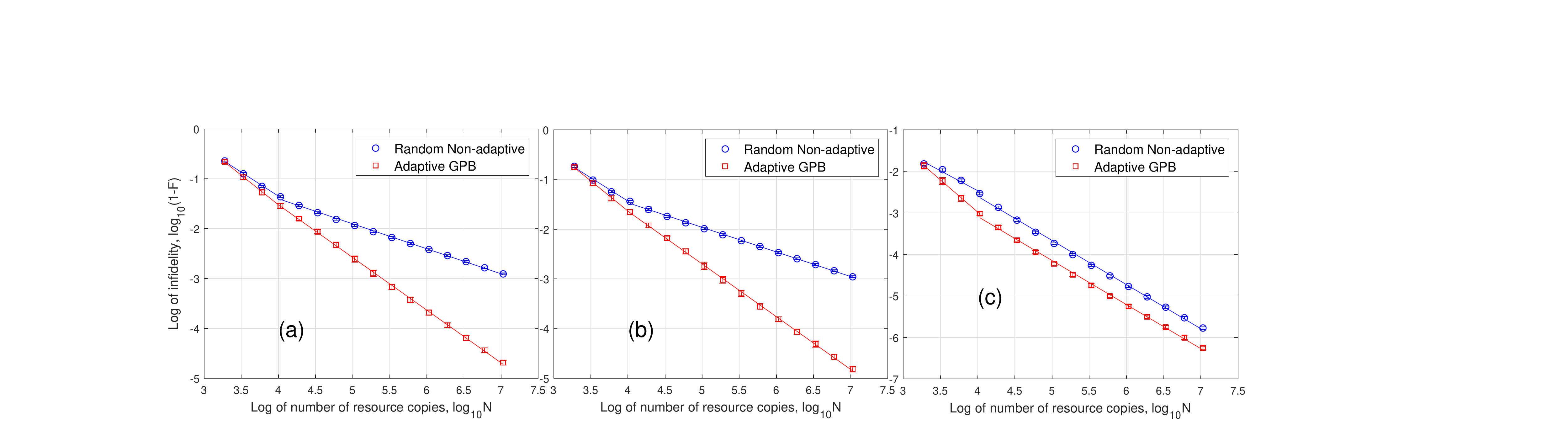}
	\caption{Infidelity for $ 200 $ different three-valued detectors by changing $ U_1 $ and $ U_2 $ in \eqref{three1}. (a) $ P_1 $; (b) $ P_2 $; (c) $ P_3 $.}
	\label{change3}
\end{figure*}
Then we consider that $ P_1 $ and $ P_2 $  have three small eigenvalues as
\begin{equation}
	\label{three0.6}
\begin{aligned}
&P_{1}+P_{2}+P_{3}=I,\\
&P_{1}=U_{1} \operatorname{diag}\left(0.4, 0.001,0.001,0.001\right) U_{1}^{\dagger},\\
&P_{2}=U_{2} \operatorname{diag}\left(0.001,0.5 ,0.001,0.001\right) U_{2}^{\dagger}.
\end{aligned}
\end{equation}

\begin{figure*}[ht]
	\centering
\includegraphics[width=1\linewidth]{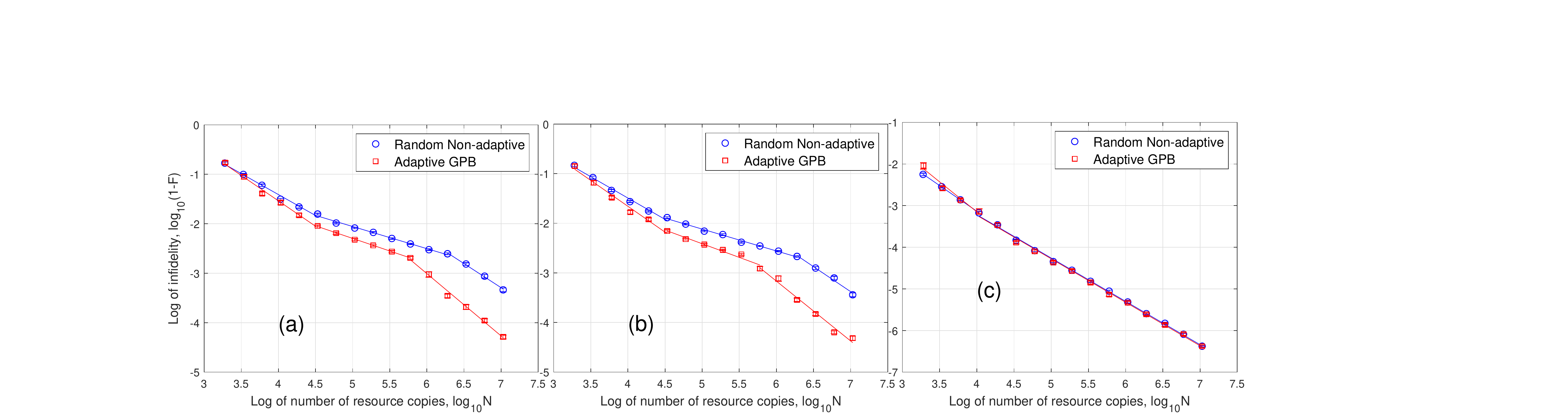}
\caption{Infidelity for three-valued detectors in \eqref{three0.6}. (a) $ P_1 $; (b) $ P_2 $; (c) $ P_3 $.}
	\label{three30.6}
\end{figure*}

The results for $ P_1 $ and $ P_2 $ can also be divided into three segments as shown in Fig. \ref{three30.6} and we have explained for binary detectors. For $ P_3 $, all the eigenvalues are far from zero and hence, both tomography can reach $ O(1/N) $.

\subsection{Adaptive QDT using coherent states}
\begin{figure}
	\centering
	\subfigure{
		\begin{minipage}[b]{0.5\textwidth}
			\includegraphics[width=1\textwidth]{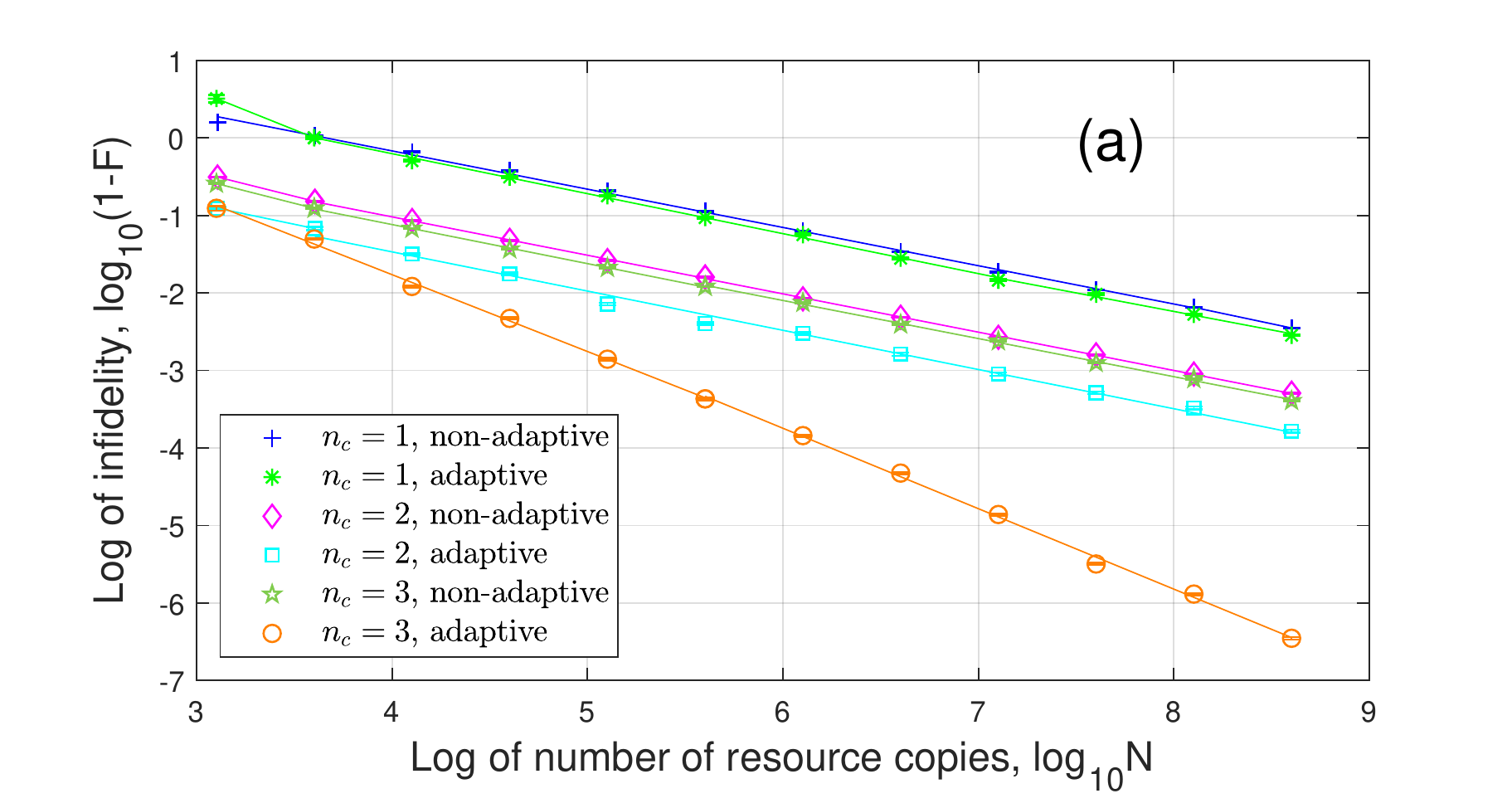}
		\end{minipage}
	}
	\subfigure{
		\begin{minipage}[b]{0.5\textwidth}
			\includegraphics[width=1\textwidth]{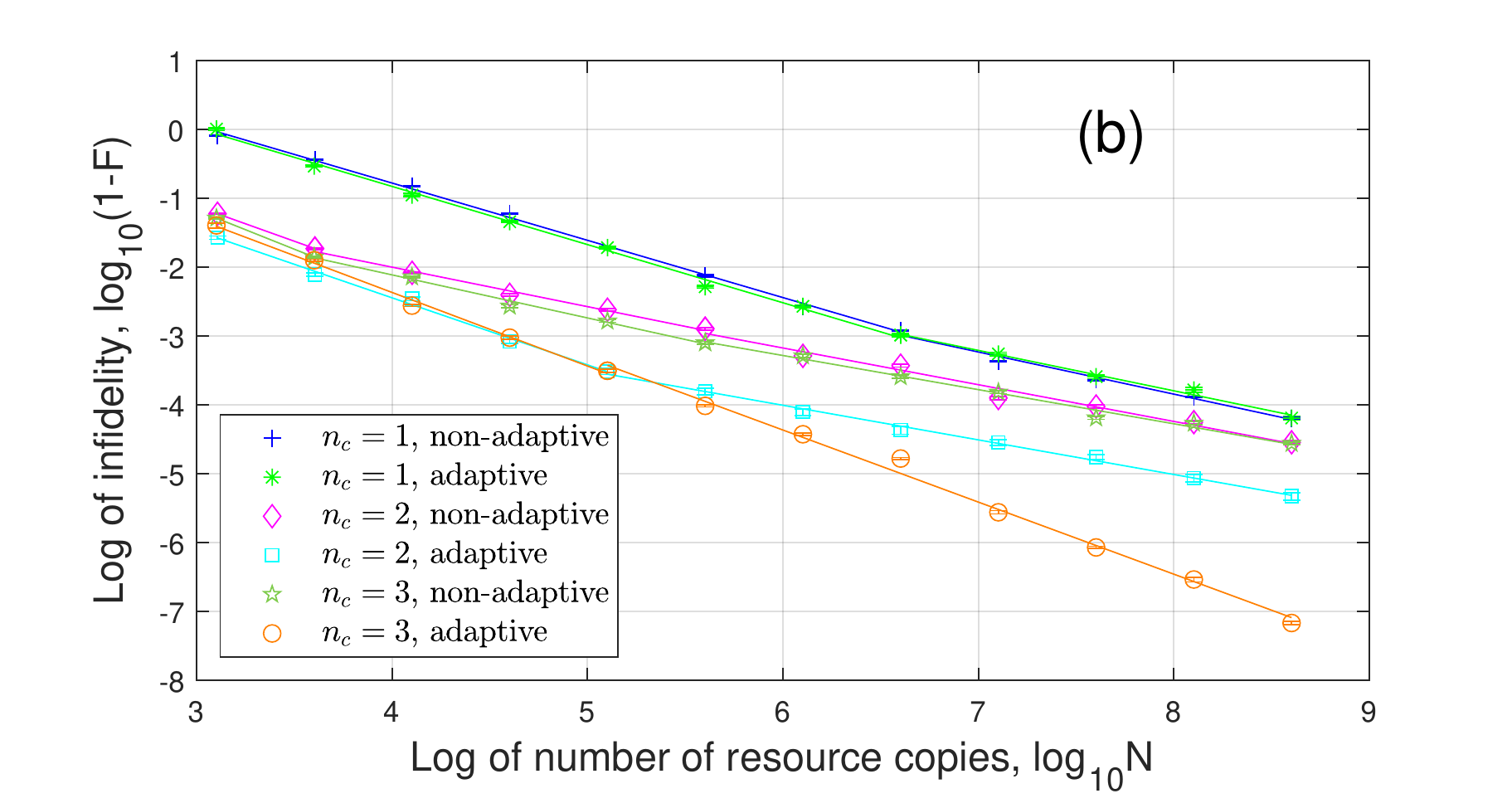}
		\end{minipage}
	}
	\caption{Superposition of $ n_c $-coherent states for \eqref{binary1} where $ n_c=1,2,3 $. (a) $ P_1 $; (b) $ P_2 $.}
		\label{co1}
\end{figure}

Since the adaptive GPB states can improve the infidelity for all detectors if they have zero or near-zero eigenvalues, in this part, we use the superposition of coherent states to approximate the ideal adaptive GPB states. We consider binary detectors as \eqref{binary1}. We use $ n_c $-coherent states as shown in Fig. \ref{co1} where $ n_c=1,2,3 $.
For given $ n_c $, the adaptive tomography performance is usually better than non-adaptive one. As $ n_c $ increases, the approximation error decreases.
When the resource number $ N $ is not large enough to distinguish the approximation error, the infidelity scales close to $ O(1/N) $, like $ n_c=2 $, adaptive for $ P_2 $ in Fig. \ref{co1}(b) when $ N\leq 10^5$. When $ N\geq 10^5$, the infidelity scales to $ O(1/\sqrt{N}) $ because of the approximation error.

\begin{figure}
	\centering
	\subfigure{
		\begin{minipage}[b]{0.5\textwidth}
			\includegraphics[width=1\textwidth]{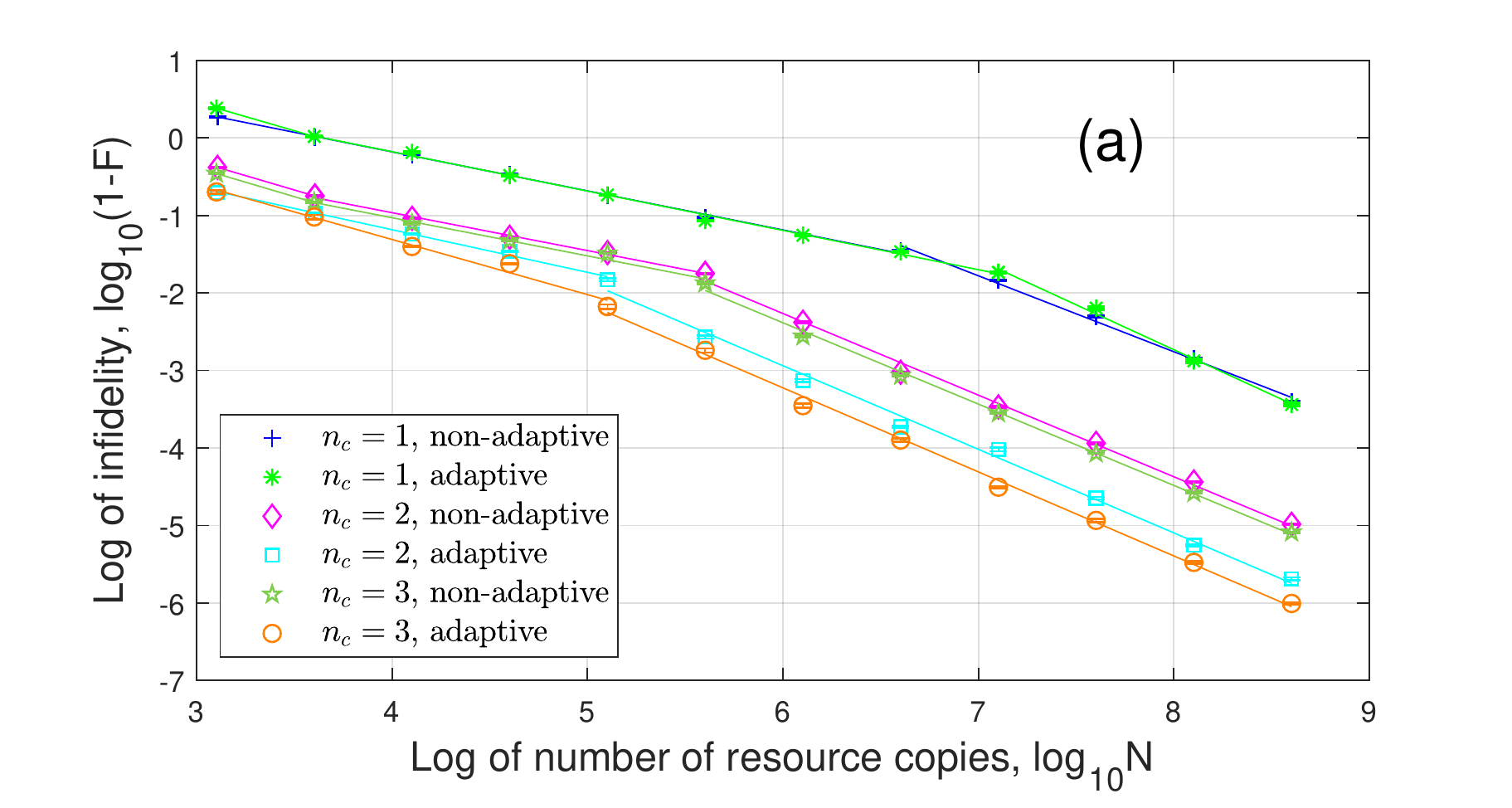}
		\end{minipage}
	}
	\subfigure{
		\begin{minipage}[b]{0.5\textwidth}
			\includegraphics[width=1\textwidth]{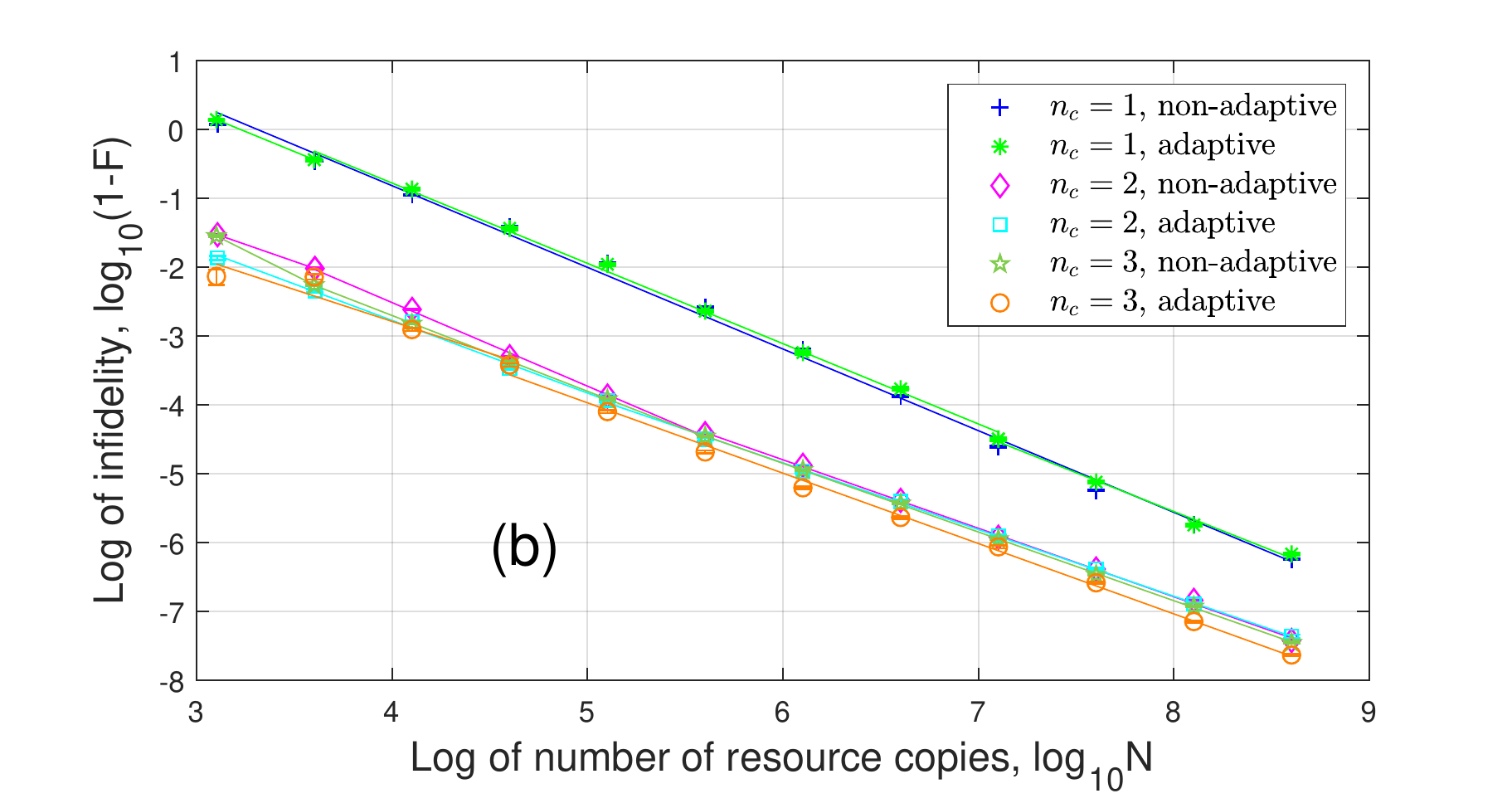}
		\end{minipage}
	}
	\caption{Superposition of $ n_c $-coherent states for \eqref{binary2} where $ n_c=1,2,3 $. (a) $ P_1 $; (b) $ P_2 $.}
	\label{co2}
\end{figure}

For binary detectors as \eqref{binary2}, the similar results are shown in Fig. \ref{co2}. 
For $ P_1 $, all the curves can be roughly divided into two segments. For 1-coherent state, the infidelity scales as $ O(1/\sqrt{N}) $ when $ N\leq10^7 $ and scales as $ O(1/N) $ when $ N\geq10^7 $. For 2,3-coherent states, the infidelity scales as $ O(1/\sqrt{N}) $ when $ N\leq10^6 $ and scales as $ O(1/N) $ when $ N\geq10^6 $. This result is similar to Fig. \ref{dis22} without the first segment. For $ P_2 $, the infidelity scales as $ O(1/N) $ because all the eigenvalues are significantly larger than zero.
\section{Conclusion}

In this paper we have investigated how to optimize the probe states in quantum detector tomography. We have characterized the optimal probe state sets based on minimizing the UMSE and minimizing the condition number. We have proven that SIC and MUB states are optimal. In the adaptive scenario we have proposed a two-step strategy to adaptively optimize the probe states, and proven that our strategy can improve the modified infidelity from $O(1/\sqrt{N})$ to $O(1/N)$ under certain conditions. Numerical examples were presented to demonstrate the effectiveness of our strategies.

\begin{ack}
	Y. Wang would like to thank Prof. Howard M. Wiseman for the helpful discussion.
	We thank the anonymous referees and the Associate Editor for the constructive comments.
\end{ack}

\appendix
\renewcommand\thesection{\appendixname~\Alph{section}}
\renewcommand\theequation{\Alph{section}.\arabic{equation}}
\section{Several two-qubit measurements and the induced probe states }\label{appenA}
Five MUB measurement sets for two-qubit quantum states \cite{PhysRevLett.105.030406}  are
\begin{small}
	\begin{equation}
		\left\{\!|\psi_{n}^{(\text{MUB})}\rangle\!\right\}\!=\!\left\{\!\left\{\left|\psi_{n}^{A}\right\rangle\!\right\}\!,\!\left\{\left|\psi_{n}^{B}\right\rangle\!\right\}\!,\!\left\{\left|\psi_{n}^{C}\right\rangle\!\right\}\!,\!\left\{\left|\psi_{n}^{D}\right\rangle\!\right\}\!,\!\left\{\left|\psi_{n}^{E}\right\rangle\!\right\}\!\right\},
	\end{equation}
\end{small}
and
\begin{equation}
\label{mubbase}
\begin{aligned}
&\left\{\left|\psi_{n}^{A}\right\rangle\right\}=\{|00\rangle,|01\rangle,|10\rangle,|11\rangle\},\\
&\left\{\left|\psi_{n}^{B}\right\rangle\right\}=\{|R\pm\rangle,|L\pm\rangle\},\\
&\left\{\left|\psi_{n}^{C}\right\rangle\right\}=\{|\pm R\rangle,|\pm L\rangle\},\\
&\left\{\left|\psi_{n}^{D}\right\rangle\right\}=\left\{\frac{1}{\sqrt{2}}(|R 0\rangle \pm \mathrm i|L 1\rangle), \frac{1}{\sqrt{2}}(|R 1\rangle \pm \mathrm i|L 0\rangle)\right\},\\
&\left\{\left|\psi_{n}^{E}\right\rangle\right\}=\left\{\frac{1}{\sqrt{2}}(|R R\rangle \pm \mathrm i|L L\rangle), \frac{1}{\sqrt{2}}(|R L\rangle \pm \mathrm i|L R\rangle)\right\}.
\end{aligned}
\end{equation}
where $|\pm\rangle=(|0\rangle \pm|1\rangle) / \sqrt{2}$, $|R\rangle=(|0\rangle-i|1\rangle) / \sqrt{2}$, and
$|L\rangle=(|0\rangle+i|1\rangle) / \sqrt{2}$.
This protocol was applied in the QST experiment in \cite{PhysRevLett.105.030406}. In this paper, $ P_n=|\psi_{n}^{(\text{MUB})}\rangle\langle\psi_{n}^{(\text{MUB})}| $ is a projection measurement and $ \rho_n=|\psi_{n}^{(\text{MUB})}\rangle\langle\psi_{n}^{(\text{MUB})}| $ is called a MUB state.

For two-qubit SIC POVM, one expression \cite{Bengtsson2010}, ignoring overall phases and normalization, is
\begin{small}
\begin{equation}
\label{sicm}
\left(\begin{array}{rrrrrrrrrrrrrrrr}
x & x & x & x & \mathrm i & \mathrm i & -\mathrm i & -\mathrm i & \mathrm i &\mathrm i & -\mathrm i & - \mathrm i & \mathrm i & \mathrm i & -\mathrm i & -\mathrm i \\
1 & 1 & -1 & -1 & x & x & x & x & \mathrm i & -\mathrm i &\mathrm i & -\mathrm i & 1 & -1 & 1 & -1 \\
1 & -1 & 1 & -1 & 1 & -1 & 1 & -1 & x & x & x & x & -\mathrm i & \mathrm i &\mathrm i & -\mathrm i \\
1 & -1 & -1 & 1 & -\mathrm i & \mathrm i & \mathrm i & -\mathrm i & -1 & 1 & 1 & -1 & x & x & x & x
\end{array}\right),
\end{equation}	
\end{small}
where $ x=\sqrt{2+\sqrt{5}} $ and $\left|\psi_{n}^{(\text{SIC})}\right\rangle$ is the $ n $-th column of \eqref{sicm}. In this paper,  the set of $ \rho_i=\frac{\left|\psi_{n}^{(\text{SIC})}\right\rangle\left\langle\psi_{n}^{(\text{SIC})}\right|}{\operatorname{Tr}\left(\left|\psi_{n}^{(\text{SIC})}\right\rangle\left\langle\psi_{n}^{(\text{SIC})}\right|\right)} $ consists of SIC states.

For Cube measurement \cite{PhysRevA.78.052122}, it is based on the projections onto all the 36 tensor
products of the eigenstates of the standard single-qubit Pauli
operators
\begin{equation}
\label{cubebase}
\left\{\left|\psi_{n}^{(\text{Cube})}\right\rangle\right\}=\{|0\rangle,|1\rangle,|+\rangle,|-\rangle,|R\rangle,|L\rangle\}^{\otimes 2}.
\end{equation}
In this paper,  $ \rho_n=|\psi_{n}^{(\text{Cube})}\rangle\langle\psi_{n}^{(\text{Cube})}| $ is a Cube state which is a product state of two single-qubit MUB states.

 For GPB (generalized Pauli basis) states, they are
\begin{equation}
\label{standard1}
\rho_{i}^{z}=|i\rangle\langle i|,
\end{equation}
where $ i=1, \ldots, d $ and 
\begin{equation}
\label{standard2}
\rho_{i j}^{x}=(|i\rangle\langle j|+| j\rangle\langle i|+| i\rangle\langle i|+| j\rangle\langle j|) / 2,
\end{equation}
\begin{equation}
\label{standard3}
\rho_{i j}^{y}=(-\mathrm i|i\rangle\langle j|+\mathrm i| j\rangle\langle i|+| i\rangle\langle i|+| j\rangle\langle j|) / 2,
\end{equation}
where  $1 \leq i<j \leq d$. The set $\{|i\rangle\}$ of states with $i=1, \ldots, d$  is an arbitrary orthonormal basis. The total number of these states is $ d^2 $. To generate new GPB states in Step 2 of adaptive QDT, we replace the set $\{|i\rangle\}_{i=1}^{d}$ by $\left\{\left|\lambda_{i}\right\rangle\right\}_{i=1}^{d}$  which are the eigenvectors of Step 1 estimation.

For random pure states, we  choose $ 32 $ random pure states using the  algorithm in \cite{Zyczkowski_1994,MISZCZAK2012118} which are uniformly distributed according to Haar measure.

\section{Optimal one-qubit states}\label{appenB}
For one-qubit state, we can expand them in Pauli matrices as
\begin{equation}
	\rho_{j}=\frac{I}{2}+\phi_j^x \frac{\sigma_{x}}{\sqrt{2}}+\phi_j^y \frac{\sigma_{y}}{\sqrt{2}}+\phi_j^z \frac{\sigma_{z}}{\sqrt{2}},
\end{equation}
where $ (\phi_j^x)^2+(\phi_j^y)^2+(\phi_j^z)^2 \leq \frac{1}{2} $.
The parameterization $ X^TX $ for $ M $ one-qubit states are 
\begin{footnotesize}
\begin{equation}
	X^{T} X=\left(\begin{array}{cccc}
		\frac{M}{2} & \frac{\sqrt{2}}{2} \sum_{j=1}^{M} \phi_j^x & \frac{\sqrt{2}}{2} \sum_{j=1}^{M} \phi_j^y & \frac{\sqrt{2}}{2} \sum_{j=1}^{M} \phi_j^z \\
		\frac{\sqrt{2}}{2} \sum_{j=1}^{M} \phi_j^x & \sum_{j=1}^{M} (\phi_j^x)^{2} & \sum_{j=1}^{M} \phi_j^x \phi_j^y & \sum_{j=1}^{M} \phi_j^x \phi_j^z \\
		\frac{\sqrt{2}}{2} \sum_{j=1}^{M} \phi_j^y & \sum_{j=1}^{M} \phi_j^x \phi_j^y & \sum_{j=1}^{M} (\phi_j^y)^{2} & \sum_{j=1}^{M} \phi_j^y \phi_j^z \\
		\frac{\sqrt{2}}{2} \sum_{j=1}^{M} \phi_j^z & \sum_{j=1}^{M} \phi_j^x \phi_j^z & \sum_{j=1}^{M} \phi_j^y \phi_j^z & \sum_{j=1}^{M} (\phi_j^z)^{2}
	\end{array}\right).
\end{equation}
\end{footnotesize}
If these $ M $ states are OPS, by Theorem \ref{theorem1}, we have
\begin{equation}
	\sum_{j=1}^{M} \phi_j^x=\sum_{j=1}^{M} \phi_j^y=\sum_{j=1}^{M} \phi_j^z=0,
\end{equation}
\begin{equation}
\sum_{j=1}^{M} \phi_j^x \phi_j^y=\sum_{j=1}^{M} \phi_j^x \phi_j^z=\sum_{j=1}^{M} \phi_j^y \phi_j^z=0,	
\end{equation}
\begin{equation}
\sum_{j=1}^{M} (\phi_j^x)^{2}=\sum_{j=1}^{M} (\phi_j^y)^{2}=\sum_{j=1}^{M} (\phi_j^z)^{2}=\frac{M}{6},
\end{equation}
and
\begin{equation}
	(\phi_j^x)^2+(\phi_j^y)^2+(\phi_j^z)^2 = \frac{1}{2},\forall 1 \leq j \leq M.
\end{equation}
We conjecture that real solutions exists for these equations (i.e.,  OPS exist) if and only if $ M=4,6,8,12 $ and $ 20 $, corresponding to the five platonic solids on the Bloch sphere. We prove this is true for $ M=5 $.

For $ M=5 $, if there exist OPS, then we construct a $ 5\times4 $ parameterization matrix of probe states  $ X$  as
\begin{equation}
X=\left(\begin{array}{cccc}
	\frac{1}{\sqrt{5}} & \sqrt{\frac{6}{5}} \phi_1^x & \sqrt{\frac{6}{5}} \phi_1^y & \sqrt{\frac{6}{5}} \phi_1^z \\
	\vdots & \vdots & \vdots & \vdots \\
	\frac{1}{\sqrt{5}} & \sqrt{\frac{6}{5}} \phi_5^x & \sqrt{\frac{6}{5}} \phi_5^y & \sqrt{\frac{6}{5}} \phi_5^z
\end{array}\right),
\end{equation}
where the columns of $ X $ are orthogonal to each other and their 2-norm is all $ 1 $. The 2-norm of each row of $ X$ is $ \sqrt{\frac{4}{5}} $. Using Gram-Schmidt process, there must exist a vector $ b $ such that $ [X,b] $ is an orthogonal matrix. By considering its row-norm, we know each element of $ b $ squares to $\frac{1}{5}$, and thus is either $ \sqrt{\frac{1}{5}} $  or $ -\sqrt{\frac{1}{5}} $. However, such $ b $ can never be  orthogonal to the first column of $ X $, contradicting the fact that $ [X,b] $ is an  orthogonal matrix. Thus, there do not exist OPS for $ M=5 $.

\section{Proof of Theorem \ref{theorem3}}\label{appenC}
\begin{pf}

\textit{a)   Preliminary calculations} 

The parameterization error in one probe state is
\begin{equation}
	\label{e1}
	\begin{array}{l}
		\left\|\rho_{j}-\hat{\rho}_{j}\right\|^{2}=\operatorname{Tr}\left[\left(\rho_{j}^{\dagger}-\hat{\rho}_{j}^{\dagger}\right)\left(\rho_{j}-\hat{\rho}_{j}\right)\right] \\
		=\operatorname{Tr}\left[\left(\sum_{b=1}^{d^{2}}\left(\phi_{j}^{b}-\hat{\phi}_{j}^{b}\right) \Omega_{b}\right)^{\dagger}\left(\sum_{b=1}^{d^{2}}\left(\phi_{j}^{b}-\hat{\phi}_{j}^{b}\right) \Omega_{b}\right)\right] \\
		=\sum_{b=1}^{d^{2}}\left|\phi_{j}^{b}-\hat{\phi}_{j}^{b}\right|^{2}=\left\|\phi_{j}-\hat{\phi}_{j}\right\|^{2}.
	\end{array}
\end{equation}
Hence, the total parameterization error in $ X $ is
\begin{equation}
	\label{e2}
	\|X-\hat{X}\|^2=\sum_{j=1}^{M}\left\|\phi_{j}-\hat{\phi}_{j}\right\|^2\leq M\varepsilon^2.
\end{equation}
For the error in $ X^TX $, we thus have
\begin{equation}
	\label{e3}
	\begin{array}{l}
		\left\|X^{T} X-\hat{X}^{T} \hat{X}\right\| \leq\left\|X^{T} X-X^{T} \hat{X}\right\|+\left\|X^{T} \hat{X}-\hat{X}^{T} \hat{X}\right\| \\
		\leq\|X\| \cdot\|X-\hat{X}\|+\|\hat{X}\| \cdot\|X-\hat{X}\| \\
		=(\|X\|+\|\hat{X}\|)\|X-\hat{X}\| \\
		\leq 2\sqrt{M}\|X-\hat{X}\|\leq 2M\varepsilon,
	\end{array}
\end{equation}
where $ \|X\|^{2}=\sum_{j=1}^{M}\left\|\phi_{j}\right\|^{2}=\sum_{j=1}^{M} \operatorname{Tr}\left(\rho^{\dagger} \rho\right)=\sum_{j=1}^{M} \operatorname{Tr}\left(\rho^{2}\right) \leq M $ and similarly $ \|\hat{X}\|\leq \sqrt{M} $.

\textit{b) Error in the UMSE} 

We introduce Weyl's perturbation theorem,
which can be found in \cite{matrix},
\begin{lemma}\label{lemma1}
	\cite{matrix} Let $A, B$ be Hermitian matrices with eigenvalues $\lambda_{1}(A) \geq \cdots \geq \lambda_{n}(A)$ and $\lambda_{1}(B) \geq \cdots \geq \lambda_{n}(B)$, respectively. Then
	\begin{equation}
		\max _{j}\left|\lambda_{j}(A)-\lambda_{j}(B)\right| \leq\|A-B\|.
	\end{equation}	
\end{lemma}
Lemma \ref{lemma1} was for the operator norm, not larger than
its Frobenius norm for any finite dimension square matrix \cite{matrix}. Therefore, it also holds for the
Frobenius norm. Suppose the eigenvalues of $ X^TX $ are $ \lambda_{1}\geq \lambda_{2}\geq \cdots \geq \lambda_{d^2}>0 $ and the eigenvalues of $ \hat X^T\hat X $ are $ \mu_{1}\geq \mu_{2}\geq \cdots \geq \mu_{d^2}>0 $. Thus, using Lemma \ref{lemma1}, we have
\begin{equation}
	\begin{array}{l}
		\left|\lambda_{1}-\mu_{1}\right| \leq\left\|X^{T} X-\hat X^{T} \hat X\right\|\leq2 M \varepsilon, \\
		\left|\lambda_{d^{2}}-\mu_{d^{2}}\right| \leq\left\|X^{T} X-\hat X^{T} \hat X\right\|\leq 2 M \varepsilon,
	\end{array}
\end{equation} 
and therefore,
\begin{equation}
	\begin{array}{l}
		\lambda_{1}-2 M \varepsilon \leq \mu_{1} \leq \lambda_{1}+2 M \varepsilon, \\
		\lambda_{d^{2}}-2 M \varepsilon \leq \mu_{d^{2}} \leq \lambda_{d^{2}}+2 M \varepsilon.
	\end{array}
\end{equation}
For UMSE, we have
\begin{equation}
	\label{e4}
	\begin{array}{l}
		\left|\operatorname{Tr}\left(X^{T} X\right)^{-1}-\operatorname{Tr}\left(\hat{X}^{T} \hat{X}\right)^{-1}\right|=\left|\sum_{i=1}^{d^{2}}\left(\frac{1}{\lambda_{i}}-\frac{1}{\mu_{i}}\right)\right| \\
		\leq \sum_{i=1}^{d^{2}}\left|\frac{1}{\lambda_{i}}-\frac{1}{\mu_{i}}\right| \leq \dfrac{\sum_{i=1}^{d^{2}}\left|\lambda_{i}-\mu_{i}\right|}{(\lambda_{d^{2}}-2 M \varepsilon)^{2}}  \\
		\leq \dfrac{d^{2}\max _{i}\left|\lambda_{i}-\mu_{i}\right|}{(\lambda_{d^{2}}-2 M \varepsilon)^{2}}  \leq \dfrac{d^{2}}{(\lambda_{d^{2}}-2 M \varepsilon)^{2}}\left\|X^{T} X-\hat{X}^{T} \hat{X}\right\|,
	\end{array}
\end{equation}
because $ \lambda_{i} \mu_{i} \geq \min \left\{\lambda_{d^{2}}^{2}, \mu_{d^{2}}^{2}\right\} \geq\left(\lambda_{d^{2}}-2 M \varepsilon\right)^{2}>0 $ for $ i=1,2,\cdots,d^2 $.

We combine \eqref{e1}, \eqref{e2}, \eqref{e3} and \eqref{e4} to  obtain
\begin{equation}
	\begin{aligned}
	&\frac{(n-1)M}{4 N}\big|\operatorname{Tr}\left(X^{T} X\right)^{-1}-\operatorname{Tr}\left(\hat{X}^{T} \hat{X}\right)^{-1}\big|\\
	\leq &\frac{2(n-1)d^2M^2\varepsilon}{4N{\left(\lambda_{d^{2}}-2 M \varepsilon\right)^{2}}}.	
	\end{aligned}
\end{equation}

\textit{c) Error in the condition number} 

For condition number, we have
\begin{equation}
	\label{conditionerror}
	\begin{aligned}
		&\quad\left|\sqrt{\frac{\lambda_{1}}{\lambda_{d^{2}}}}-\sqrt{\frac{\mu_{1}}{\mu_{d^{2}}}}\right|=\frac{\left|\sqrt{\lambda_{1} \mu_{d^{2}}}-\sqrt{\mu_{1} \lambda_{d^{2}}}\right|}{\sqrt{\lambda_{d^{2}} \mu_{d^{2}}}}\\
		&=\frac{\left|\lambda_{1} \mu_{d^{2}}-\lambda_{d^{2}} \mu_{1}\right|}{\sqrt{\lambda_{d^{2}} \mu_{d^{2}}}\left(\sqrt{\lambda_{1} \mu_{d^{2}}}+\sqrt{\lambda_{d^{2}} \mu_{1}}\right)}\\
		& \leq \frac{\left|{\lambda_{1}\left(\lambda_{d^{2}}+2 M \varepsilon\right)}-{\lambda_{d^{2}}\left(\lambda_{1}-2 M \varepsilon\right) }\right|}{2\left(\lambda_{d^{2}}-2 M \varepsilon\right)^2}\\
		&=\frac{ M \left(\lambda_{1}+\lambda_{d^{2}}\right)\varepsilon}{\left(\lambda_{d^{2}}-2 M \varepsilon\right)^2}.
	\end{aligned}
\end{equation}\hfill $\Box$
\end{pf}

\section{On the distortion of $ F_0 $ in \eqref{infide}}\label{appenD}

According to the definition, distortion happens when there exists $ \left\{\hat{P}_{i}\right\}_{i=1}^{n} \neq\left\{P_{i}\right\}_{i=1}^{n} $ such that $ F_0\left(\hat{P}_{i}, P_{i}\right)=1 $ for all $ i $, which means $ \frac{\hat{P}_{i}}{\operatorname{Tr}\left(\hat{P}_{i}\right)}=\frac{P_{i}}{\operatorname{Tr}\left(P_{i}\right)} $ for all $ i $. This indicates that the vector $ \hat{a}_{0}=\left[\frac{\operatorname{Tr}\left(\hat{P}_{1}\right)}{\operatorname{Tr}\left(P_{1}\right)}, \cdots, \frac{\operatorname{Tr}\left(\hat{P}_{n}\right)}{\operatorname{Tr}\left(P_{n}\right)}\right]^{T} $ and $a_0=[1,\cdots,1]^T$ are different. Note that $ \sum_{i=1}^{n} \hat{P}_{i}=\sum_{i=1}^{n} \hat{a}_{0i} P_{i}=I $ and $ \sum_{i=1}^{n} P_{i}=I $. Hence, the following linear equation
\begin{equation}
	\label{c1}
	\left[\operatorname{vec}\left( P_{1}\right), \cdots, \operatorname{vec}\left( P_{n}\right)\right]a=\operatorname{vec}(I),
\end{equation}
has at least two different solutions $ a=\hat{a}_0 $ and $ a=a_0 $.  Therefore, $\operatorname{Rank}\left[\operatorname{vec}\left(P_{1}\right), \cdots, \operatorname{vec}\left( P_{n}\right)\right]<n$, and thus $ \left\{ P_{i}\right\}_{i=1}^{n} $ are linearly dependent.

Conversely, if $ \left\{ P_{i}\right\}_{i=1}^{n} $ are linearly dependent, \eqref{c1} will have non-trivial homogeneous general solutions, among which we randomly pick a non-zero solution $a=[v_1,\cdots,v_n]^T$. Since  \eqref{c1} already has a trivial solution $a_0$,  we choose $w>0$ small enough, such that $a'=[1+wv_1,\cdots,1+wv_n]^T$ has all the elements strictly positive. Hence, $a'$ is another set of positive numbers not equal to $[1,\cdots,1]^T$ such that $\sum_{i=1}^n a'_i P_i=I$. Let the estimation be $ \hat{P}_i=a'_i P_i$,  and distortion happens.

Notice that there are $ d^2 $ real degrees of freedom in $ {P_i} $, and hence the row rank of LHS of \eqref{c1} is at most $ d^2 $. If $d^2<n $, we must have $\operatorname{Rank}\left[\operatorname{vec}\left( P_{1}\right), \cdots, \operatorname{vec}\left( P_{n}\right)\right]<n$ and there must exist distortion. 

\section{Lower bound of the new fidelity \eqref{infide2}}\label{appenE}
 We show a tight lower bound of our new fidelity \eqref{infide2} as $ \frac{1}{d}-1 $. One can arbitrarily approximate this lower bound but cannot reach it.
 
We assume that the spectral decomposition of $ P_i $ and $ \hat P_i $ are $P_{i}=U_{i} \Delta_i U_{i}^{\dagger}$ where $\Delta_{i}=$
$\operatorname{diag}\left(\delta_{1}, \cdots, \delta_{d}\right)$, $0\leq\delta_{1}\leq\cdots\leq \delta_{d}\leq1   $  and $\hat P_{i}=\hat U_{i} \hat\Delta_i \hat U_{i}^{\dagger}$
where $\hat{\Delta}_i=\operatorname{diag}\left(\hat{\delta}_{1}, \cdots, \hat{\delta}_{d}\right)$, $1\geq\hat\delta_{1}\geq\cdots\geq \hat\delta_{d}\geq0   $. Define ${\Delta}_{i}^{\downarrow}=\operatorname{diag}\left({\delta}_{d}, \cdots, {\delta}_{1}\right)$. According to Theorem 2.1 in \cite{Zhang_2014}, we have 
\begin{equation}
	F_0\left(\hat{\Delta}_{i}, \Delta_{i}\right) \leq F_0\left(\hat{P}_{i}, P_{i}\right) \leq F_0\left(\hat{\Delta}_{i}, \Delta_{i}^{\downarrow}\right).
\end{equation}
Therefore, 
\begin{equation}
	F\left(\hat{\Delta}_{i}, \Delta_{i}\right) \leq F\left(\hat{P}_{i}, P_{i}\right) \leq F\left(\hat{\Delta}_{i}, \Delta_{i}^{\downarrow}\right),
\end{equation}
and we only need to consider the lower bound of $F\left(\hat{\Delta}_{i}, \Delta_{i}\right)$. We have
	\begin{equation}\label{fi}
		\begin{aligned}
			F\left(\hat{\Delta}_{i}, \Delta_{i}\right)&\!=\!\frac{\left(\sum_{j=1}^{d} \sqrt{\delta_{j} \hat{\delta}_{j}}\right)^{2}}{\left(\!\sum_{j=1}^{d} \delta_{j}\!\right)\left(\!\sum_{j=1}^{d} \hat{\delta}_{j}\!\right)}\!-\!\frac{\left(\!\sum_{j=1}^{d}\left(\delta_{j}-\hat{\delta}_{j}\right)\!\right)^{2}}{d^{2}}\\
			&\!=\!\frac{\sum_{j=1}^{d} \delta_{j} \hat{\delta}_{j}+\sum_{j \neq k} \sqrt{\delta_{j} \hat{\delta}_{j} \delta_{k} \hat{\delta}_{k}}}{\left(\sum_{j=1}^{d} \delta_{j}\right)\left(\sum_{j=1}^{d} \hat{\delta}_{j}\right)}\\
			&\!\quad-\frac{\left(\sum_{j=1}^{d}\left(\delta_{j}-\hat{\delta}_{j}\right)\right)^{2}}{d^{2}} \\
			&\!\geq\!\frac{\sum_{j=1}^{d} \delta_{j} \hat{\delta}_{j}}{\left(\!\sum_{j=1}^{d} \delta_{j}\!\right)\left(\!\sum_{j=1}^{d} \hat{\delta}_{j}\!\right)}-\frac{\left(\!\sum_{j=1}^{d}\left(\delta_{j}-\hat{\delta}_{j}\right)\!\right)^{2}}{d^{2}}. \\
		\end{aligned}
	\end{equation}
If $ \sum_{j=1}^{d} \delta_{j}\geq \sum_{j=1}^{d} \hat\delta_{j} $, we have
\begin{equation}\label{da1}
	\begin{aligned}
		F\left(\hat{\Delta}_{i}, \Delta_{i}\right)&\!\geq\!\frac{\sum_{j=1}^{d} \delta_{j} \hat{\delta}_{j}}{\left(\!\sum_{j=1}^{d} \delta_{j}\!\right)\left(\!\sum_{j=1}^{d} \hat{\delta}_{j}\!\right)}\!-\!\frac{\left(\!\sum_{j=1}^{d}\left(\delta_{j}-\hat{\delta}_{j}\right)\!\right)^{2}}{d^{2}} \\
		&\!\geq\! \frac{\delta_{1}\left(\sum_{j=1}^{d} \hat{\delta}_{j}\right)}{\left(\sum_{j=1}^{d} \delta_{j}\right)\left(\sum_{j=1}^{d} \hat{\delta}_{j}\right)}-\frac{\left(\sum_{j=1}^{d} \delta_{j}\right)^{2}}{d^{2}}\\
		&\!=\!\frac{\delta_{1}}{\sum_{j=1}^{d} \delta_{j}}-\frac{\left(\sum_{j=1}^{d} \delta_{j}\right)^{2}}{d^{2}}\\
		&\!\geq\!\frac{\delta_{1}}{\delta_{1}+d-1}-\frac{\left(\delta_{1}+d-1\right)^{2}}{d^{2}} \geq \frac{1}{d}-1,
	\end{aligned}
\end{equation}
where the last inequality can be obtained by analyzing the derivative on $\delta_1$.
Similarly, if $ \sum_{j=1}^{d} \delta_{j}< \sum_{j=1}^{d} \hat\delta_{j} $, we have
\begin{equation}\label{da2}
	\begin{aligned}
		F\left(\hat{\Delta}_{i}, \Delta_{i}\right)&\geq\frac{\hat\delta_{d}}{\sum_{j=1}^{d} \hat\delta_{j}}-\frac{\left(\sum_{j=1}^{d} \hat\delta_{j}\right)^{2}}{d^{2}}\geq\frac{1}{d}-1.
	\end{aligned}
\end{equation}
The minimum is achieved if and only if one of $ P_{i} $ and $ \hat P_{i} $ is an identity matrix and the other is a zero matrix, which cannot happen in practice. Hence, $ \frac{1}{d}-1 $ cannot be achieved. Finally, consider for example $ P_i=\operatorname{diag}(1-\tau,1,1,\cdots,1) $ and $\hat P_{i}=\operatorname{diag}({\tau},0,0, \cdots,0)  $ where $ \tau>0 $. Then as $ \tau $ tends to zero, $ F\left(\hat P_i,  P_{i}\right) $ can be arbitrarily close to $ \frac{1}{d}-1 $. Hence, $ \frac{1}{d}-1 $ is a tight but unattainable lower bound.

\section{Detailed calculation of the  first-order term in infidelity}\label{appenF}
Firstly, we analyze the error between real coefficients $ \left\{S_{i j}^{x}, S_{i j}^{y}, S_{k}^{z}\right\} $ and estimated coefficients $ \left\{\hat{S}_{i j}^{x}, \hat{S}_{i j}^{y}, \hat{S}_{k}^{z}\right\} $.

Using a reconstruction algorithm whose MSE scales as $ O(1/N) $ such as  two-stage QDT reconstruction algorithm in \cite{wang2019twostage} and MLE in \cite{PhysRevA.64.024102}, the mean squared error from Step 2 for the detector $ P $ is bounded by $ O\left(\frac{1}{N-N_0} \right)$, and thus
\begin{equation}
\begin{aligned}
&\quad	\mathbb{E}\left(\left\|\hat{P}-P\right\|^{2}\right)\\
=&\sum_{1 \leq i<j \leq d}\mathbb{E}\left(\hat S_{i j}^{x}-S_{i j}^{x}\right)^{2}+\sum_{1 \leq i<j \leq d}\mathbb{E}\left(\hat S_{i j}^{y}-S_{i j}^{y}\right)^{2}\\
&+\sum_{k=1}^{d}\mathbb{E}\left(\hat S_{k}^{z}-S_{k}^{z}\right)^{2}\sim O\left(\frac{1}{N-N_{0}}\right).
\end{aligned}
\end{equation}
Therefore, the errors between real coefficients $ \left\{S_{i j}^{x}, S_{i j}^{y}, S_{k}^{z}\right\} $ and estimated coefficients $ \left\{\hat{S}_{i j}^{x}, \hat{S}_{i j}^{y}, \hat{S}_{k}^{z}\right\} $ all scale as $ O(\frac{1}{\sqrt{N-N_0}}) $. Also, 
\begin{equation}
\mathbb{E}\left|\operatorname{Tr}(\hat{P})-\operatorname{Tr}(P)\right|=\mathbb{E}\left|\sum_{k=1}^{d}\left(\hat S_{k}^{z}-S_{k}^{z}\right)\right| \sim O\left(\frac{1}{\sqrt{N-N_{0}}}\right),
\end{equation}
and thus,
\begin{scriptsize}
	\begin{equation}
	\label{var}
	\begin{aligned}	 	
	&\mathbb{E}\left|\frac{\hat{S}_{i j}^{x}}{\operatorname{Tr}(\hat{P})}-\frac{S_{i j}^{x}}{\operatorname{Tr}(P)}\right|\\
	&=\mathbb{E}\frac{\left|S_{i j}^{x}\operatorname{Tr}(P)+\operatorname{Tr}(P) O\left(\frac{1}{\sqrt{N-N_{0}}}\right)-S_{i j}^{x}\operatorname{Tr}(P)-S_{i j}^{x} O\left(\frac{1}{\sqrt{N-N_{0}}}\right)\right|}{\operatorname{Tr}(P)\left[\operatorname{Tr}(P)+O\left(\frac{1}{\sqrt{N-N_{0}}}\right)\right]}\\
	&\sim O\left(\frac{1}{\sqrt{N-N_{0}}}\right).
	\end{aligned}
	\end{equation}	
\end{scriptsize}

Similarly, $\mathbb{E}\left|\frac{\hat{S}_{i j}^{y}}{\operatorname{Tr}(\hat{P})}-\frac{S_{i j}^{y}}{\operatorname{Tr}(P)}\right| \sim O\left(\frac{1}{\sqrt{N-N_{0}}}\right) $ and\\ $\mathbb{E}\left| \frac{\hat{S}_{k}^{z}}{\operatorname{Tr}(\hat{P})}-\frac{S_{k}^{z}}{\operatorname{Tr}(P)}\right|\sim O\left(\frac{1}{\sqrt{N-N_{0}}}\right) $.

For the first term in RHS of \eqref{first}, if $ t\neq i$ and $t\neq j $, using \eqref{erroreig}, we have
\begin{equation}
\label{sx1}
\begin{aligned}
&\mathbb{E}\left|\frac{\hat{S}_{i j}^{x}}{\operatorname{Tr}(\hat P)}-\frac{S_{i j}^{x}}{\operatorname{Tr}( P)}\right|\left|\left\langle\lambda_{t}\left|\tilde{\sigma}_{i j}^{x}\right| \lambda_{t}\right\rangle\right|\\
&\leq\mathbb{E}\Bigg[\left|\frac{\hat{S}_{i j}^{x}}{\operatorname{Tr}(\hat P)}-\frac{S_{i j}^{x}}{\operatorname{Tr}( P)}\right|\left(\left|\left\langle\lambda_{t} |\tilde{\lambda}_{i}\right\rangle\left\langle\tilde{\lambda}_{j}| \lambda_{t}\right\rangle\right|\right.\\
&+\left.\left|\left\langle\lambda_{t} | \tilde{\lambda}_{j}\right\rangle\left\langle\tilde{\lambda}_{i} | \lambda_{t}\right\rangle\right|\right)\Bigg]\\
&=O\left(\frac{1}{\sqrt{N-N_{0}}}\right)\left[O\left(\frac{1}{{N_{0}}}\right)+O\left(\frac{1}{{N_{0}}}\right)\right]\\
&=O\left(\frac{1}{ N_{0}\sqrt{N-N_{0}}}\right).
\end{aligned}
\end{equation}	
Thus, $ \mathbb{E}\left|\frac{\hat{S}_{i j}^{x}}{\operatorname{Tr}(\hat P)}-\frac{S_{i j}^{x}}{\operatorname{Tr}( P)}\right|\left\langle\lambda_{t}\left|\tilde{\sigma}_{i j}^{x}\right| \lambda_{t}\right\rangle=O\left(\frac{1}{ N_{0}\sqrt{N-N_{0}}}\right) $.

Otherwise, if $ t=i $ or $ t=j $, without loss of generality, we assume $ t=i $. Using \eqref{erroreig}, we have
\begin{equation}
\label{sx2}
\begin{aligned}
&\mathbb{E}\left|\frac{\hat{S}_{i j}^{x}}{\operatorname{Tr}(\hat P)}-\frac{S_{i j}^{x}}{\operatorname{Tr}( P)}\right|\left|\left\langle\lambda_{i}\left|\tilde{\sigma}_{i j}^{x}\right| \lambda_{i}\right\rangle\right|\\
&\leq\mathbb{E}\Bigg[\left|\frac{\hat{S}_{i j}^{x}}{\operatorname{Tr}(\hat P)}-\frac{S_{i j}^{x}}{\operatorname{Tr}( P)}\right|\left(\left|\left\langle\lambda_{i} |\tilde{\lambda}_{i}\right\rangle\left\langle\tilde{\lambda}_{j}| \lambda_{i}\right\rangle\right|\right.\\
&+\left.\left|\left\langle\lambda_{i} | \tilde{\lambda}_{j}\right\rangle\left\langle\tilde{\lambda}_{i} | \lambda_{i}\right\rangle\right|\right)\Bigg]\\
&=O\left(\frac{1}{\sqrt{N-N_{0}}}\right)\left(\left(1+O\left(\frac{1}{N_{0}}\right)\right)O\left(\frac{1}{\sqrt{N_{0}}}\right)\right.\\
&+\left.O\left(\frac{1}{\sqrt {N_{0}}}\right)\left(1+O\left(\frac{1}{N_{0}}\right)\right)\right)\\
&=O\left(\frac{1}{ \sqrt{N_{0}(N-N_{0})}}\right).
\end{aligned}
\end{equation}
Thus, $ \mathbb{E}\left|\frac{\hat{S}_{i j}^{x}}{\operatorname{Tr}(\hat P)}-\frac{S_{i j}^{x}}{\operatorname{Tr}( P)}\right|\left\langle\lambda_{i}\left|\tilde{\sigma}_{i j}^{x}\right| \lambda_{i}\right\rangle=O\left(\frac{1}{ \sqrt{N_{0}(N-N_{0})}}\right) $.

Therefore, the total error for the first term in RHS of \eqref{first} scales as $ O\left(\frac{1}{ \sqrt{N_{0}(N-N_{0})}}\right) $. Similarly, the same scaling holds for the second term in RHS of \eqref{first}. 

Finally, we calculate  the third term in RHS of \eqref{first} and also divide it into two cases. For the first case, we consider  $ 1\leq k\leq r, t\geq r+1 $ and  have 
\begin{equation}
\begin{aligned}
&\mathbb{E}\left|\frac{\hat{S}_{k}^{z}}{\operatorname{Tr}(\hat{P})}-\frac{S_{k}^{z}}{\operatorname{Tr}(P)}\right|\left\langle\lambda_{t}\left|\tilde\sigma_{k}^{z}\right| \lambda_{t}\right\rangle\\
&=O\left(\frac{1}{\sqrt{N-N_{0}}}\right)O\left(\frac{1}{{N_{0}}}\right)\\
&=O\left(\frac{1}{N_{0} \sqrt{N-N_{0}}}\right).	
\end{aligned}
\end{equation}

For the second case where $ r+1\leq k\leq d $, from \cite{PhysRevA.98.012339}  we have
\begin{equation}\label{varsz}
\operatorname{var}\left(\hat{S}_{k}^{z}\right)=\frac{1}{N_{k}^{z}}\left(\tilde{p}_{k}^{z}\left(1-\tilde{p}_{k}^{z}\right)\right),
\end{equation}
where $ {N_{k}^{z}} \sim O\left({{N-N_0}}\right)$  is the resource number used in Step 2 to estimate $ \hat{S}_{k}^{z} $, and ${\tilde{p}_{k}^{z}}$ is
\begin{equation}
\label{sz}
\begin{aligned}
{\tilde{p}_{k}^{z}}&=\operatorname{Tr}\left(P|\tilde{\lambda}_{k}\rangle\langle\tilde{\lambda}_{k}|\right)=\operatorname{Tr}\left(\sum_{i=1}^{d} \lambda_{i}\left|\lambda_{i}\right\rangle\langle\lambda_{i}|\tilde{\lambda}_{k}\rangle\langle\tilde{\lambda}_{k}|\right)\\
&=\sum_{t \neq k} \lambda_{t} \operatorname{Tr}\left(\left|\lambda_{t}\right\rangle\langle\lambda_{t} | \tilde{\lambda}_{k}\rangle\langle\tilde{\lambda}_{k}|\right)+\lambda_{k} \operatorname{Tr}\left(\left|\lambda_{k}\right\rangle\langle\lambda_{k} | \tilde{\lambda}_{k}\rangle\langle\tilde{\lambda}_{k}|\right)\\
&=\sum_{t \neq k}\lambda_{t}\left(O\left(\frac{1}{N_{0}}\right)\right)\\
&=O\left(\frac{1}{N_{0}}\right),
\end{aligned}
\end{equation}
because $ \lambda_{k}=0 $.
Hence, we have 
\begin{equation}
\begin{aligned}
&\operatorname{var}\left(\frac{\hat{S}_{k}^{z}}{\operatorname{Tr}(\hat{P})}\right)=\operatorname{var}\left(\frac{\hat{S}_{k}^{z}}{\operatorname{Tr}(P)+O\left(1 /\left(N-N_{0}\right)\right)}\right)\\
&=\operatorname{var}\left(\frac{\hat{S}_{k}^{z}}{\operatorname{Tr}(P)}-\frac{\hat{S}_{k}^{z} O\left(1 /\left(N-N_{0}\right)\right)}{\operatorname{Tr}(P)\left(\operatorname{Tr}(P)+O\left(1 /\left(N-N_{0}\right)\right)\right)}\right)\\
&\sim \operatorname{var}\left(\frac{\hat{S}_{k}^{z}}{\operatorname{Tr}(P)}\right) \sim \operatorname{var}\left(\hat{S}_{k}^{z}\right) \sim O\left(\frac{1}{N_{0}\left(N-N_{0}\right)}\right).
\end{aligned}
\end{equation}
Therefore, the third term in RHS of \eqref{first} scales as $ O\left(\frac{1}{ \sqrt{N_{0}(N-N_{0})}}\right) $. Hence, the first-order term \eqref{first} scales as 
\begin{equation}
\begin{aligned}
\mathbb{E}\left(\sum_{t=r+1}^{d}\left\langle\lambda_{t}|\Delta_1| \lambda_{t}\right\rangle\right)=O\left(\frac{1}{ \sqrt{N_{0}(N-N_{0})}}\right).
\end{aligned}
\end{equation}

\bibliographystyle{ieeetr}         
\bibliography{detector}

\end{document}